\documentclass[a4paper,fleqn]{cas-dc}
\usepackage[numbers,sort&compress]{natbib}
\usepackage{amsmath,amssymb}
\usepackage{booktabs}
\usepackage{graphicx}
\usepackage[caption=false]{subfig}
\usepackage{textcomp}
\usepackage{xcolor}
\usepackage{siunitx}

\begin{document}

\shortauthors{W. Cheng et al.}
\shorttitle{}
\title[size={\fontsize{18}{23}\selectfont},weight=\bfseries]{Coherent collective amplification of terahertz microbunching\texorpdfstring{\\}{ }seeded by laser frequency beating in relativistic electron beams}

\author[1]{Wencai Cheng}
\author[1]{Yin Kang}
\author[1]{Kaiqing Zhang}
\cormark[1]
\ead{zhangkq@sari.ac.cn}
\author[1]{Zhen Wang}
\author[1]{Duan Gu}
\author[2]{Guangling Chen}
\author[1]{Chao Feng}
\cormark[1]
\ead{fengc@sari.ac.cn}
\author[1]{Haixiao Deng}
\cormark[1]
\ead{denghx@sari.ac.cn}
\cortext[1]{Corresponding author}
\affiliation[1]{organization={Shanghai Advanced Research Institute, Chinese Academy of Sciences},city={Shanghai},postcode={201210},country={China}}
\affiliation[2]{organization={Extreme Light Infrastructure -- Nuclear Physics (ELI-NP), Horia Hulubei National Institute for R\&D in Physics and Nuclear Engineering (IFIN-HH)},addressline={Str. Reactorului No. 30},city={Magurele},postcode={077125},country={Romania}}

\begin{abstract}
High-power, continuously tunable terahertz sources based on free-electron lasers require precise control of electron-beam microstructures. Quantitative prediction of the modulation amplitude across a broad frequency range remains challenging because the laser-induced distribution and its collective evolution must be treated together. To describe this coupled evolution, a nonlinear model is developed for microbunching seeded by a frequency-beating laser heater. The non-Gaussian heater-exit distribution is obtained by optical phase averaging and propagated through multistage compression in six-dimensional phase space. Source-induced correlations are retained, with space charge, coherent synchrotron radiation, and radio-frequency wakefields evaluated self-consistently. The wavelength-dependent bunching response is thereby connected to the laser beat frequency, laser power, and compression partition. The short-wavelength double peak is found to be governed mainly by longitudinal space charge, while bunching near a selected wavelength can be enhanced by redistributing compression at fixed total compression. The predicted source modulation and downstream response are benchmarked against Elegant and IMPACT-Z, respectively. Experimental measurements of the wavelength-dependent bunching factor at different laser pulse energies are found to agree well with the trends predicted by theory and simulation. The resulting framework provides a computationally efficient and predictive route to optimizing electron-beam microbunching for tunable, high-power terahertz generation at free-electron-laser facilities.
\end{abstract}


\maketitle

\section{Introduction}

Terahertz (THz) radiation directly accesses low-energy excitations in matter,  providing powerful tools for ultrafast spectroscopy, nonlinear light--matter studies, and quantum-material control~\cite{Kubacka2014,Mitrano2016,Li2019,Basini2024,Salen2019}. The development of THz applications imposes new requirements on the characteristics of terahertz sources---including high pulse energy, narrow bandwidth, and broad frequency tunability---which remain challenging for most existing sources.
Recently, the new developed coherent THz free-electron lasers (FELs) sources driven by precise modulated electron beams provides a new route for the generation of high power THz with great tunability ~\cite{Kang2025,Liang2026Superradiant}. Laser-imprinted longitudinal phase-space structures can be converted into density modulation via magnetic compression, thereby enhancing emission at selected frequencies~\cite{Zhang2017THz,Liang2023}. Building upon this scheme, high peak and average power, continuous spectral coverage, millijoule-level THz emissions have been demonstrated~\cite{Krasilnikov2025THz,Kang2025,Liang2026Superradiant,Liang2026Waveform}. Furthermore, intrinsically synchronized X-ray and narrow-band THz FEL pulses have been generated from a single pre-bunched beam, where the long-wavelength density modulation enhances X-ray output, survives the interaction, and subsequently drives a downstream THz radiator~\cite{Kang2026SelfSynchronized}. To achieve continuous frequency tuning, a frequency-beating laser heater can be implemented, where   imprints a variable-period slice-energy distribution for subsequent compression-driven density modulation~\cite{Brynes2020a,Kang2025}.

The bunching spectrum is shaped by both the laser-induced distribution and collective fields during compression. Longitudinal space charge (LSC) and coherent synchrotron radiation (CSR) can amplify small density perturbations and drive microbunching instability. In x-ray FELs~\cite{Yang2023SelfModulation,Huang2023MING,Huang2026ModeLocking}, this growth is usually suppressed by laser heating~\cite{PhysRevSTAB.7.074401,PhysRevSTAB.13.020703}. Controlled amplification can instead be initiated by a temporally modulated heater pulse~\cite{Brynes2020a}. For THz generation, the useful density modulation must therefore be determined together with the collective response and phase mixing. Its amplitude cannot be inferred from the laser beat frequency and total compression alone.

Collective gain and Landau damping can be calculated with linearized Vlasov and semianalytical models~\cite{Tsai2015Vlasov,DiMitri2025Models}. Nonlinear longitudinal evolution can also be resolved by direct Vlasov integration~\cite{Venturini2007}, while multidimensional beam dynamics can be followed by particle tracking~\cite{Borland2000Elegant,PhysRevSTAB.9.044204}. For a frequency-beating heater, a non-Gaussian energy distribution is produced, with correlations between the transverse laser profile and longitudinal modulation. A quantitative connection between the laser settings and final bunching requires these source properties to be retained during collective transport. The source formation and downstream response must therefore be described within a common model.

In this paper, a nonlinear model is developed for the coupled formation and evolution of laser-seeded microbunching. The heater-exit distribution is constructed from the incident beam and laser field by optical phase averaging. Its non-Gaussian structure and transverse--longitudinal correlations are retained through multistage compression in six-dimensional phase space. LSC, CSR, and radio-frequency (RF) wakefields are evaluated self-consistently from the evolving distribution. Reusable external transport maps allow the laser and compression settings to be scanned efficiently. The predicted heater response is benchmarked against Elegant, and the downstream bunching is compared with IMPACT-Z and experimental measurements at the Shanghai soft x-ray free-electron laser (SXFEL).

LSC is identified as the dominant collective effect responsible for the short-wavelength double peak in the investigated configuration. The bunching spectrum is strongly reshaped by LSC, whereas smaller changes are produced by RF wakefields and CSR. At fixed nominal total compression, bunching near a selected wavelength is enhanced when more compression is assigned to the first stage. The compression partition thus provides control of the bunching amplitude beyond the nominal wavelength scaling with total compression. Experimental measurements at different laser pulse energies exhibit wavelength-dependent bunching trends that agree well with theory and simulation. By directly linking beat frequency, laser strength, and compression partition to the downstream bunching spectrum, the framework enables efficient identification of accelerator operating points for enhanced coherent THz emission. The resulting parameter dependence can be used to guide experimental tuning of narrow-band, continuously tunable THz sources.

\section{Six-dimensional phase-space transport model}
\label{sec:theory}
The bunching spectrum is calculated by following the electron distribution from the laser heater to the linac exit. The heater-induced distribution is obtained from the incident beam and prescribed laser field. Its evolution through magnetic compression is then calculated with collective fields. Correlations among the six phase-space coordinates are retained throughout the transport.

\subsection{Frequency-beating laser-heater interaction}

For two chirped pulse replicas separated by $\tau$, the beating power is
\begin{equation}
 \begin{aligned}
 P(t;\tau)&=P_+(t)+P_-(t)\\
 &\quad+2\sqrt{P_+(t)P_-(t)}\cos(\Omega_b t+\phi_0),\\
 \Omega_b&=|\mu|\tau.
 \end{aligned}
 \label{eq:beating_power}
\end{equation}
where $\mu=2\pi\alpha$ is the angular-frequency chirp and $f_b=\Omega_b/(2\pi)$. The power envelopes of the two replicas are denoted by $P_\pm$, and $\phi_0$ is their relative phase. The applied field is determined by their temporal overlap. For Gaussian pulses with intensity FWHM $T_L$,
\begin{equation}
 \frac{P_{\rm pk}(\tau)}{P_{\rm pk}(\tau_{\rm ref})}
 =\exp\!\left[-\ln2\,
 \frac{\tau^2-\tau_{\rm ref}^2}{T_L^2}\right].
 \label{eq:delay_attenuation}
\end{equation}
Let $\mathbf X=(x,x',y,y',z,\delta)$, with $z=c(t-t_{\rm ref})$ and $\delta=(p-p_{\rm ref})/p_{\rm ref}$. The six-dimensional entrance distribution is propagated to the heater interaction plane, including the prescribed upstream wake and space-charge transport. The resonant heater is represented by
\begin{equation}
 \begin{aligned}
 \gamma^+&=\gamma^-+A(\mathbf X;\tau,P_{\rm pk})\cos\psi,\\
 A&=A_0 p_L(t;\tau)e^{-(x^2+y^2)/w_0^2}.
 \end{aligned}
 \label{eq:heater_kick}
\end{equation}
where $p_L$ is the peak-normalized laser field envelope and $\psi$ is the optical phase. The on-axis kick amplitude is calculated from the laser-heater interaction~\cite{PhysRevSTAB.7.074401}, with a finite-focus and termination-pole correction,
\begin{equation}
 \begin{aligned}
 A_0&=\sqrt{\frac{P_{\rm pk}}{P_0}}
 \frac{K L_u}{\gamma_r\sigma_L}[J_0(\xi)-J_1(\xi)]\,|F_u|,\\
 F_u&=\frac{1}{L_u}\int_0^{L_u}
 \frac{g_u(s)\,ds}{1+i(s-L_u/2)/z_R},\\
 \xi&=\frac{K^2}{4+2K^2}.
 \end{aligned}
 \label{eq:heater_amplitude}
\end{equation}
Here $P_0=I_A mc^2/e$, where $I_A$ is the Alfv\'en current. The laser size is $\sigma_L=w_0/2$, $z_R$ is the Rayleigh length, and $g_u$ describes the undulator termination poles. The energy kick is converted to momentum using $p=mc\sqrt{\gamma^2-1}$. Particle flight times and transport through the heater chicane are included up to the exit of the laser-heater chicane. The interaction is approximated by a thin resonant kick.

Optical phase averaging is performed on the entrance ensemble. With $u^\pm=(\gamma^\pm-\gamma_r)/\gamma_r$, the conditional characteristic function after the kick is
\begin{equation}
 \Phi(q\mid z)=\left\langle
 e^{-iqu^-}J_0\!\left(qA/\gamma_r\right)
 \right\rangle_{\mathbf X\mid z}.
 \label{eq:characteristic_function}
\end{equation}
Here $q$ is conjugate to $u$, and $\Phi(q\mid z)=\langle e^{-iqu^+}\rangle_z$. The non-Gaussian energy distribution and periodic energy variance are retained by this phase average. In the numerical evaluation, each entrance particle with weight $w_i$ is represented by $Q_\psi$ optical phases, $\psi_{i\ell}=\psi_i+2\pi\ell/Q_\psi$, with weights $w_i/Q_\psi$. The phase offsets $\psi_i$ are distributed along the bunch. All phase samples are included in the density used to calculate the subsequent collective fields.

\subsection{Natural propagation and collective feedback}

Let $d\nu_0$ denote the normalized electron distribution at the exit of the laser-heater chicane. In the absence of downstream collective fields, the complex bunching is
\begin{equation}
 \widetilde b^{(0)}(k,s)=\int
 \exp\!\left[-ik\,z\!\left(\mathcal T_{s\leftarrow0}\mathbf X\right)\right]
 d\nu_0(\mathbf X),
 \label{eq:source_term}
\end{equation}
where $\mathcal T$ is the external transport map. Dispersive conversion and phase mixing are included through the transported distribution. This limit is referred to as natural propagation. Collective amplification can be described by an impedance-based Volterra equation in the small-signal limit~\cite{Venturini2007,Tsai2015Vlasov}. Here the finite-amplitude distribution is evolved directly.

The distribution is advanced through external transport and collective kicks. Outside RF sections, the transport is represented by linear and quadratic maps,
\begin{equation}
 (\mathcal M_j\mathbf X)_a
 =\sum_b R^{(j)}_{ab}X_b
 +\sum_{b,c}T^{(j)}_{abc}X_bX_c.
 \label{eq:external_map}
\end{equation}
Phase-dependent acceleration and focusing are included in RF sections, with the reference energy updated along the lattice. Geometric and edge transport are included in dipoles. At each collective step, the charge density is calculated from the particle coordinates. The resulting fields are interpolated to the particles. A midpoint update is written as
\begin{equation}
 \mathbf X_i^{j+1}=
 \mathcal M_{j,2}\circ
 \mathcal K_j[\nu_{j+1/2}]\circ
 \mathcal M_{j,1}(\mathbf X_i^j),
 \label{eq:finite_amplitude_kick}
\end{equation}
where $\mathcal M_{j,1}$ and $\mathcal M_{j,2}$ describe external transport before and after the collective kick $\mathcal K_j$. The kick is evaluated from the midpoint distribution $\nu_{j+1/2}$.

Downstream space charge is obtained from the quasistatic three-dimensional Poisson problem,
\begin{equation}
 \nabla'^2\phi'=-\rho'/\epsilon_0,
 \qquad \mathbf E'=-\nabla'\phi',
 \label{eq:collective_field}
\end{equation}
where primes denote the beam rest frame. Both transverse and longitudinal self-forces are retained~\cite{PhysRevSTAB.9.044204}. RF wakefields are calculated from the longitudinal charge density and transverse dipole moments. CSR is evaluated with an unshielded, finite-energy, one-dimensional integrated Green function. Entrance, in-bend, and exit contributions are included for the specified dipole and drift geometry~\cite{Tsai2015CSR}. The one-dimensional approximation applies to CSR, whereas space charge is evaluated in three dimensions. Density--energy conversion and harmonic generation are obtained from the evolving distribution.

\subsection{Bunching spectrum}

The distribution-level bunching is
\begin{equation}
 \begin{aligned}
 \widetilde b(k,s)&=\sum_i w_i e^{-ikz_i(s)},\\
 b(k,s)&=|\widetilde b(k,s)|,\qquad \sum_i w_i=1.
 \end{aligned}
 \label{eq:bunching_definition}
\end{equation}
For comparisons within the bunch core, the same interval selection, current detrending, and Hann window are applied to theory and simulation. This local modulation estimator differs from the full-bunch Fourier moment in Eq.~\eqref{eq:bunching_definition}. In the response scans below, the local spectral maximum is reported together with its wavelength, $\lambda_{\rm out}=2\pi/k_{\rm peak}$.

The external transport maps and RF parameters are calculated once for each machine setting. They are reused when the laser parameters are varied. The heater distribution and collective fields are recalculated for each laser setting. The accuracy of this evaluation depends on optical phase sampling, particle sampling, and field resolution.

\section{Experimental platform and numerical methods}
\label{sec:methods}

The model is formulated in terms of standard machine inputs: external transport maps, beam and laser parameters, and longitudinal impedances. The diagnostic response must also be specified for comparison with measurements. In this study, the formulation is applied to SXFEL, where a frequency-beating laser heater, two-stage magnetic compression, and a linac-exit transverse deflecting diagnostic allow the evolution from source formation to measured bunching to be examined~\cite{app7060607,Kang2025}. Its applicability extends to other linacs with multistage compression. With the corresponding machine inputs, laser-induced modulation can be followed through source formation, dispersive conversion, and collective amplification within the same framework.

\subsection{SXFEL linac}

\begin{figure*}[width=\textwidth,pos=tp,align=\centering]
\centering
\includegraphics[width=0.9\textwidth]{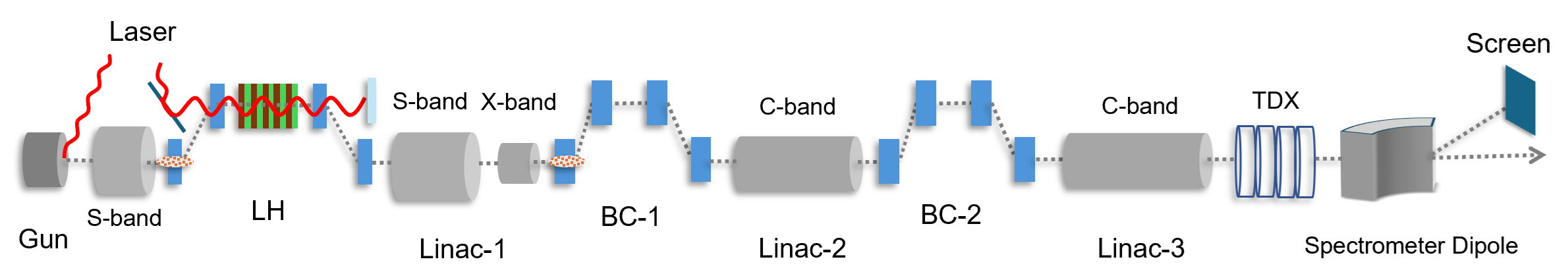}
\caption{SXFEL linac layout for microbunching induced by laser frequency beating. The laser-heater modulation is amplified by collective effects and two-stage compression; the exit phase space is measured with the TDX, spectrometer dipole, and screen (not to scale).}
\label{fig1}
\end{figure*}

At SXFEL, the electron beam is delivered from the S-band injector to the laser heater, as shown in Fig.~\ref{fig1}. The slice energy spread is modulated by the frequency-beating laser. An energy chirp is then established and linearized by the S- and X-band structures upstream of BC1. Density modulation is produced in BC1 and is converted into energy modulation by LSC and RF wakefields. Further density modulation is generated in BC2, with CSR contributions in the bends. The beam is then transported through the final C-band section to the linac-exit diagnostic.

At this station, the longitudinal coordinate is mapped onto one screen axis by a transverse deflecting structure (TDX), while energy is mapped onto the orthogonal axis by a spectrometer dipole. Both the longitudinal phase space and the projected current profile used in Sec.~\ref{sec:experiment} are contained in the resulting two-dimensional distribution.

The beam and lattice parameters used in the calculations are listed in Table~\ref{tab:beam_parameters}. The beam energy is increased from $\SI{129.7}{MeV}$ at the LH exit to $\SI{1.3}{GeV}$ at the linac exit. The RF phases are varied to redistribute an approximately tenfold total compression between BC1 and BC2; for every setting, the achieved compression and transport matrices are recalculated.

\begin{table}[width=\columnwidth,pos=tbp,align=\centering]
\centering
\caption{Beam and longitudinal-lattice parameters used in the SXFEL calculations. Energies denote the unperturbed reference trajectory.}
\label{tab:beam_parameters}
\begin{tabular*}{\columnwidth}{@{\extracolsep{\fill}}ll}
\toprule
Parameter & Value \\
\midrule
Bunch charge & $\SI{500}{pC}$ \\
Beam energy at the LH exit & $\SI{129.7}{MeV}$ \\
Beam energy before BC1 & $\SI{252.7}{MeV}$ \\
Beam energy before BC2 & $\SI{541.5}{MeV}$ \\
Beam energy at the linac exit & $\SI{1.3}{GeV}$ \\
Peak current before compression & $\SI{60}{A}$ \\
Normalized transverse emittance & $\SI{1.3}{\micro\metre}$ \\
Intrinsic slice energy spread & $\SI{1.407}{keV}$ \\
Initial energy chirp & $h_0=\SI{1.179}{m^{-1}}$ \\
Overall compression, $C$ & $\simeq 10$ \\
\bottomrule
\end{tabular*}
\end{table}

The laser-heater parameters are summarized in Table~\ref{tab:heater_parameters}. Two equally intense linearly chirped pulses with a variable delay are recombined, so that $f_b=|\alpha|\tau$ in the overlap region~\cite{Brynes2020a}. A delay range from $\SI{1.72}{ps}$ to $\SI{8.38}{ps}$ is considered, corresponding to an initial laser beat-frequency range of $\SIrange{0.783}{3.817}{THz}$. The seed is subsequently mapped by compression to the downstream beam modulation; continuous THz emission over $\SIrange{7.8}{30.8}{THz}$ was demonstrated in the corresponding SXFEL experiment~\cite{Kang2025}.

\begin{table}[width=\columnwidth,pos=tbp,align=\centering]
\centering
\caption{Laser-heater and frequency-beating laser parameters. Power settings and the extended theory scan refer to the calculations. The experimental ranges refer to the available laser settings.}
\label{tab:heater_parameters}
\footnotesize
\newcommand{\tablabel}[1]{\parbox[t]{0.60\columnwidth}{\raggedright #1}}
\newcommand{\tabvalue}[1]{\parbox[t]{0.35\columnwidth}{\raggedright #1}}
\begin{tabular}{@{}ll@{}}
\toprule
\tablabel{Parameter} & \tabvalue{Value} \\
\midrule
\multicolumn{2}{l}{\textit{Laser-heater chicane}} \\
\tablabel{Dipole bend angle} & \tabvalue{$\SI{5.0}{\degree}$} \\
\tablabel{Horizontal dispersion, $\eta_x$} & \tabvalue{$\SI{39.6}{mm}$} \\
\midrule
\multicolumn{2}{l}{\textit{Undulator}} \\
\tablabel{Undulator period, $\lambda_u$} & \tabvalue{$\SI{50}{mm}$} \\
\tablabel{Number of undulator periods} & \tabvalue{$10$} \\
\tablabel{Undulator magnetic length, $L_u$} & \tabvalue{$\SI{0.5}{m}$} \\
\tablabel{Peak undulator field, $B_u$} & \tabvalue{$\SI{0.3118}{T}$} \\
\tablabel{Undulator parameter, $K$} & \tabvalue{$1.456$} \\
\midrule
\multicolumn{2}{l}{\textit{Frequency-beating laser}} \\
\tablabel{Laser central wavelength} & \tabvalue{$\SI{800}{nm}$} \\
\tablabel{Pulse length (FWHM)} & \tabvalue{$\SI{22}{ps}$} \\
\tablabel{Chirp coefficient, $\alpha=df/dt$} & \tabvalue{$\SI{4.555e23}{s^{-2}}$} \\
\tablabel{Delay between pulses, $\tau$} & \tabvalue{$\SIrange{1.72}{8.38}{ps}$} \\
\tablabel{Experimental beat-frequency range} & \tabvalue{$\SIrange{0.783}{3.817}{THz}$} \\
\tablabel{Source-benchmark power scan, $P_L$} & \tabvalue{$\SIrange{10}{446.2}{\kilo\watt}$} \\

\tablabel{Delivered pulse energy (experiment), $E_L$} & \tabvalue{$\SIrange{1.2}{9.8}{\micro\joule}$} \\
\tablabel{Laser waist, $w_0$} & \tabvalue{$\SI{0.711}{mm}$} \\
\bottomrule
\end{tabular}
\end{table}

\begin{figure*}[width=\textwidth,pos=tp,align=\centering]
\centering
\includegraphics[width=0.9\textwidth]{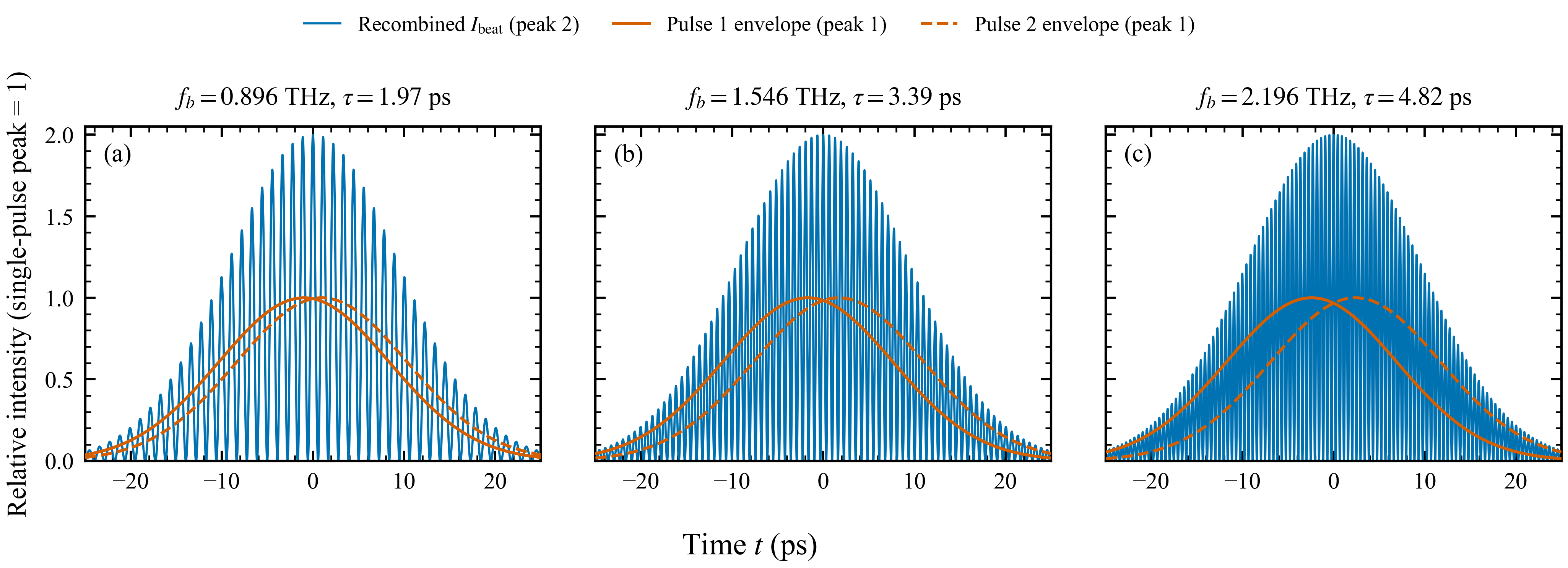}
\caption{Frequency-beating laser profiles for three delays. The recombined intensity $I_{\rm beat}$ (blue) and the two delayed envelopes (solid and dashed orange) are normalized to the individual-pulse peak.}
\label{fig:laser_profiles}
\end{figure*}

Representative frequency-beating laser profiles at three delays are shown in Fig.~\ref{fig:laser_profiles}. Panels (a)--(c) correspond to $(f_b,\tau)=(0.896~\mathrm{THz},1.97~\mathrm{ps})$, $(1.546~\mathrm{THz},3.39~\mathrm{ps})$, and $(2.196~\mathrm{THz},4.82~\mathrm{ps})$, respectively. As $\tau$ is increased, $f_b$ is increased and the beat period is shortened, so that denser oscillations are produced within the common pulse envelope.

\subsection{Elegant and IMPACT-Z simulations}

Three stages are used in the numerical simulations. First, each beating profile is tracked through the laser heater with the Elegant LSRMDLTR element~\cite{Borland2000Elegant}, and the resulting heater-exit particle distribution is supplied to IMPACT-Z~\cite{QIANG2000434,PhysRevSTAB.9.044204}. The analytical source is evaluated independently from the same beam, laser, and undulator parameters in Eqs.~\eqref{eq:heater_amplitude}--\eqref{eq:characteristic_function}. For the peak-normalized field envelope $p_L(t;\tau)$, the effective pulse energy is
\begin{equation}
 U_{\rm eff}(\tau)=P_{\rm pk}(\tau)T_{\rm eff}(\tau),
 \qquad T_{\rm eff}(\tau)=\int |p_L(t;\tau)|^2\,dt .
 \label{eq:effective_laser_energy}
\end{equation}

The nominal laser power setting is denoted by $P_L$; the delay-dependent peak power $P_{\rm pk}(\tau)$ includes the temporal-overlap factor in Eq.~\eqref{eq:delay_attenuation}. For the six energy comparisons in Fig.~\ref{fig:three_way_comparison}, nominal powers of $446.2$, $407.7$, $347.4$, $273.1$, $192.3$, and $117.7\,\mathrm{kW}$ are used. These settings are obtained by dividing the calibrated delivered energies by $T_L=22\,\mathrm{ps}$. This nominal conversion is distinct from the envelope integral in Eq.~\eqref{eq:effective_laser_energy}. The theoretical beat-frequency scan for this comparison extends from $0.577$ to $7.495\,\mathrm{THz}$, corresponding to input wavelengths of $520$ to $40\,\mu\mathrm{m}$. The extended range is used to resolve short-wavelength structure beyond the experimental coverage.

Second, a map-only Elegant calculation is performed for every RF and compression setting. The linear and quadratic transport coefficients, reference energy, compressor geometry, and accelerating-structure parameters supply the external transport from the heater exit to the linac exit. LSC, CSR, and longitudinal wakefields are disabled in this calculation because they are evaluated explicitly through the self-consistent update in Eq.~\eqref{eq:finite_amplitude_kick}. The resulting background map can then be reused for power and beat-frequency scans at fixed machine settings.

Third, the same heater-exit particles are tracked with IMPACT-Z, including three-dimensional space charge, steady-state and transient CSR, and the S-, X-, and C-band longitudinal wakefields. A longitudinal mesh with $N_Z=128$ is used for the comparisons.

For each linac-exit distribution, the spectrum is evaluated with the same longitudinal window and wave-number sampling used for the analytical result. The local estimator defined in Sec.~\ref{sec:theory} is used for the reported core bunching. When a one-sided FFT routine returns the sinusoidal modulation depth $2|b_1|$, the result is divided by two before comparison with the standard bunching factor $|b_1|$.

\section{Theory and simulation results}
\label{sec:simulation}

\subsection{Laser-heater source: theory and Elegant benchmark}

The analytical source model is compared with Elegant at the laser-modulator exit using scans of the laser peak power and beat frequency, with the parameters listed in Tables~\ref{tab:beam_parameters} and~\ref{tab:heater_parameters}. The source-plane observable is the Hann-weighted slice-energy-variance modulation amplitude $\sqrt{A_{\sigma,H}}$. For a slice centered at $z_i$, the variance is calculated as $V_i=N_i^{-1}\sum_{j\in i}(E_{ij}-\overline{E}_i)^2$ and fitted to
\begin{equation}
 \begin{aligned}
 V_i&=V_{\rm bg}(z_i)+C_{\sigma,H}\cos(k_bz_i)\\
 &\quad+S_{\sigma,H}\sin(k_bz_i),\\
 A_{\sigma,H}&=\sqrt{C_{\sigma,H}^2+S_{\sigma,H}^2},
 \end{aligned}
 \label{eq:lh_variance_modulation}
\end{equation}
where $N_i$ and $\overline{E}_i$ are the particle number and centroid energy in slice $i$, and $V_{\rm bg}$ is a cubic background. The distribution is divided into 720 equal-population slices, and the central 90\% of slice centers is retained. The fit minimizes the squared residuals with weights $N_i H_i$, where
\begin{equation}
 H_i=\sin^2\!\left[\pi\frac{z_i-z_{\min}}{z_{\max}-z_{\min}}\right].
 \label{eq:lh_hann_weight}
\end{equation}
Here $z_{\min}$ and $z_{\max}$ delimit the retained interval, determined separately for each distribution using the same selection rule. The quantity $\sqrt{A_{\sigma,H}}$ measures the periodic variance component with emphasis on the bunch center. The same estimator is applied to the independently generated analytical source and the Elegant distribution.

\begin{figure}[width=\columnwidth,pos=tbp,align=\centering]
\centering
\includegraphics[width=0.92\columnwidth]{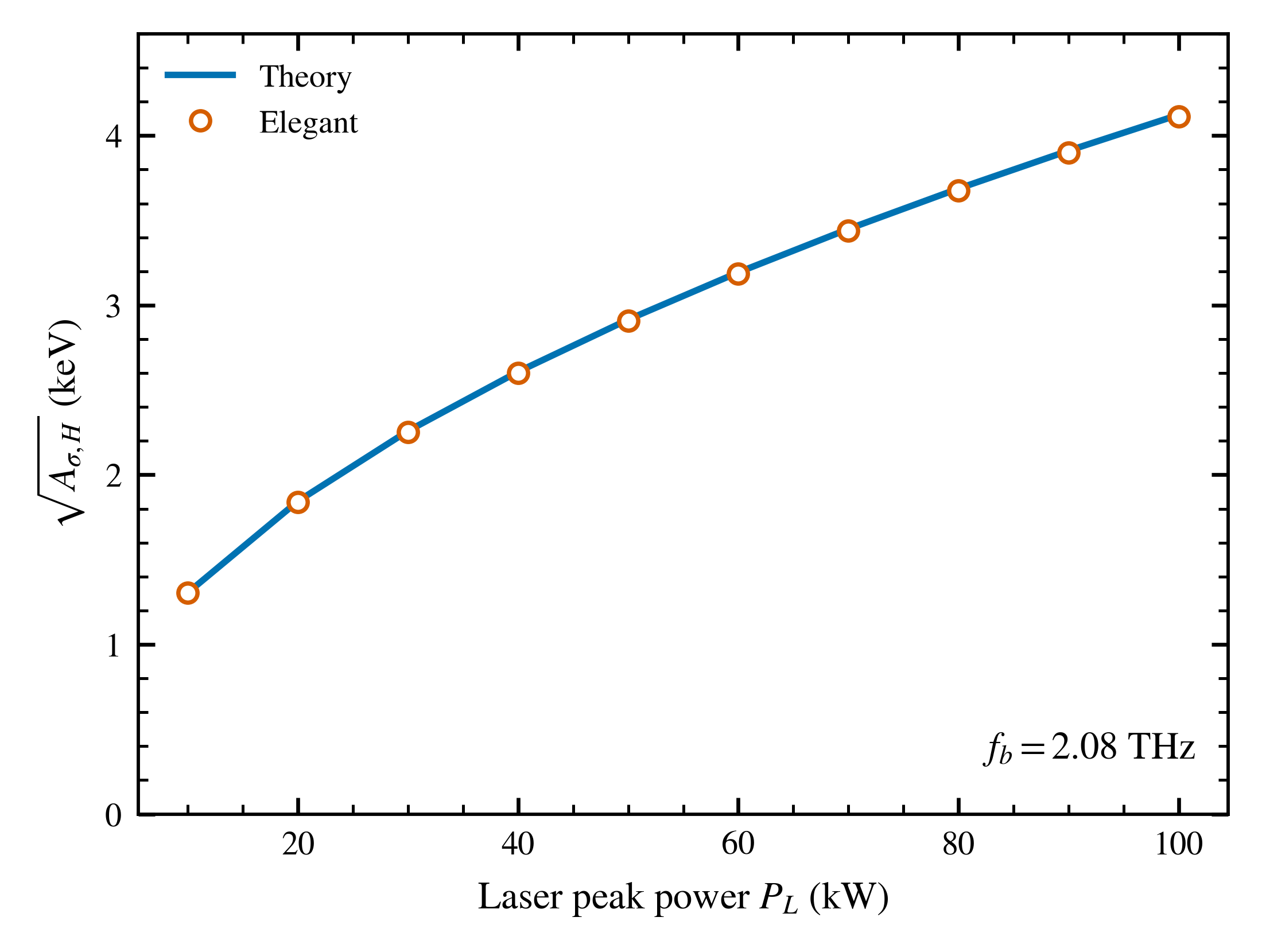}
\caption{Hann-weighted source-plane modulation amplitude $\sqrt{A_{\sigma,H}}$ versus $P_L$ at $f_b=2.08\,\mathrm{THz}$; theory (blue solid line) and Elegant (orange open circles).}
\label{fig:lh_source_power}
\end{figure}

\begin{figure}[width=\columnwidth,pos=tbp,align=\centering]
\centering
\includegraphics[width=0.92\columnwidth]{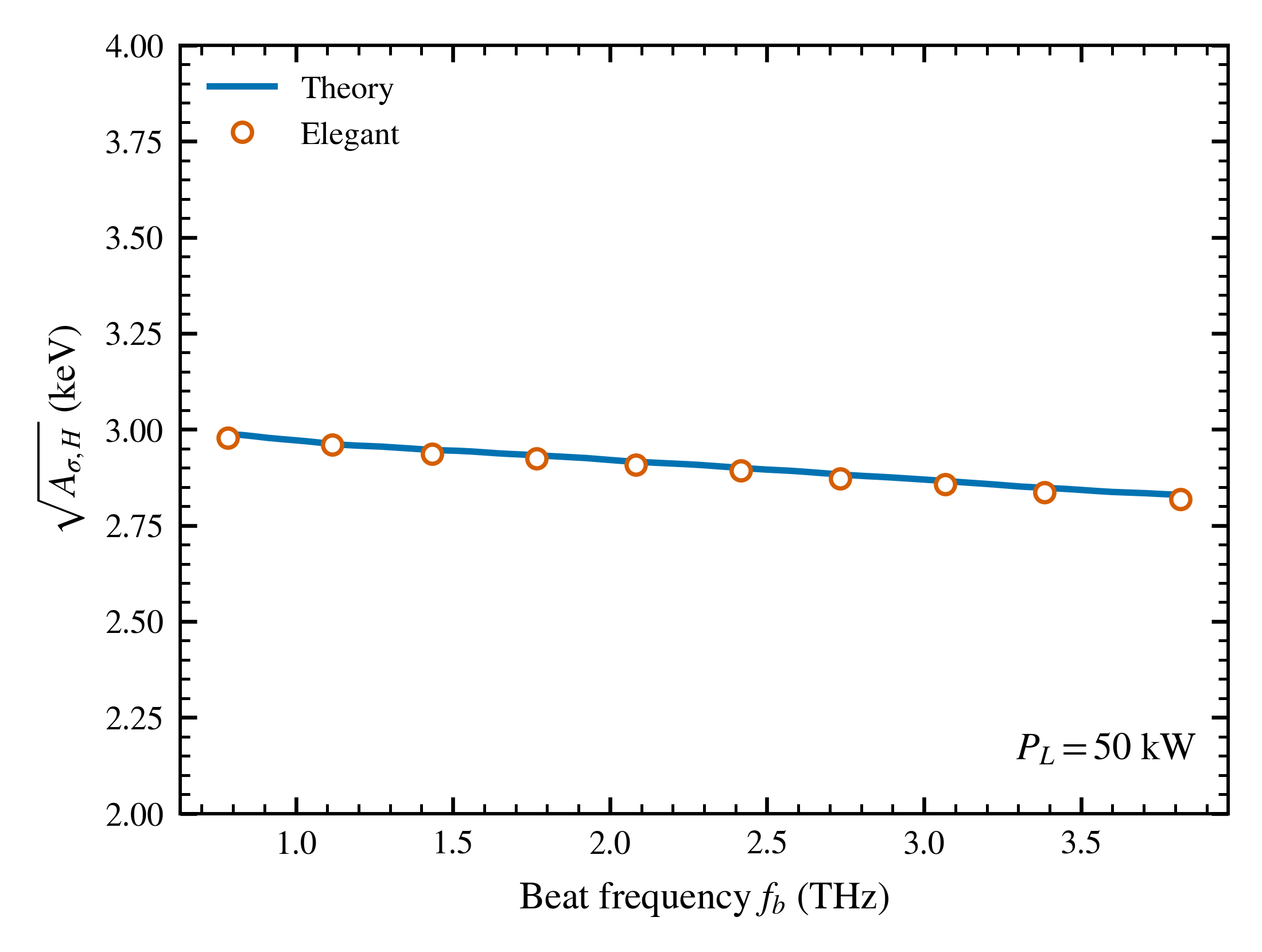}
\caption{Hann-weighted source-plane modulation amplitude $\sqrt{A_{\sigma,H}}$ versus $f_b$ at $P_L=50\,\mathrm{kW}$; theory (blue solid line) and Elegant (orange open circles).}
\label{fig:lh_source_frequency}
\end{figure}

At $f_b=2.08\,\mathrm{THz}$, the power dependence over $10$--$100\,\mathrm{kW}$ is shown in Fig.~\ref{fig:lh_source_power}. The theoretical modulation amplitude increases from $1.306$ to $4.123\,\mathrm{keV}$, compared with $1.303$ to $4.112\,\mathrm{keV}$ in Elegant, approximately following the square-root power dependence of the heater kick. At fixed $P_L=50\,\mathrm{kW}$, the frequency dependence over $0.783$--$3.817\,\mathrm{THz}$ is shown in Fig.~\ref{fig:lh_source_frequency}. A weak monotonic decrease from $2.987$ to $2.829\,\mathrm{keV}$ is obtained in theory, compared with $2.978$ to $2.819\,\mathrm{keV}$ in Elegant, consistent with the reduced temporal overlap of the delayed laser replicas.

The mean absolute relative deviations, evaluated as $\langle|a_{\rm th}/a_{\rm Elegant}-1|\rangle$ with $a=\sqrt{A_{\sigma,H}}$, are $0.250\%$ and $0.296\%$ for the power and frequency scans, respectively; the corresponding maxima are $0.263\%$ and $0.427\%$. These comparisons test the source-plane variance modulation over both laser control variables.

\begin{figure*}[width=\textwidth,pos=tp,align=\centering]
\centering
\includegraphics[width=0.9\textwidth]{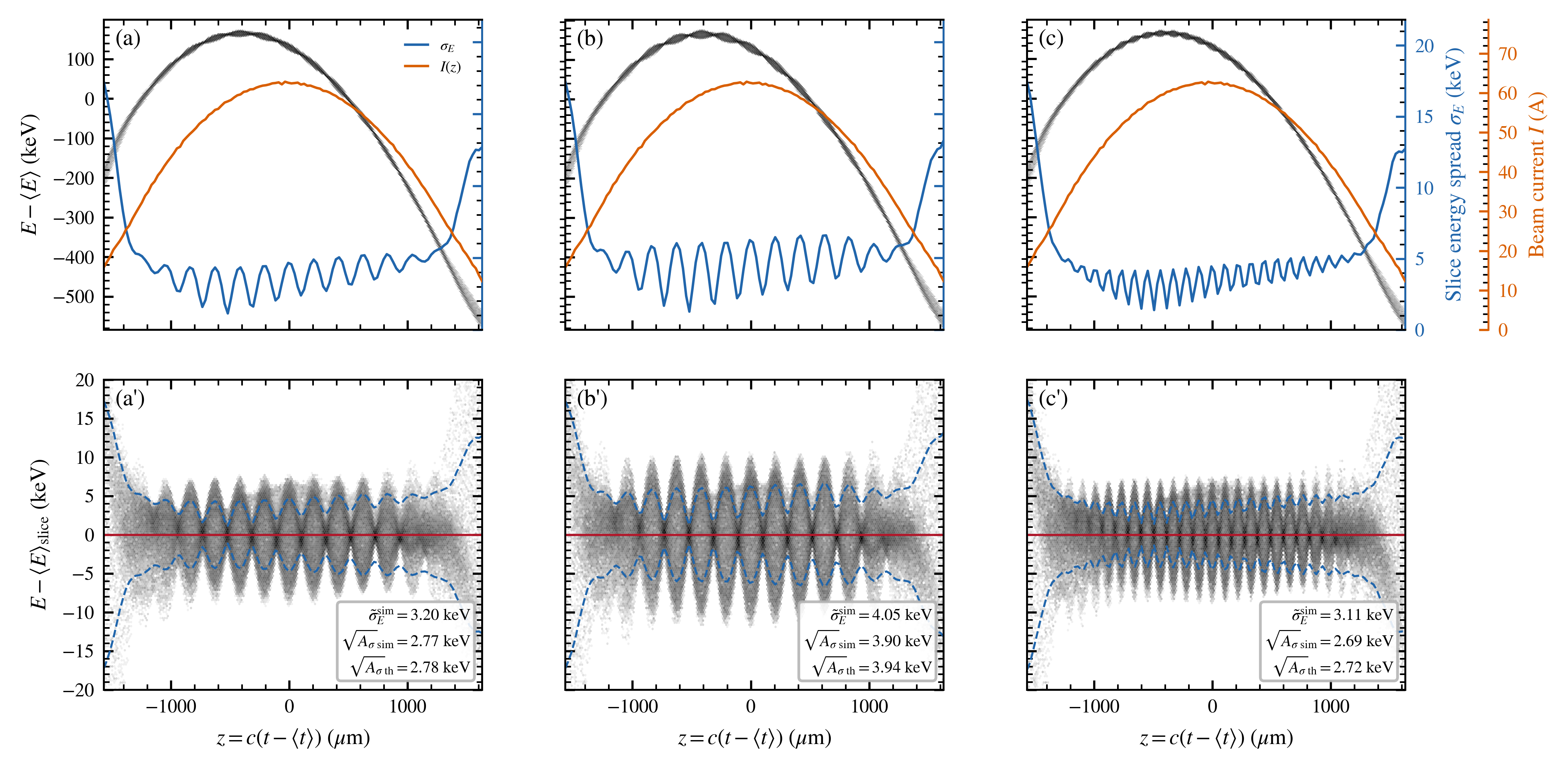}
\caption{Source-plane longitudinal phase spaces for $(P_L,f_b)=(50\,\mathrm{kW},1.43\,\mathrm{THz})$, $(100\,\mathrm{kW},1.43\,\mathrm{THz})$, and $(50\,\mathrm{kW},2.73\,\mathrm{THz})$ in columns (a)--(c), respectively. The upper row shows the RF-correlated phase space with $\sigma_E$ (blue) and $I(z)$ (orange); panels (a$'$)--(c$'$) show the centroid-subtracted phase space and spread envelopes. Annotations give the median slice energy spread and the unwindowed variance-modulation amplitudes from simulation and theory.}
\label{fig:lh_phase_frequency}
\end{figure*}

The source-plane longitudinal phase spaces are shown in Fig.~\ref{fig:lh_phase_frequency} for $(P_L,f_b)=(50\,\mathrm{kW},1.43\,\mathrm{THz})$, $(100\,\mathrm{kW},1.43\,\mathrm{THz})$, and $(50\,\mathrm{kW},2.73\,\mathrm{THz})$. The upper row shows the RF-correlated phase space, slice energy spread, and current; the lower row shows the phase space after subtraction of the slice-energy centroid. Between panels (a) and (b), the energy-spread modulation is increased at the same beat frequency, while the longitudinal period remains unchanged. Between panels (a) and (c), the modulation period is shortened at fixed laser power. The annotations report the unwindowed variance-modulation amplitude $\sqrt{A_\sigma}$; the Hann-weighted comparison is given separately in Figs.~\ref{fig:lh_source_power} and~\ref{fig:lh_source_frequency}.

\subsection{Collective amplification and spectral control}
\label{sec:collective_amplification}

The heater-induced distribution is propagated through the SXFEL linac to evaluate the downstream density modulation. The same bunch-core interval and spectral estimator are used for theory and IMPACT-Z. The reported $|b_1|$ is the local modulation amplitude defined in Sec.~\ref{sec:theory}.

The downstream response is determined by successive conversions between energy and density modulation. The periodic slice-energy-spread modulation generated by the laser heater is converted into density modulation by the first compressor. Coherent energy modulation is subsequently generated by LSC and accelerating-structure wakes, while its amplitude and phase are modified by CSR in the bends. The evolved energy modulation is converted back into density bunching by the second compressor. Along the lattice, the local wave number and current are approximately described by $k(s)=C(s)k_0$ and $I(s)=I_0C(s)$, respectively. The first-order output-frequency scaling is consequently given by $f_{\rm out}\simeq C_{\mathrm{tot}}f_b$, whereas the bunching amplitude and the precise spectral maximum are determined by the collective phase evolution and the associated Landau damping.

\begin{figure}[width=\columnwidth,pos=tbp,align=\centering]
\centering
\subfloat[]{\includegraphics[width=0.49\columnwidth]{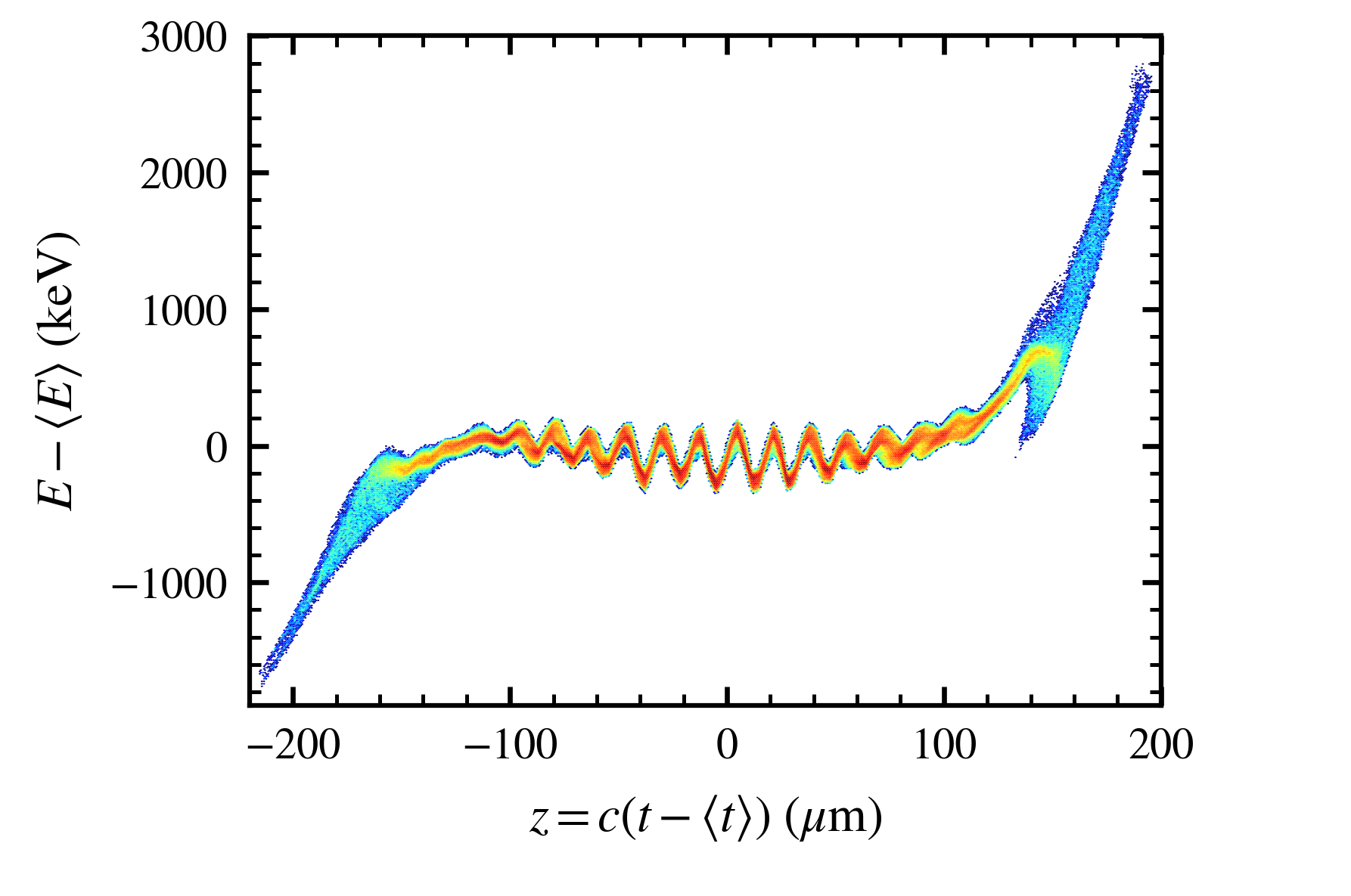}}%
\hspace{0.01\columnwidth}%
\subfloat[]{\includegraphics[width=0.49\columnwidth]{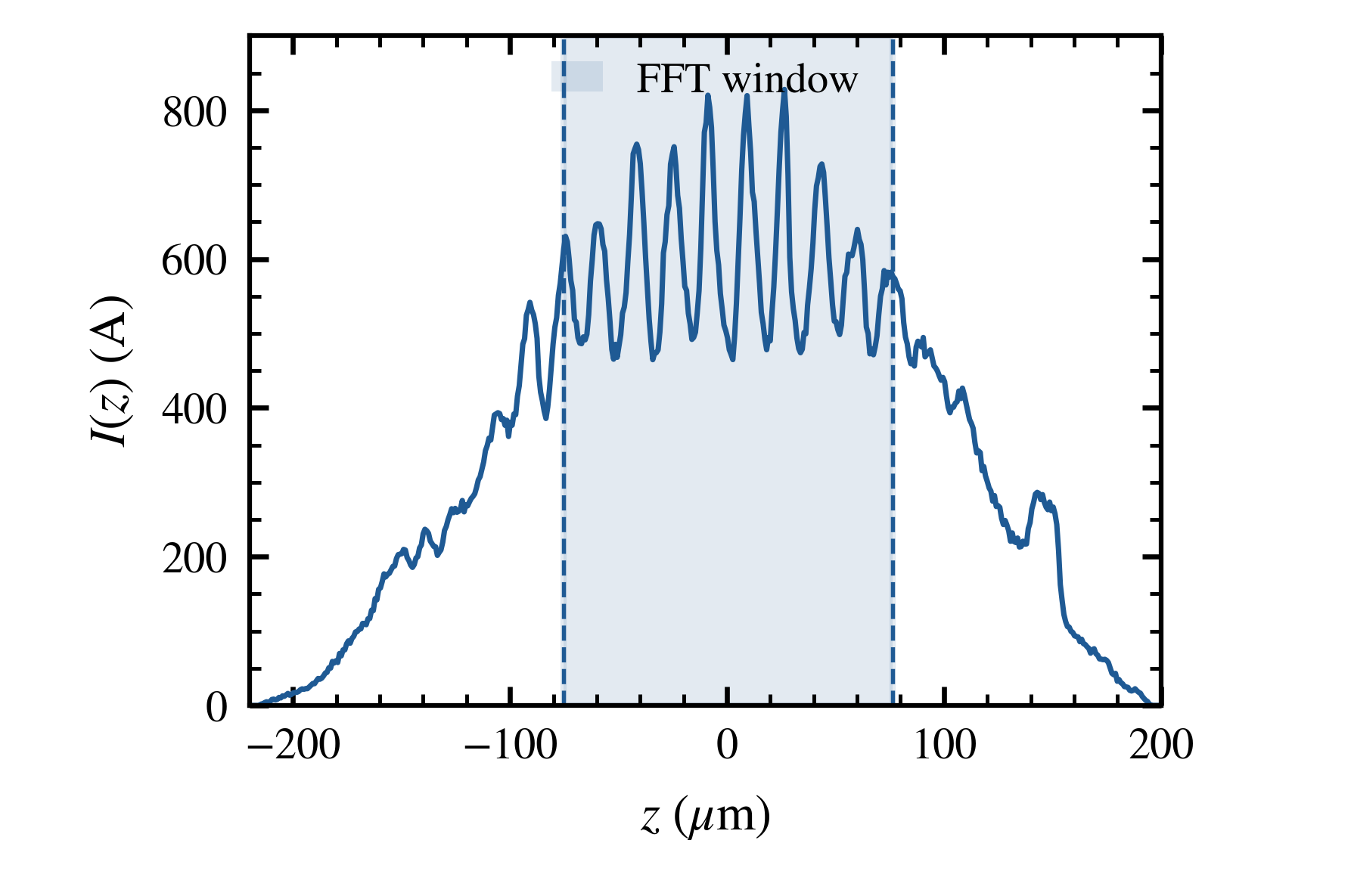}}\\[-0.25em]
\subfloat[]{\includegraphics[width=0.49\columnwidth]{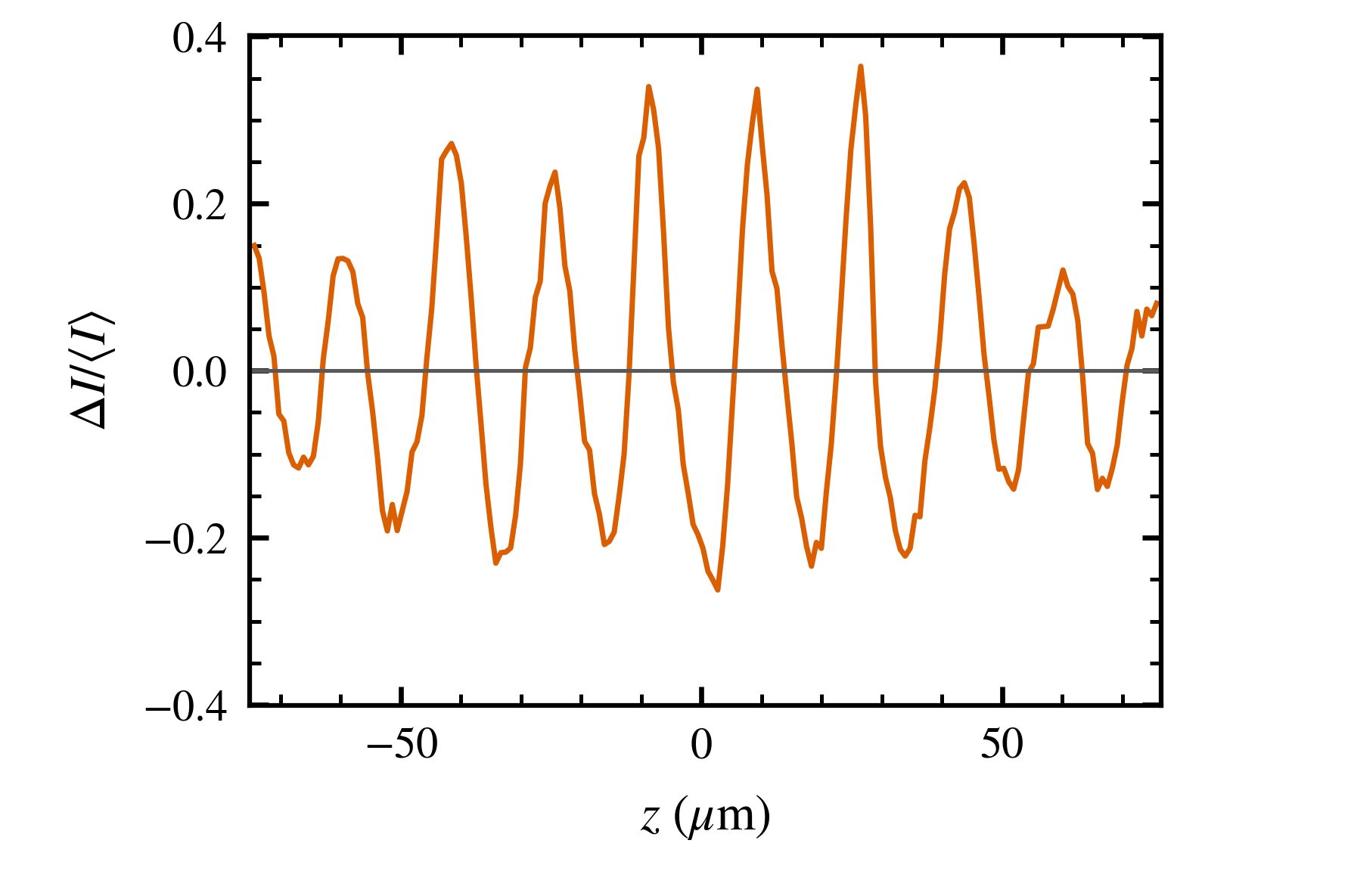}}%
\hspace{0.01\columnwidth}%
\subfloat[]{\includegraphics[width=0.49\columnwidth]{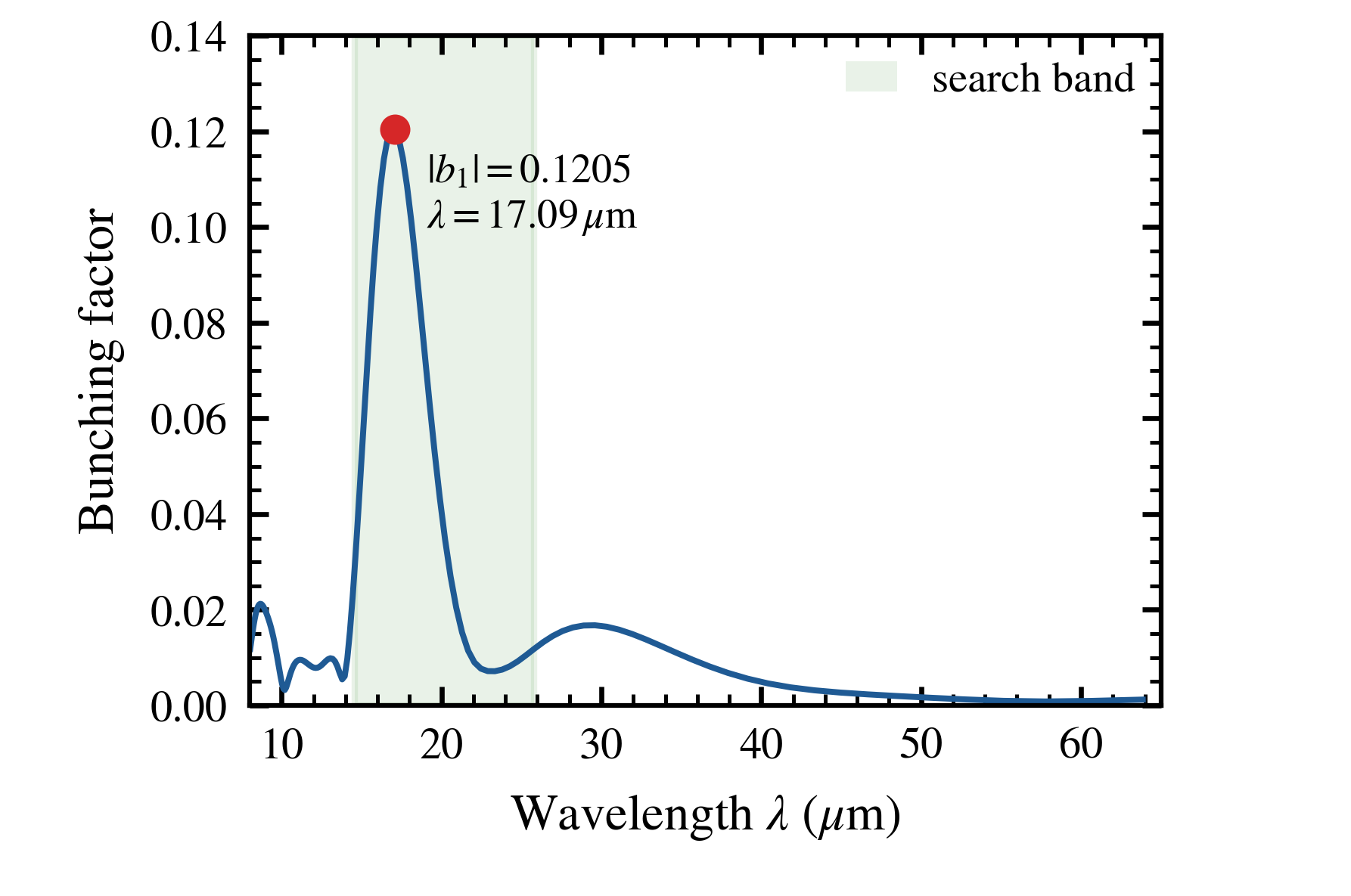}}
\caption{Linac-exit diagnostics for $P_L=50\,\mathrm{kW}$, $C_1=8$, $C_2=1.25$, and $f_b=1.767\,\mathrm{THz}$. Panels (a)--(d) show the phase-space density, current profile, detrended relative-current modulation, and bunching spectrum $|b_1|$; the shaded interval and red marker indicate the spectral-analysis window and maximum.}
\label{fig:exit_case_diagnostic}
\end{figure}

The linac-exit longitudinal phase-space density, current profile, relative-current modulation, and bunching spectrum are shown in Fig.~\ref{fig:exit_case_diagnostic} for $P_L=50\,\mathrm{kW}$, $C_1=8$, $C_2=1.25$, and $f_b=1.767\,\mathrm{THz}$. A coherent energy modulation and the associated current modulation are resolved. After removal of the smooth current baseline with a third-order polynomial, the relative fluctuation is windowed for spectral analysis. A peak standard bunching factor of $|b_1|=0.1205$ is obtained at $\lambda=17.09\,\mu\mathrm{m}$, close to the first-order estimate $c/(C_{\rm core}f_b)=16.76\,\mu\mathrm{m}$ for the achieved core compression $C_{\rm core}=10.122$. The spectral maximum is therefore close to, but distinct from, the nominal compressed frequency.

Natural propagation is defined by Eq.~\eqref{eq:source_term}. Collective transport is evaluated with Eq.~\eqref{eq:finite_amplitude_kick}, including space charge, CSR, and RF wakefields. Their separate contributions are compared at fixed source conditions and external transport.

\begin{figure*}[width=\textwidth,pos=tp,align=\centering]
\centering
\subfloat[]{\includegraphics[width=0.32\textwidth]{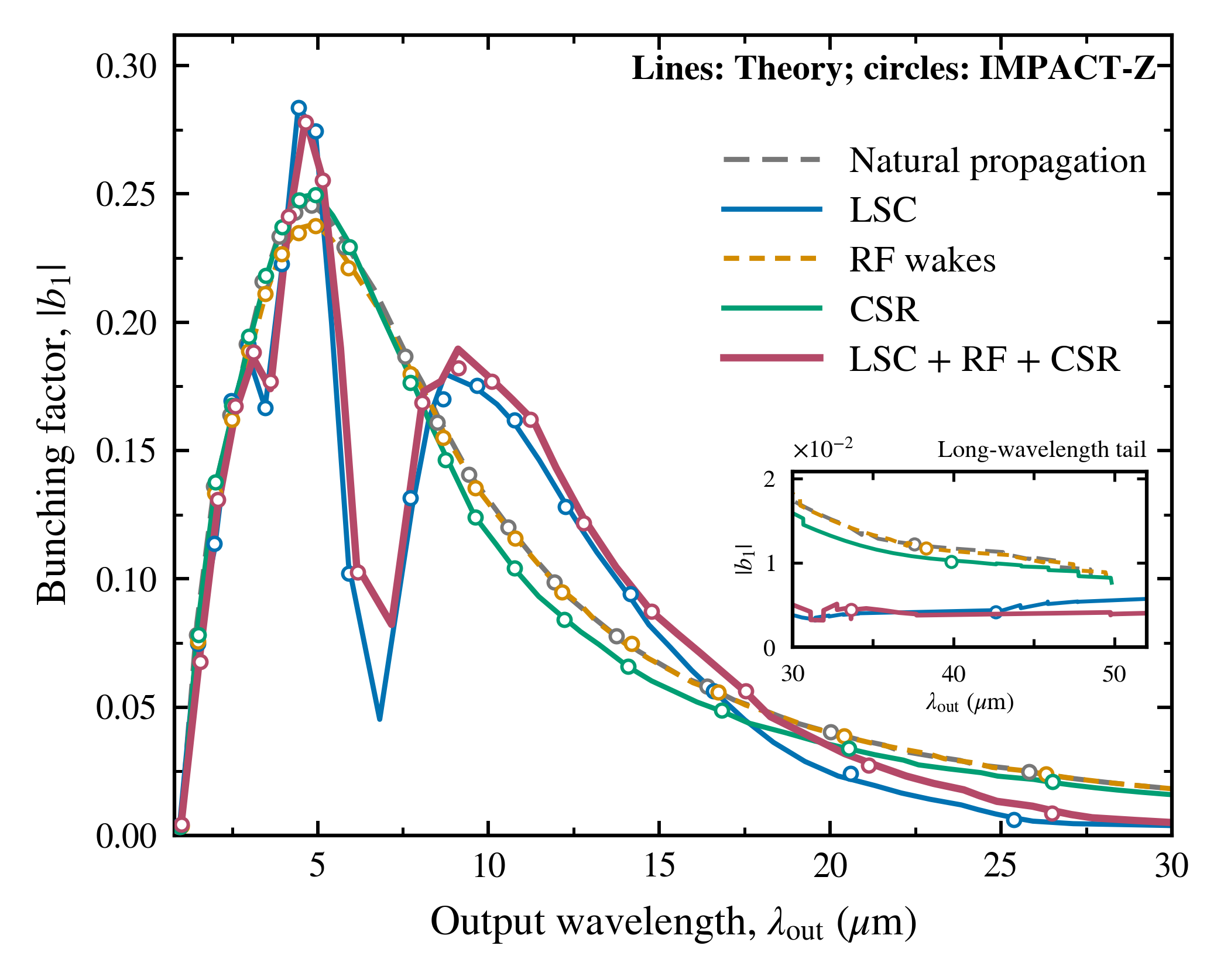}}\hspace{0.01\textwidth}%
\subfloat[]{\includegraphics[width=0.32\textwidth]{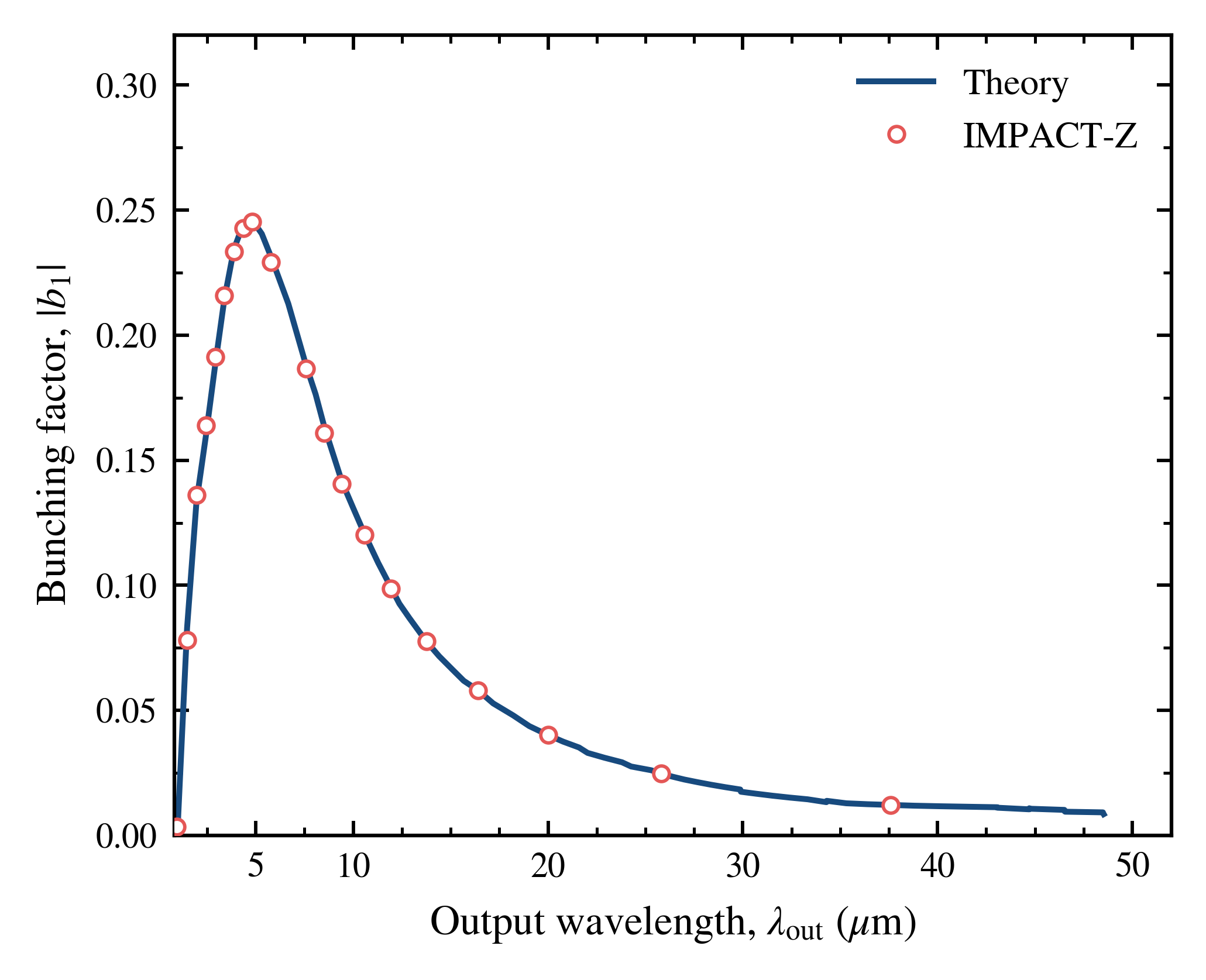}}\hspace{0.01\textwidth}%
\subfloat[]{\includegraphics[width=0.32\textwidth]{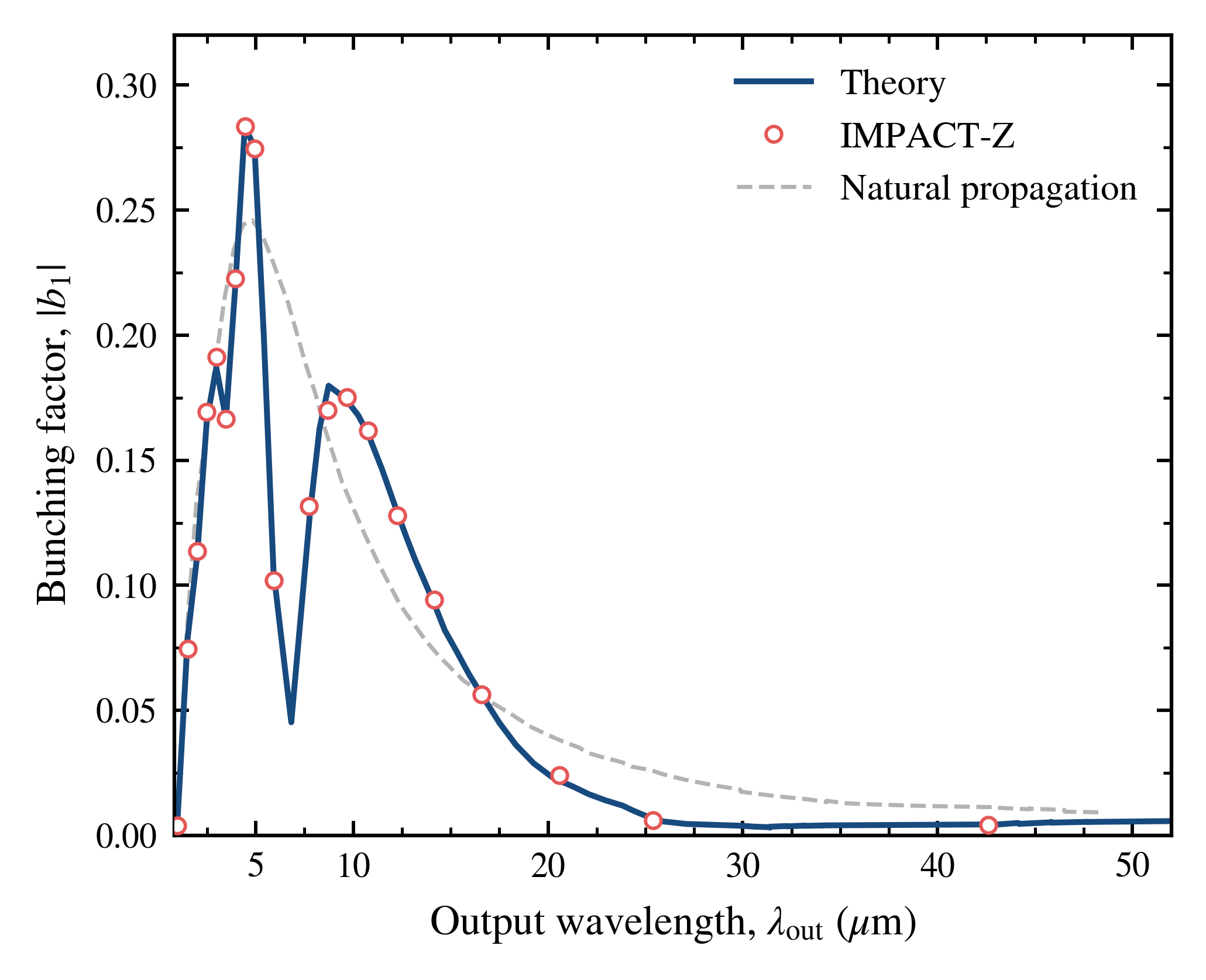}}\\[0.5em]
\subfloat[]{\includegraphics[width=0.32\textwidth]{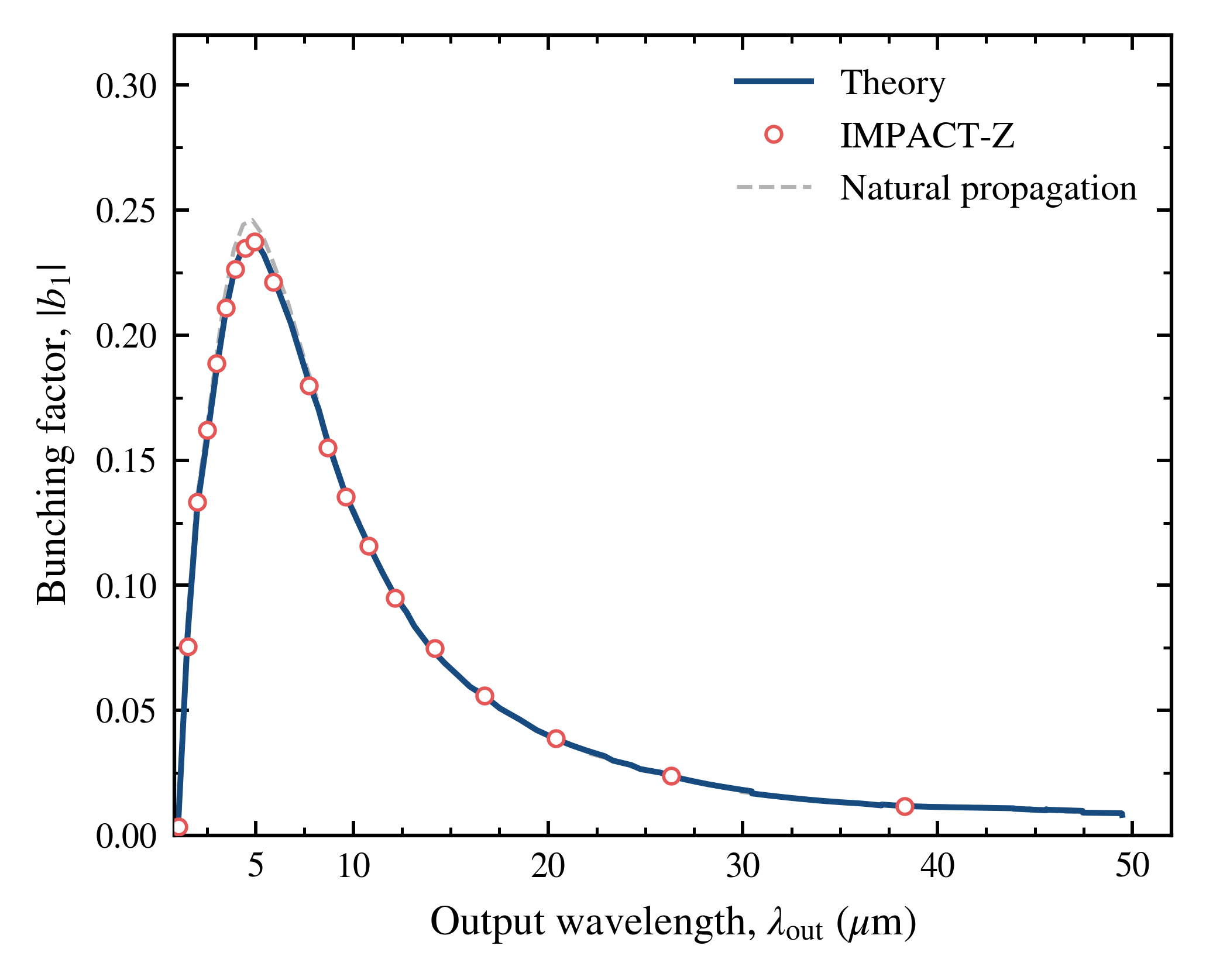}}\hspace{0.01\textwidth}%
\subfloat[]{\includegraphics[width=0.32\textwidth]{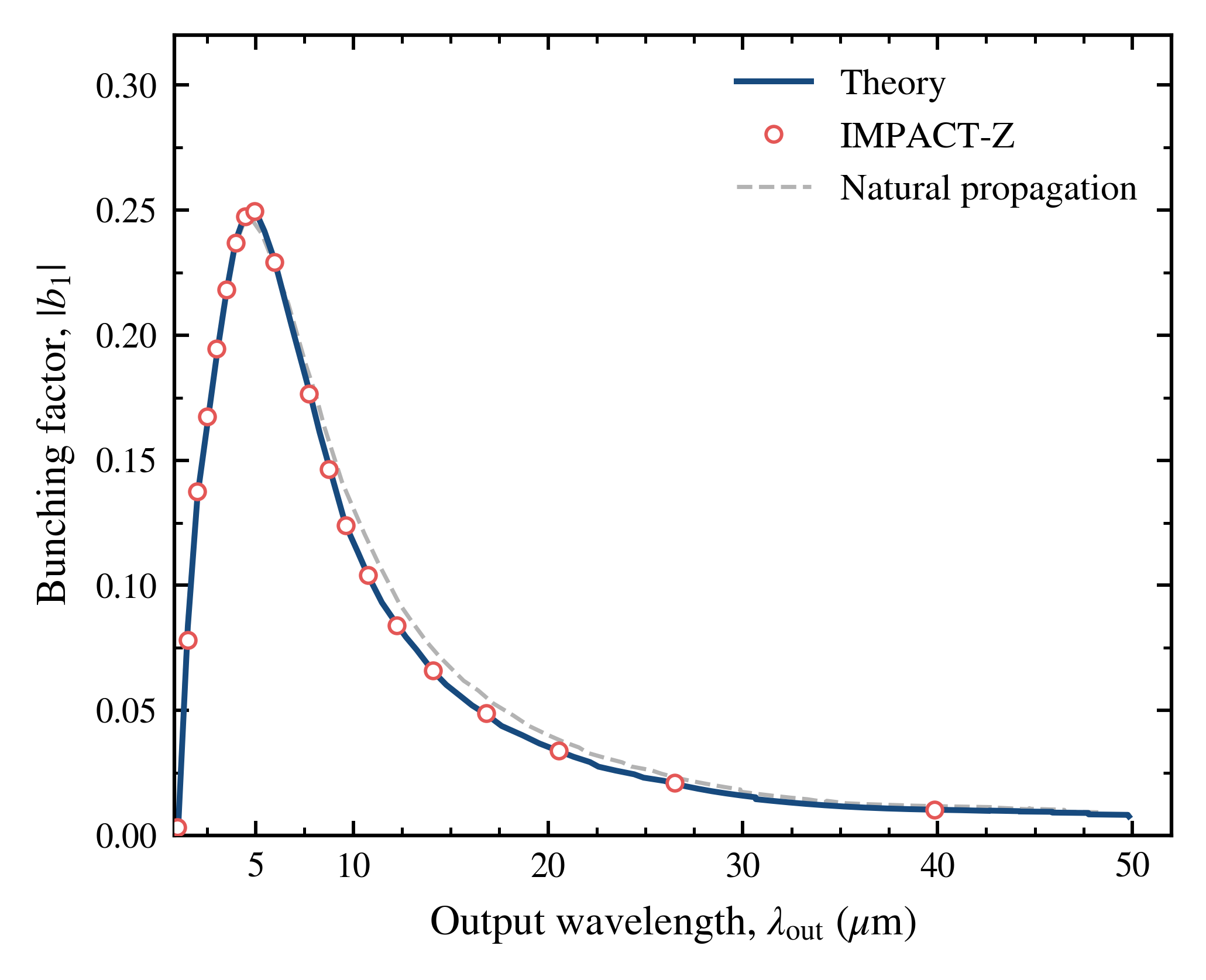}}\hspace{0.01\textwidth}%
\subfloat[]{\includegraphics[width=0.32\textwidth]{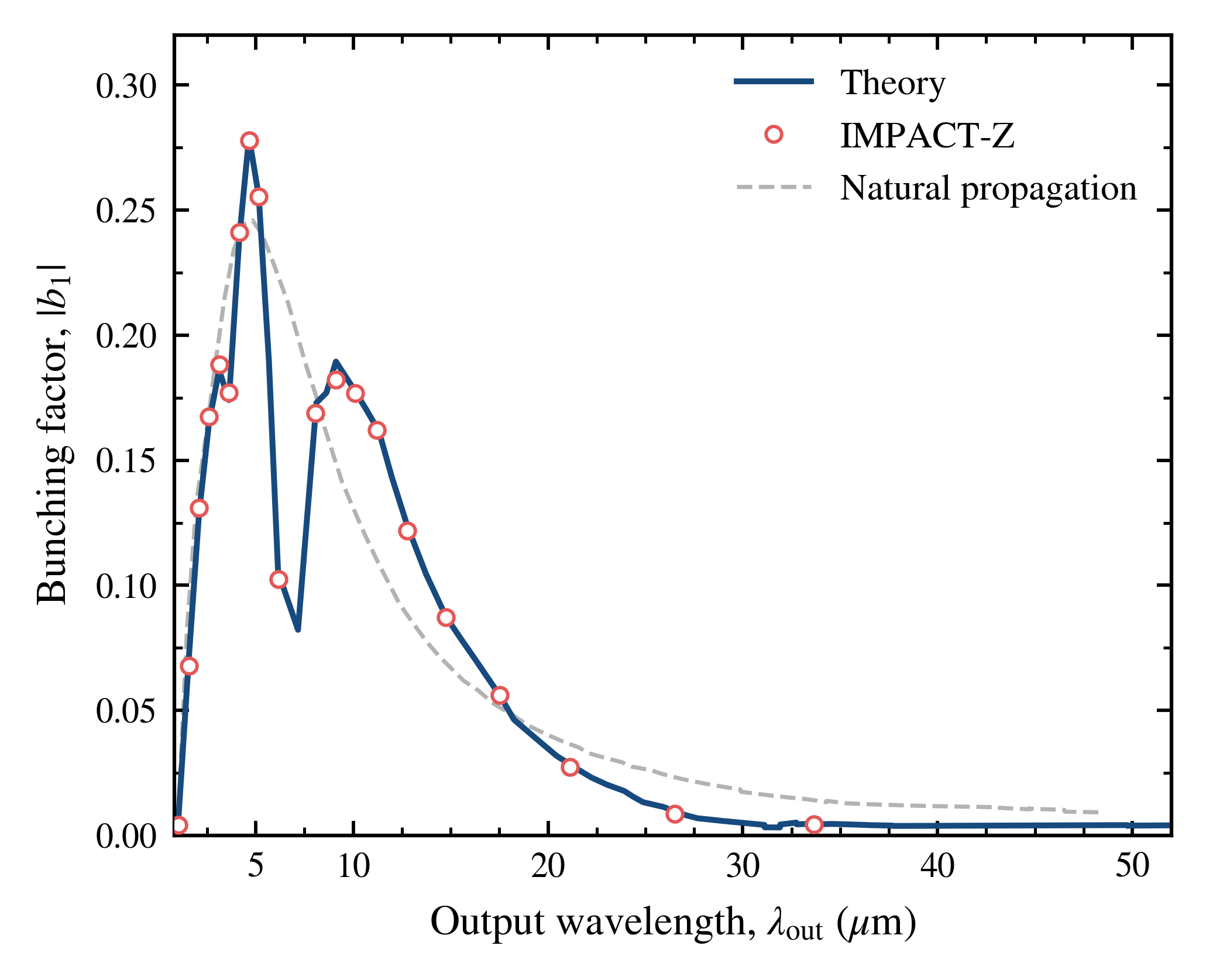}}
\caption{Collective-effect contributions to the linac-exit bunching response at $P_L=117.692\,\mathrm{kW}$ and nominal compression settings $C_1=6.5$ and $C_2=10/6.5$. Local spectral maxima of $|b_1|$ are plotted against their extracted output wavelengths as the input beat frequency is varied. Panel (a) compares all five configurations; colors identify the effects, lines denote theory, and open circles denote IMPACT-Z. Its inset shows the long-wavelength tail. Panels (b)--(f) show natural propagation, LSC only, RF wakes only, CSR only, and all three effects, respectively. In these panels, blue solid lines denote theory and red open circles denote IMPACT-Z; gray dashed lines in (c)--(f) show the natural-propagation theory.}
\label{fig:collective_decomposition}
\end{figure*}

The contributions of the collective effects are compared in Fig.~\ref{fig:collective_decomposition}(a) at $P_L=117.692\,\mathrm{kW}$ and nominal compression settings $C_1=6.5$ and $C_2=10/6.5$. The source conditions and external lattice are held fixed while downstream LSC, RF wakes, and CSR are enabled separately or together. Natural propagation denotes transport with all three downstream collective effects disabled. For each input beat frequency, the local spectral maximum of $|b_1|$ is plotted against its extracted linac-exit wavelength. The individual configurations are resolved in panels (b)--(f), with the natural-propagation theory retained as a common reference. The amplitudes and wavelength dependence obtained with IMPACT-Z are closely reproduced by the theory for all five configurations, including the main peaks and the long-wavelength tails.

The largest change from natural propagation in panel (b) is produced by LSC in panel (c). The single broad maximum of the natural-propagation response is replaced by two short-wavelength maxima near $4$--$5$ and $9\,\mu\mathrm{m}$, separated by a pronounced minimum near $7\,\mu\mathrm{m}$. This structure is retained when all collective effects are included in panel (f). RF wakes alone in panel (d) produce a smaller change from the reference response, whereas the longer-wavelength side of the main maximum is mainly reduced by CSR in panel (e). The pronounced double peak is absent in both cases. The formation of the short-wavelength double-peak structure is therefore governed primarily by LSC under these conditions, with RF wakes and CSR modifying its amplitudes and spectral shape.

The model provides a computationally efficient alternative to end-to-end tracking with Pelegant for heater-source generation and IMPACT-Z for subsequent collective transport, each run on 96 cores. Laser and compression settings can be scanned efficiently while the wavelength-dependent bunching response is retained. Accelerator operating points for enhanced THz emission can thereby be identified to guide experimental tuning.

\begin{figure*}[width=\textwidth,pos=tp,align=\centering]
\centering
\subfloat[]{\includegraphics[width=0.24\textwidth]{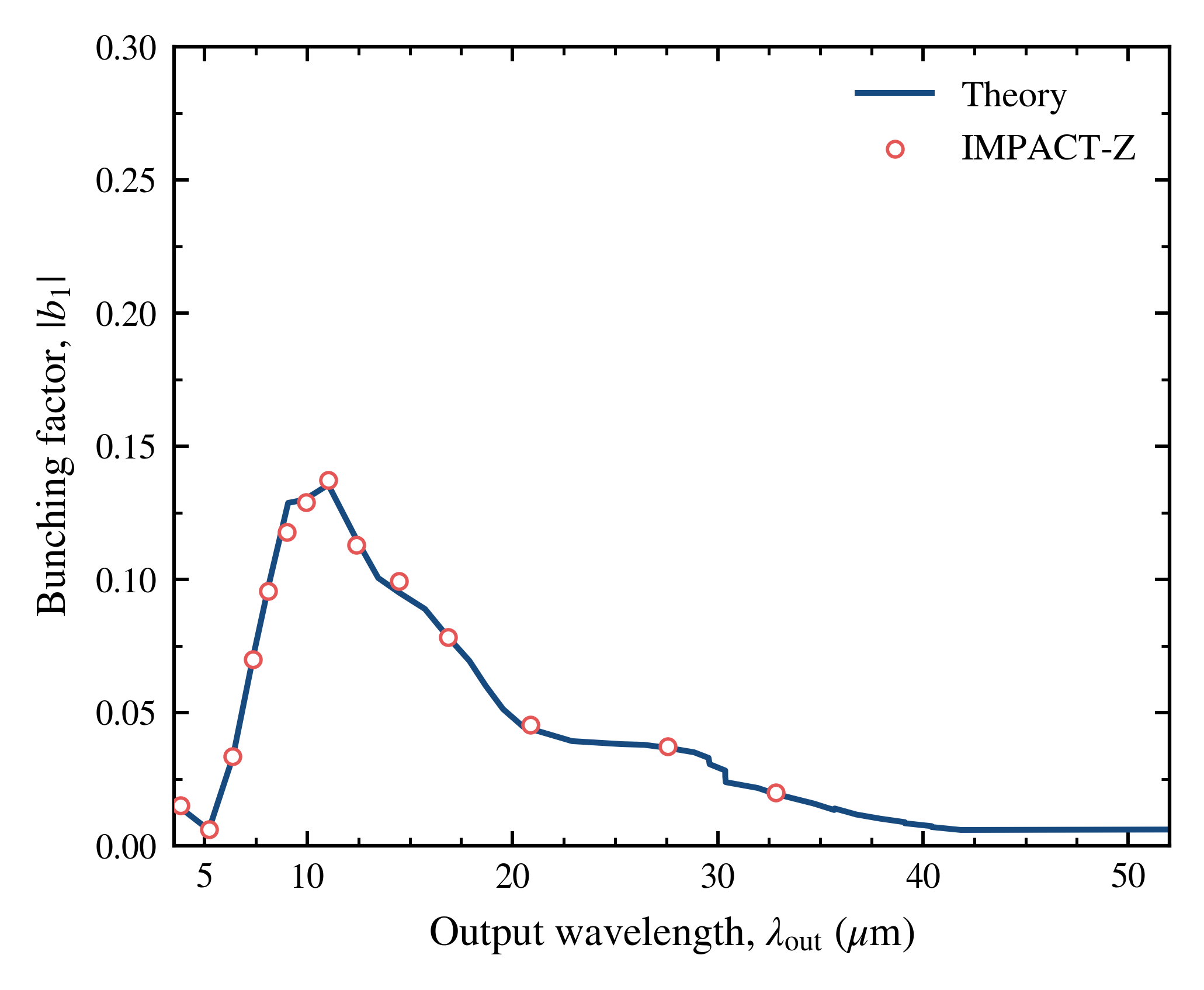}}\hspace{0.01\textwidth}%
\subfloat[]{\includegraphics[width=0.24\textwidth]{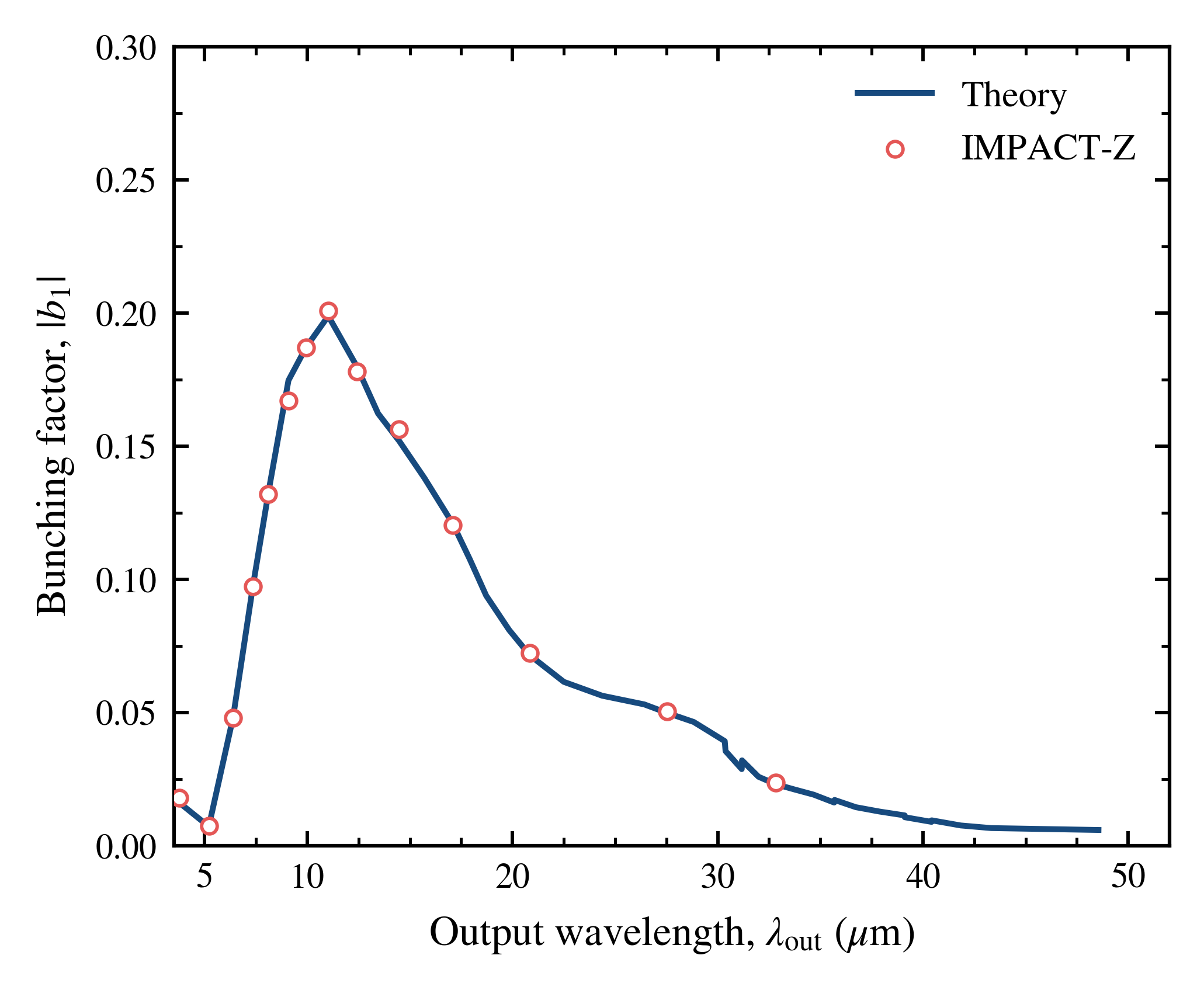}}\hspace{0.01\textwidth}%
\subfloat[]{\includegraphics[width=0.24\textwidth]{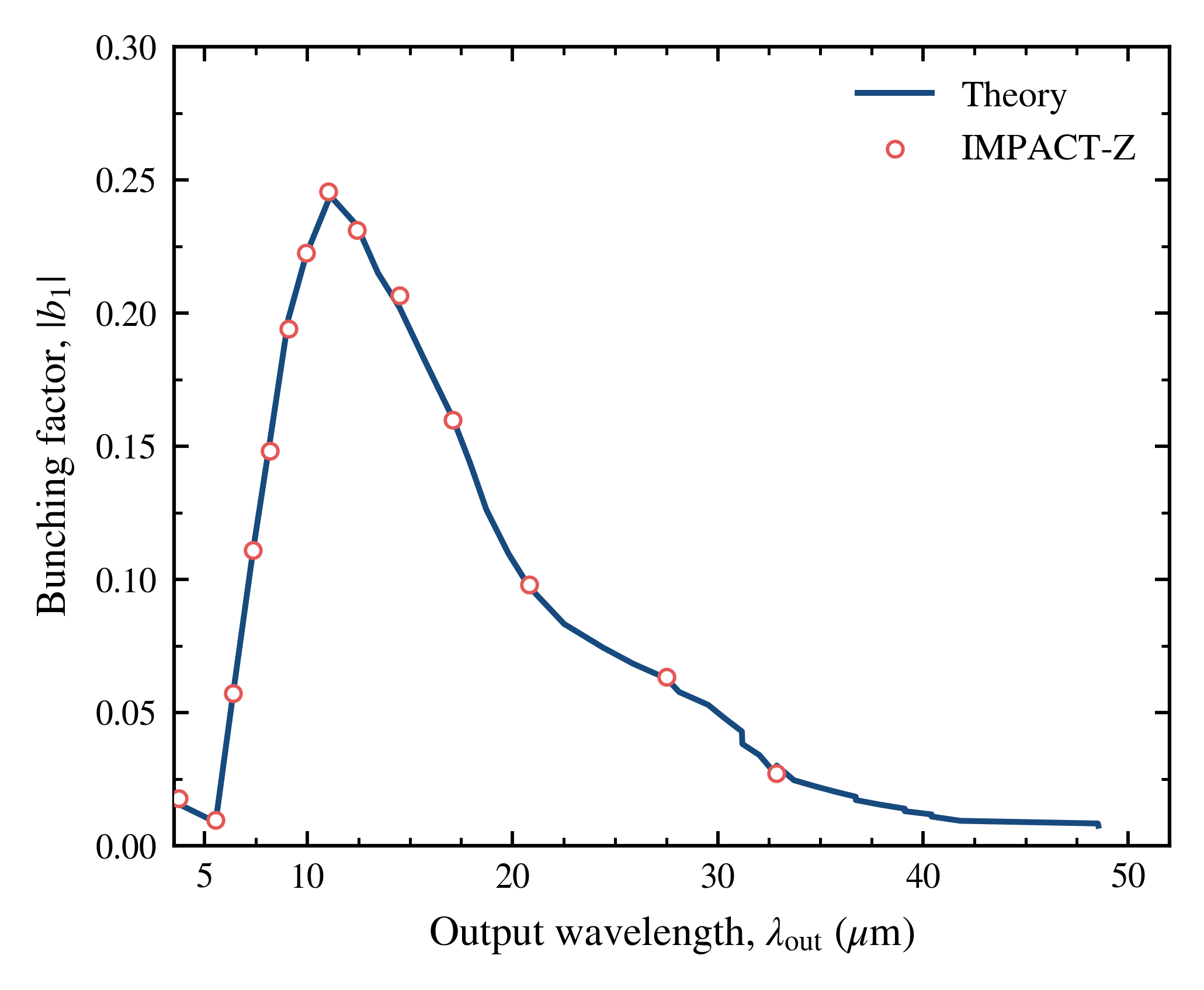}}\hspace{0.01\textwidth}%
\subfloat[]{\includegraphics[width=0.24\textwidth]{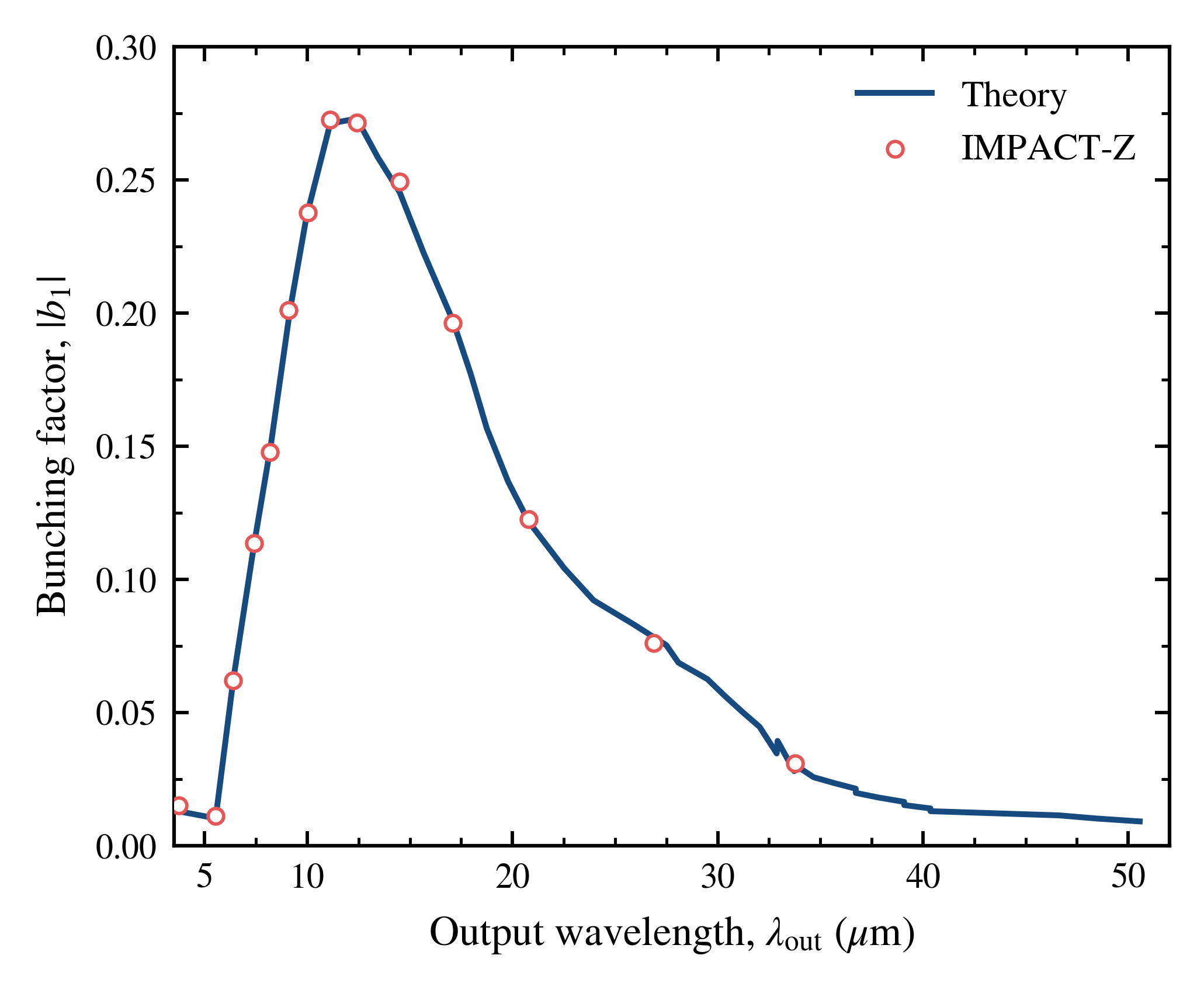}}
\caption{Linac-exit bunching response at nominal compression settings $C_1=8$ and $C_2=1.25$ for $P_L=30$, $50$, $70$, and $90\,\mathrm{kW}$ in panels (a)--(d), respectively. Local spectral maxima of the standard bunching factor $|b_1|$ are plotted against their extracted output wavelengths as the input beat frequency is varied. Blue solid lines denote theory and red open circles denote IMPACT-Z.}
\label{fig:power_scan}
\end{figure*}

The strength of the laser-imprinted source provides an additional control of the collective response. For a fixed temporal profile, the effective pulse energy is proportional to the peak power through Eq.~\eqref{eq:effective_laser_energy}. With LSC, RF wakes, and CSR included, the laser-power dependence of the linac-exit bunching response is shown in Fig.~\ref{fig:power_scan} for $P_L=30$, $50$, $70$, and $90\,\mathrm{kW}$ at nominal compression settings $C_1=8$ and $C_2=1.25$. The theoretical peak bunching factors are $0.136$, $0.199$, $0.244$, and $0.273$, respectively, compared with $0.137$, $0.201$, $0.246$, and $0.273$ in IMPACT-Z. The growth toward the dominant maximum and the subsequent long-wavelength decay are reproduced at all four powers. At $50\,\mathrm{kW}$, both maxima occur near $11.03\,\mu\mathrm{m}$, with a relative amplitude difference of approximately $1\%$.

The dominant response is concentrated near $11$--$12.5\,\mu\mathrm{m}$. Its amplitude increases with laser power, whereas the peak region broadens at $90\,\mathrm{kW}$. At this highest power, the sampled theoretical maximum occurs at $12.42\,\mu\mathrm{m}$ and the IMPACT-Z maximum at $11.12\,\mu\mathrm{m}$, with nearly equal peak amplitudes. The maximum absolute difference in $|b_1|$ at matched input frequencies remains below $0.011$ across the four scans. These comparisons support the predicted power dependence of the bunching response while resolving the residual differences in peak location and spectral shape.

It was previously demonstrated at SXFEL that the final bunching factor can be modified through compression-ratio adjustment while the final beam current is kept approximately constant~\cite{Cheng2023MBI}. In the present analysis, compression partition is treated as a practical control parameter for enhancing bunching near a prescribed output wavelength, and its spectral response is calculated with the finite-amplitude theory.

\begin{figure*}[width=\textwidth,pos=tp,align=\centering]
\centering
\subfloat[]{\includegraphics[width=0.32\textwidth]{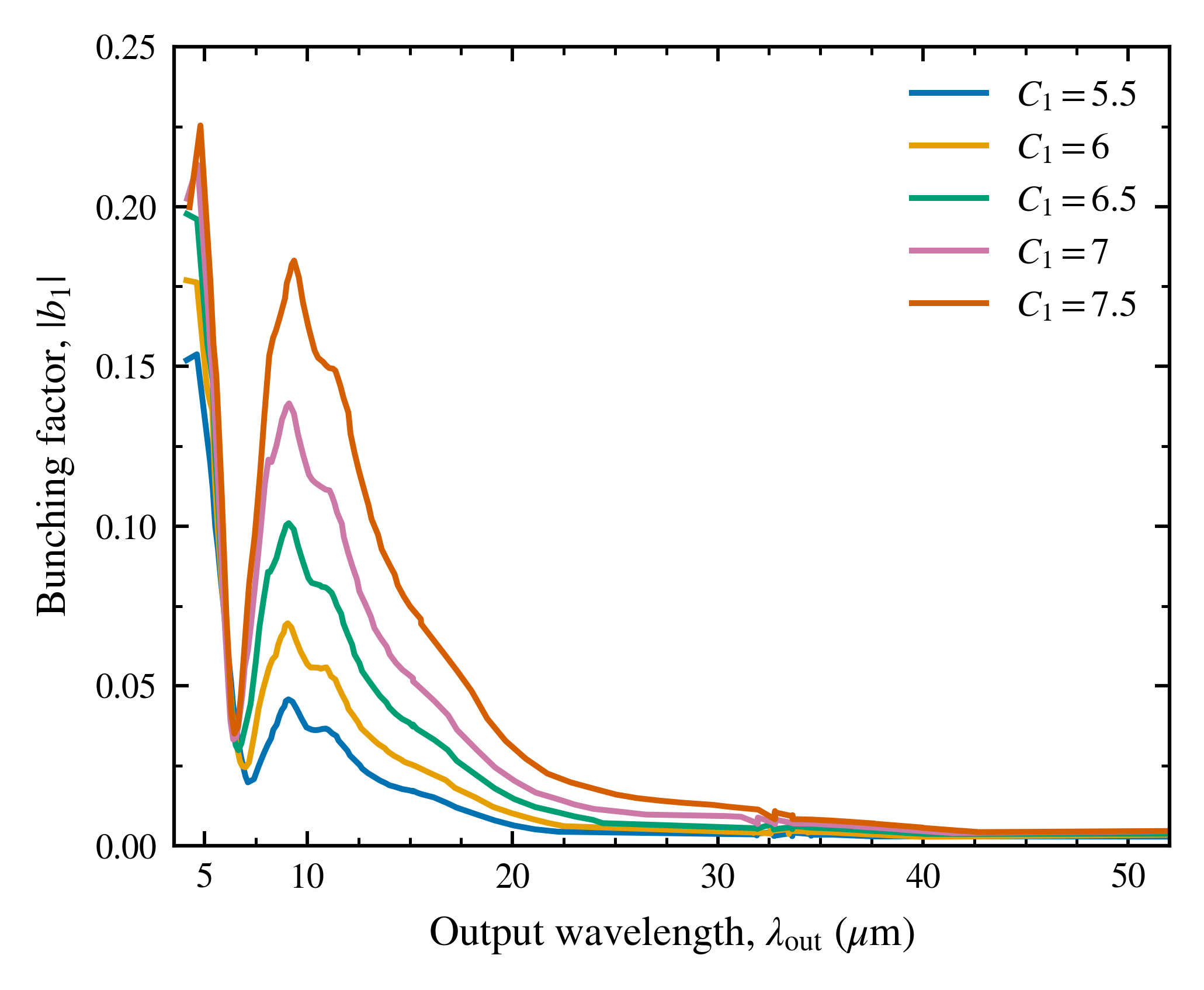}}\hspace{0.01\textwidth}%
\subfloat[]{\includegraphics[width=0.32\textwidth]{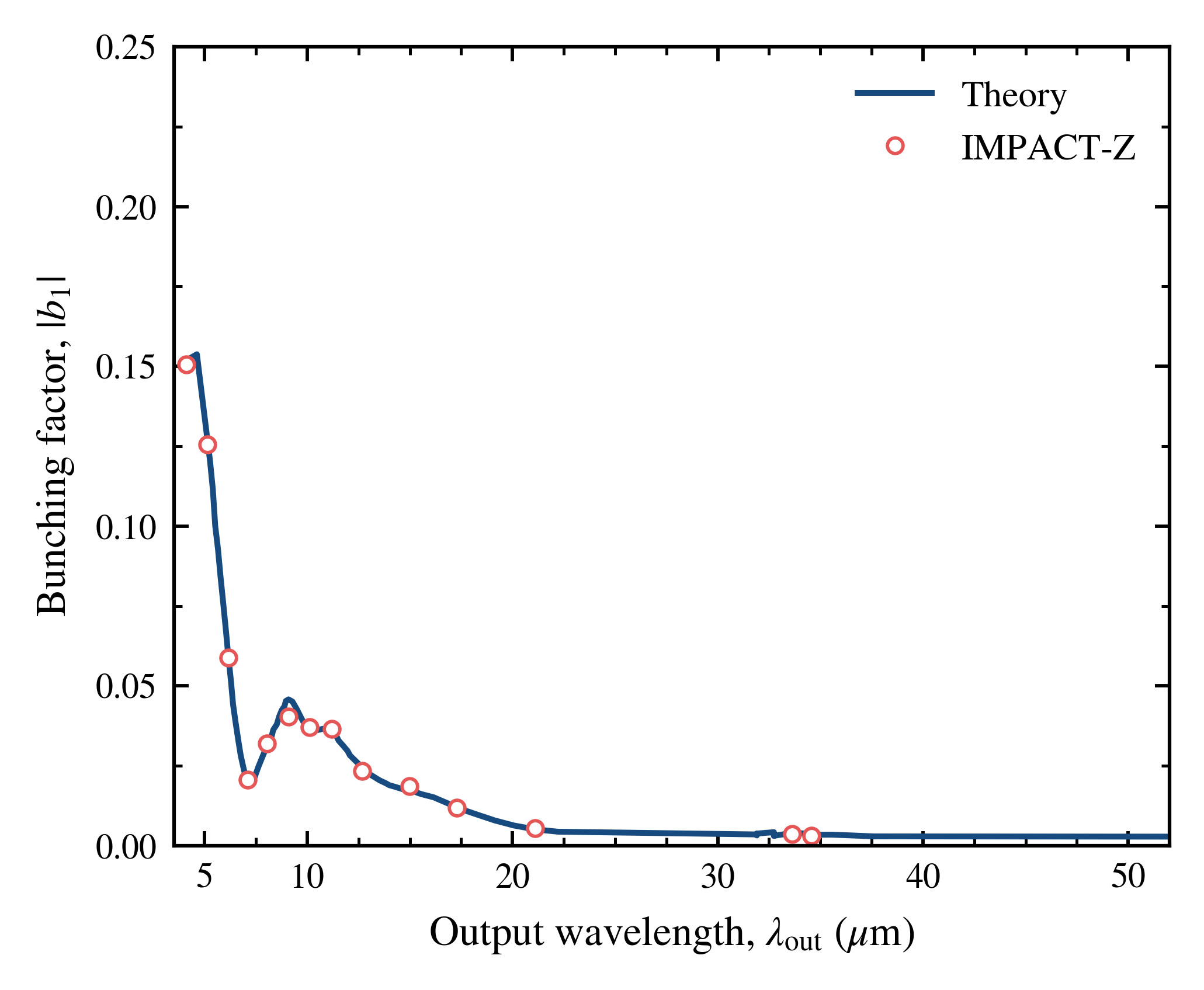}}\hspace{0.01\textwidth}%
\subfloat[]{\includegraphics[width=0.32\textwidth]{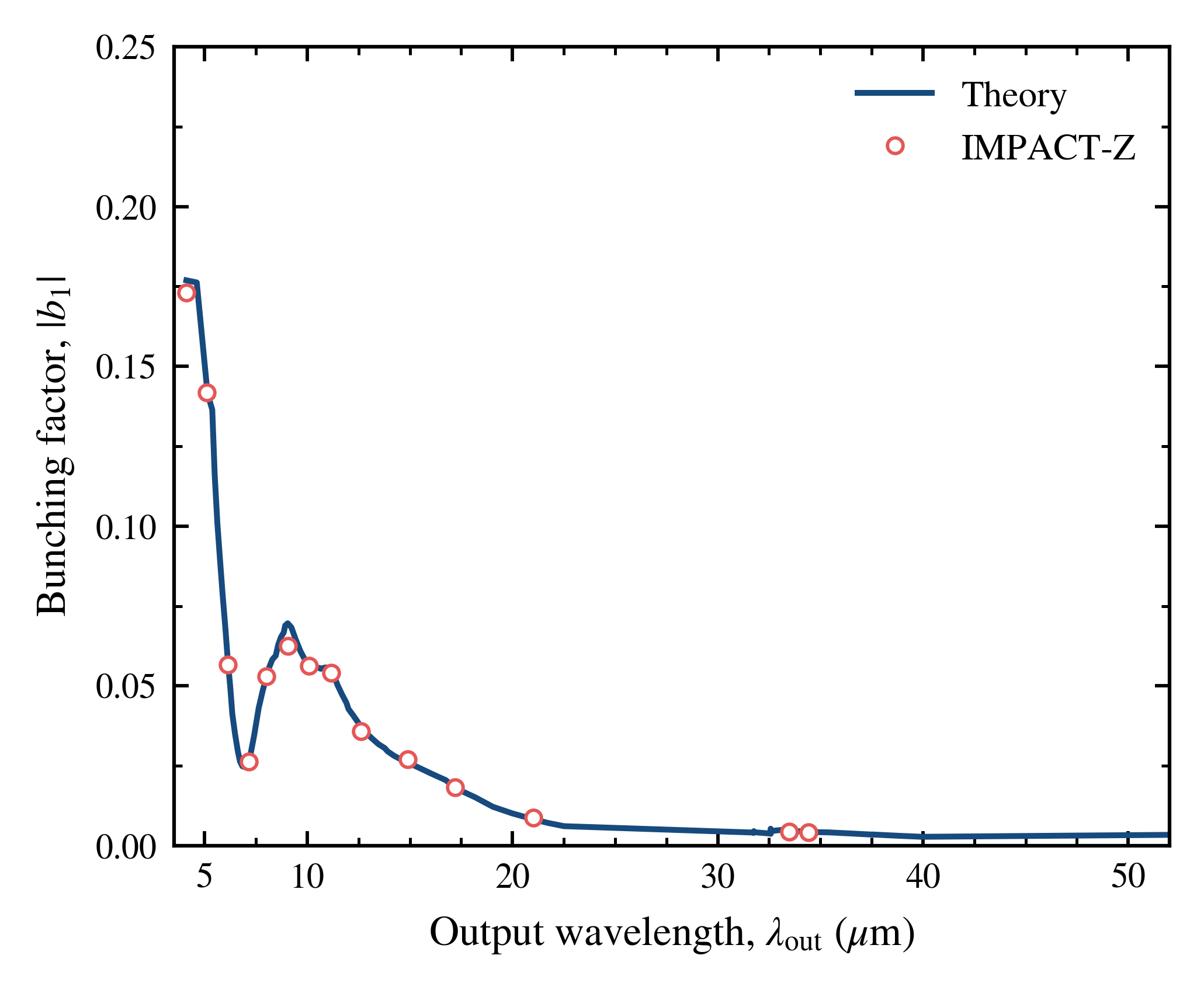}}\\[0.5em]
\subfloat[]{\includegraphics[width=0.32\textwidth]{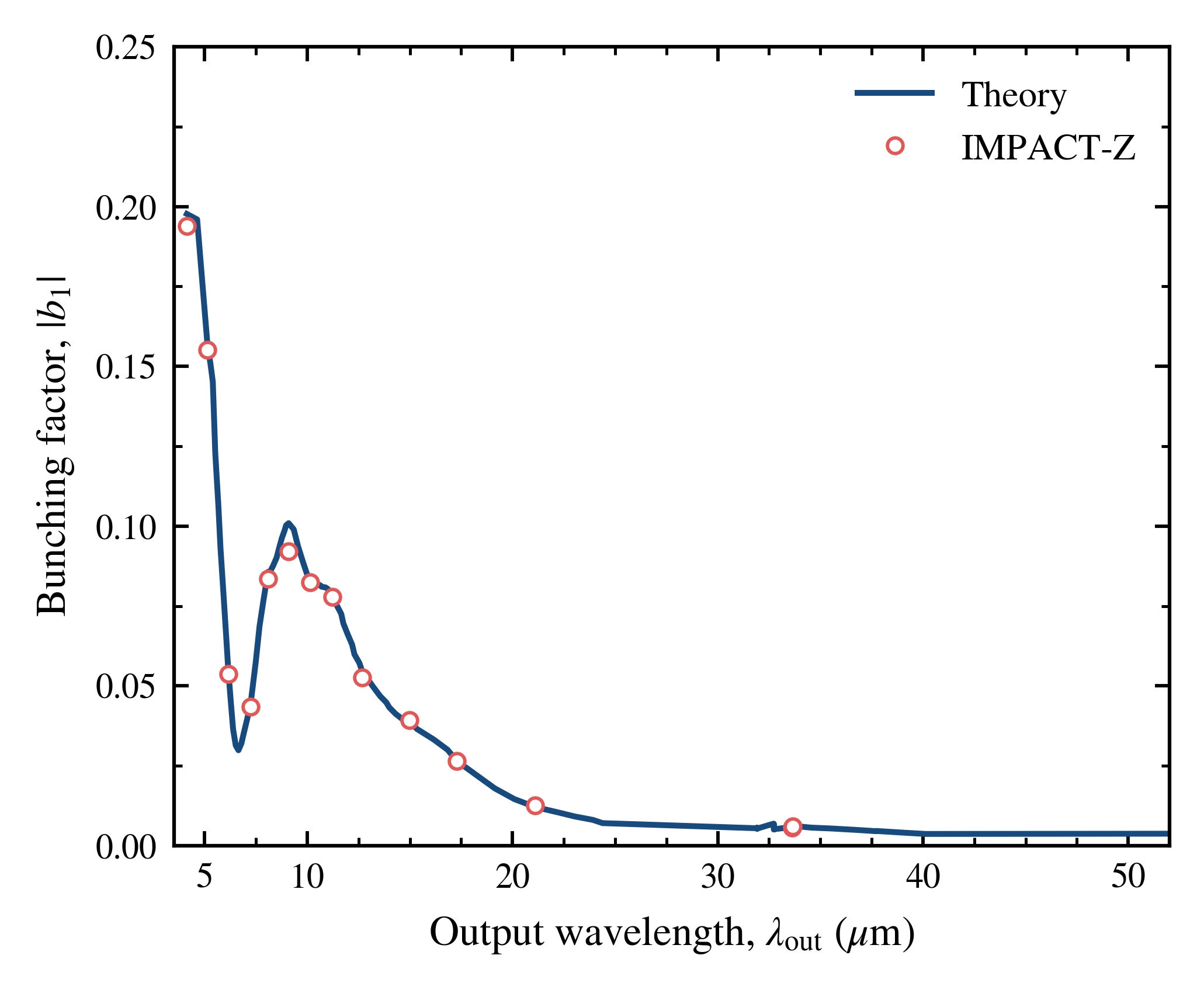}}\hspace{0.01\textwidth}%
\subfloat[]{\includegraphics[width=0.32\textwidth]{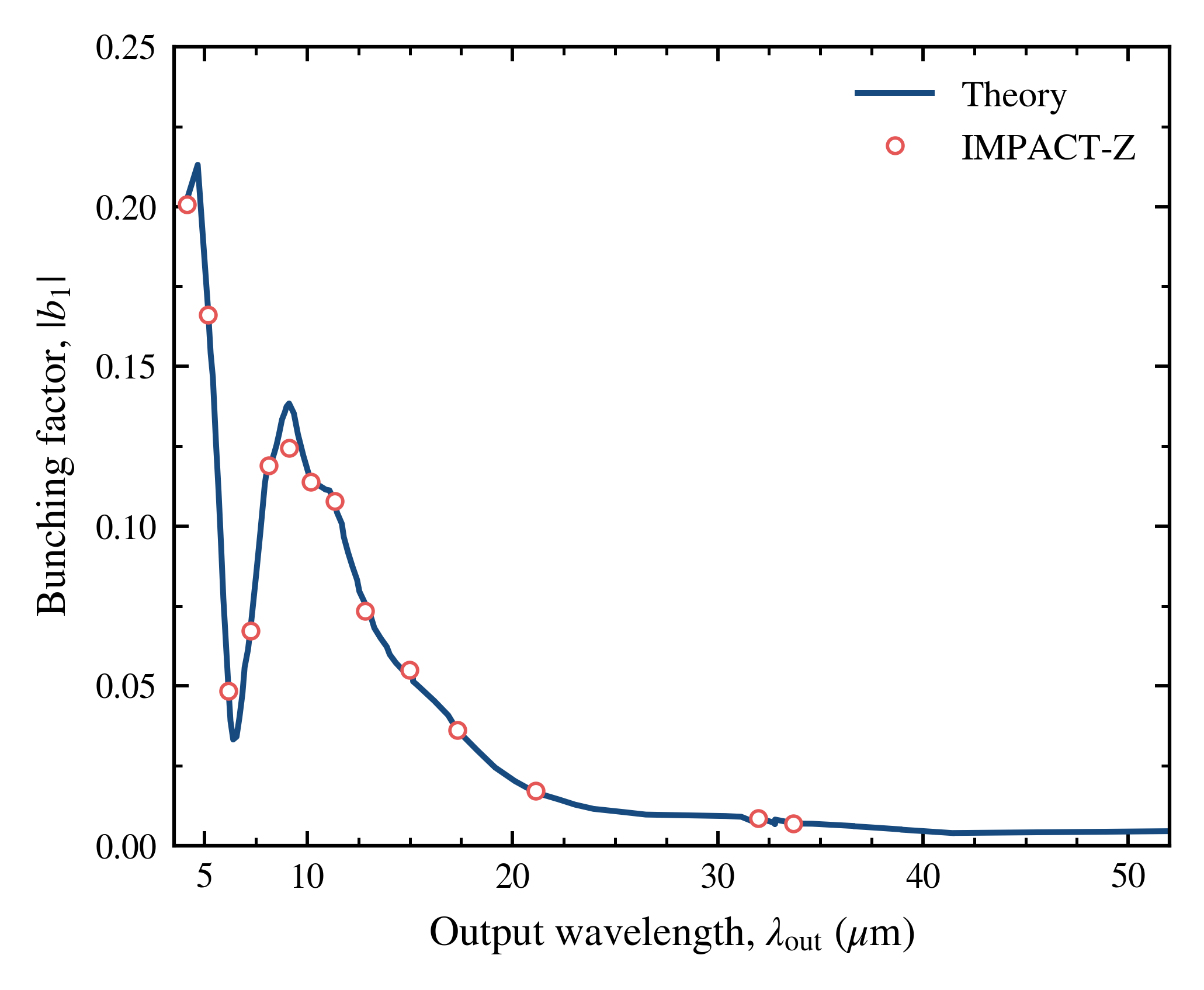}}\hspace{0.01\textwidth}%
\subfloat[]{\includegraphics[width=0.32\textwidth]{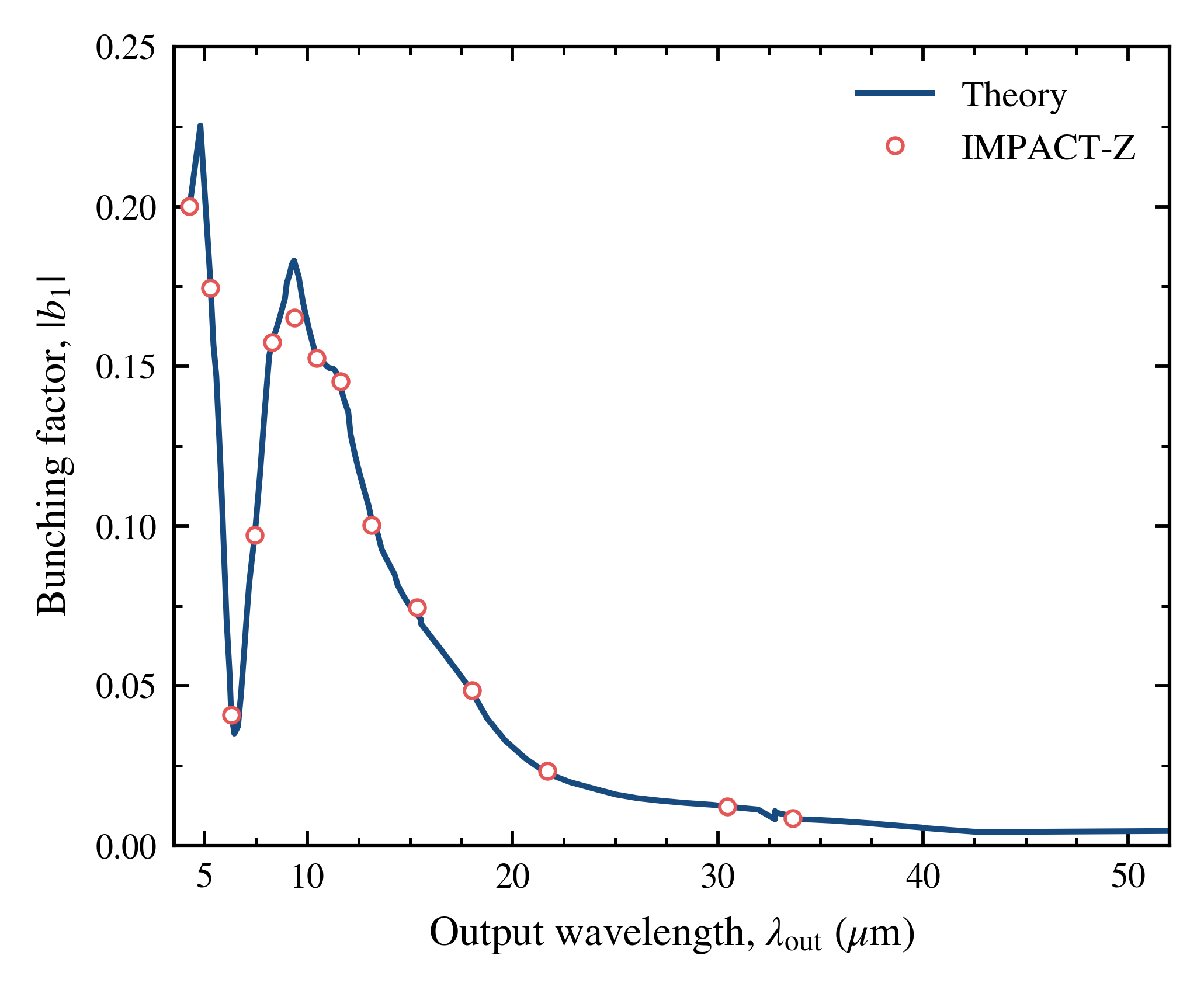}}
\caption{Compression-partition dependence of the linac-exit bunching response at $P_L=50\,\mathrm{kW}$, with nominal $C_1=5.5$--$7.5$ and $C_2=10/C_1$. Local spectral maxima of $|b_1|$ are plotted against their extracted output wavelengths as the input beat frequency is varied. Panel (a) compares the five theoretical responses. Panels (b)--(f) correspond to $C_1=5.5$, $6$, $6.5$, $7$, and $7.5$, respectively; blue solid lines denote theory and red open circles denote IMPACT-Z. LSC, RF wakes, and CSR are included in all cases.}
\label{fig:compression_partition}
\end{figure*}

The compression-partition dependence at $P_L=50\,\mathrm{kW}$ is shown in Fig.~\ref{fig:compression_partition}. The nominal first-stage compression is varied from $C_1=5.5$ to $7.5$, with $C_2=10/C_1$ and all collective effects included. Panel (a) compares the theoretical responses, and panels (b)--(f) compare theory and IMPACT-Z at each setting. The intermediate current and collective phase evolution are changed by the compression partition, so the output bunching is not determined by the total compression alone.

A short-wavelength maximum near $4$--$5\,\mu\mathrm{m}$ and a second peak near $9\,\mu\mathrm{m}$ are resolved. In the $8$--$12\,\mu\mathrm{m}$ interval, the theoretical peak bunching factors are $0.046$, $0.070$, $0.101$, $0.138$, and $0.183$ for increasing $C_1$, compared with $0.040$, $0.062$, $0.092$, $0.124$, and $0.165$ in IMPACT-Z. The peak wavelength changes from approximately $9.09$ to $9.36\,\mu\mathrm{m}$ across the scan. Enhanced bunching in this wavelength interval is therefore obtained when a larger share of compression is assigned to BC1.

The double-peak response and long-wavelength decay are reproduced across the five settings. The theoretical amplitude exceeds the IMPACT-Z result near the $9\,\mu\mathrm{m}$ peak, with the absolute difference increasing from approximately $0.005$ to $0.018$ as $C_1$ is increased. Over the full sampled wavelength range, the theoretical maxima increase from $0.154$ to $0.225$, whereas the sampled IMPACT-Z maxima remain near $0.20$ at $C_1=7$ and $7.5$. The dependence on compression partition is thus wavelength dependent; the enhancement near $9\,\mu\mathrm{m}$ does not imply uniform growth of the spectral maximum throughout the scan.

The scans distinguish the roles of laser strength and compression partition. The initial modulation and phase mixing are changed by the laser power. The intermediate current and collective evolution are changed by the compression partition. Both controls affect the final bunching, with a response that depends on wavelength.

\section{Experimental comparison}
\label{sec:experiment}

The experiment was performed at SXFEL with the beamline in Fig.~\ref{fig1}. The slice energy spread was modulated in the laser heater and converted into density modulation during two-stage compression. At the linac exit, the longitudinal coordinate was mapped onto one screen axis by the TDX. Energy was mapped onto the orthogonal axis by the spectrometer dipole, allowing the longitudinal phase space to be measured.

The measurements were performed at approximately tenfold total compression. The incoming peak current of about $\SI{60}{A}$ was increased to about $\SI{400}{A}$ after BC1. This corresponds to a first-stage compression near $6.5$, with a second-stage compression near $1.5$. The other beam parameters are listed in Table~\ref{tab:beam_parameters}.

As described in Sec.~\ref{sec:theory}, the input beat frequency was controlled by varying the delay $\tau$ between the two chirped laser replicas. The available delay range of $\SIrange{1.72}{8.38}{ps}$ corresponds through $f_b=|\alpha|\tau$ to $f_b=\SIrange{0.783}{3.817}{THz}$. The delivered laser pulse energy $E_L$ was controlled independently by varying the optical attenuation and covered approximately $\SIrange{1.2}{9.8}{\micro\joule}$. For each laser setting, the beam was transported to the linac exit, where its longitudinal phase space was recorded with the TDX diagnostic.

\begin{figure*}[width=\textwidth,pos=tp,align=\centering]
\centering
\subfloat[]{\includegraphics[width=0.24\textwidth]{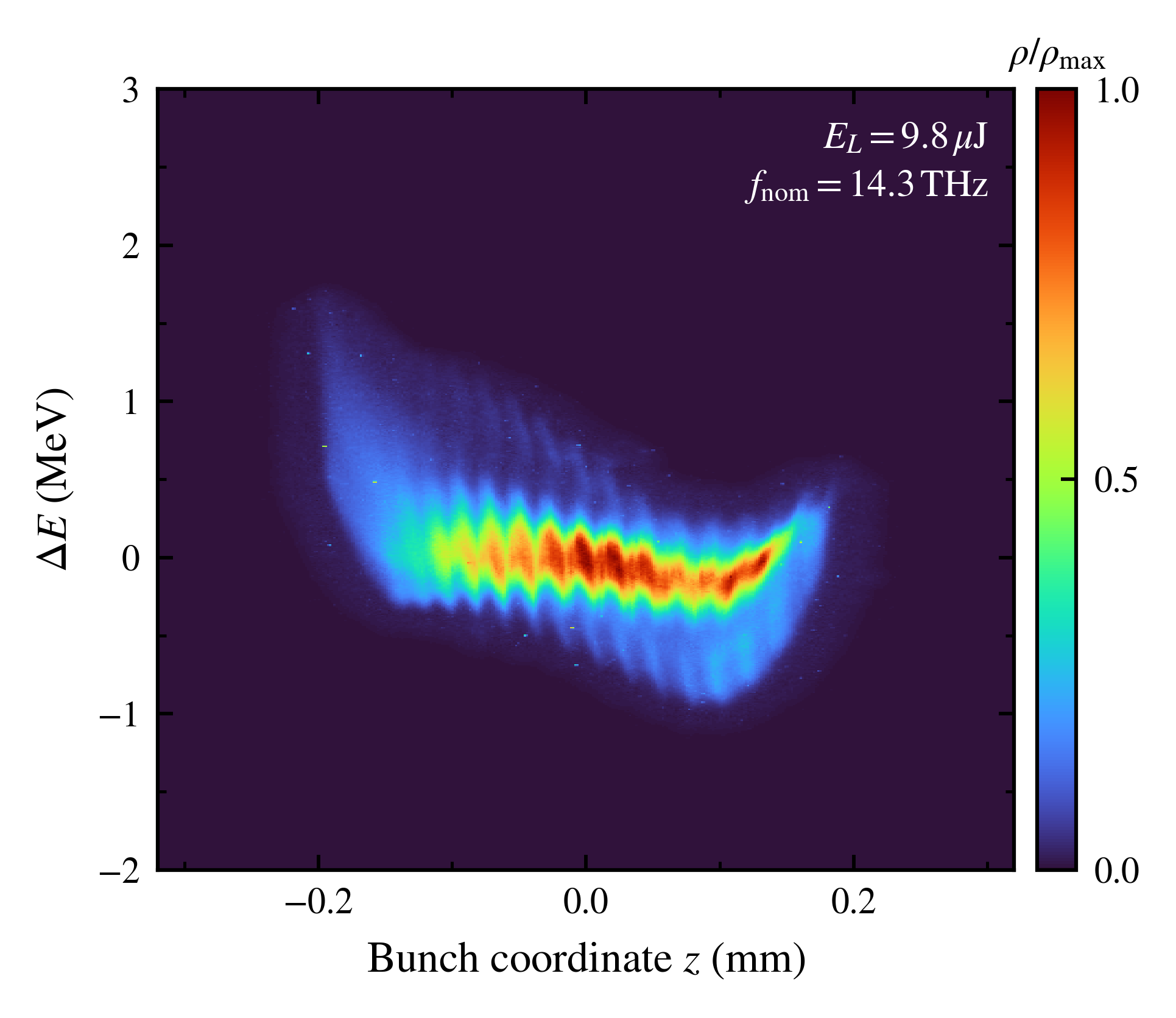}}\hspace{0.01\textwidth}%
\subfloat[]{\includegraphics[width=0.24\textwidth]{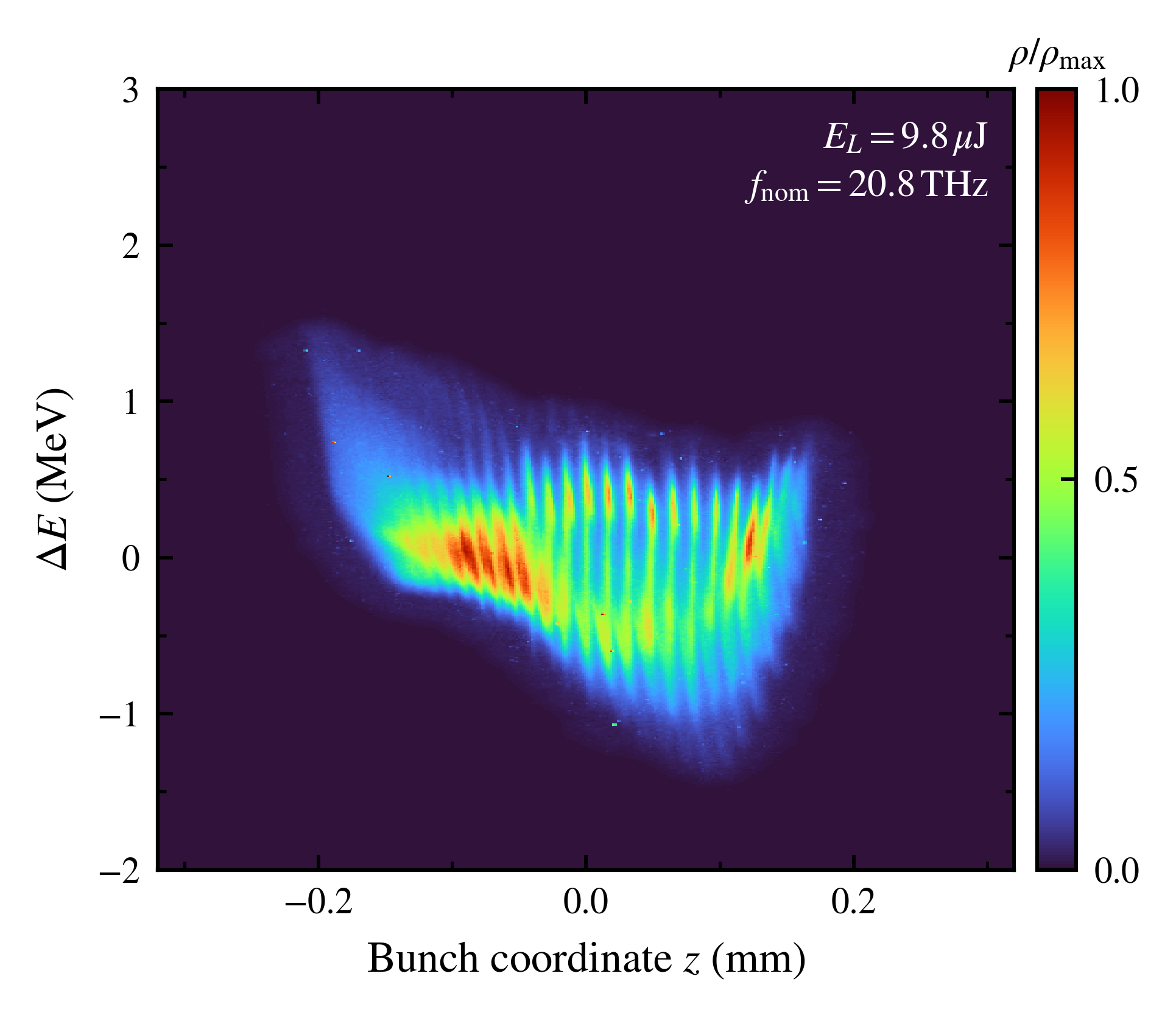}}\hspace{0.01\textwidth}%
\subfloat[]{\includegraphics[width=0.24\textwidth]{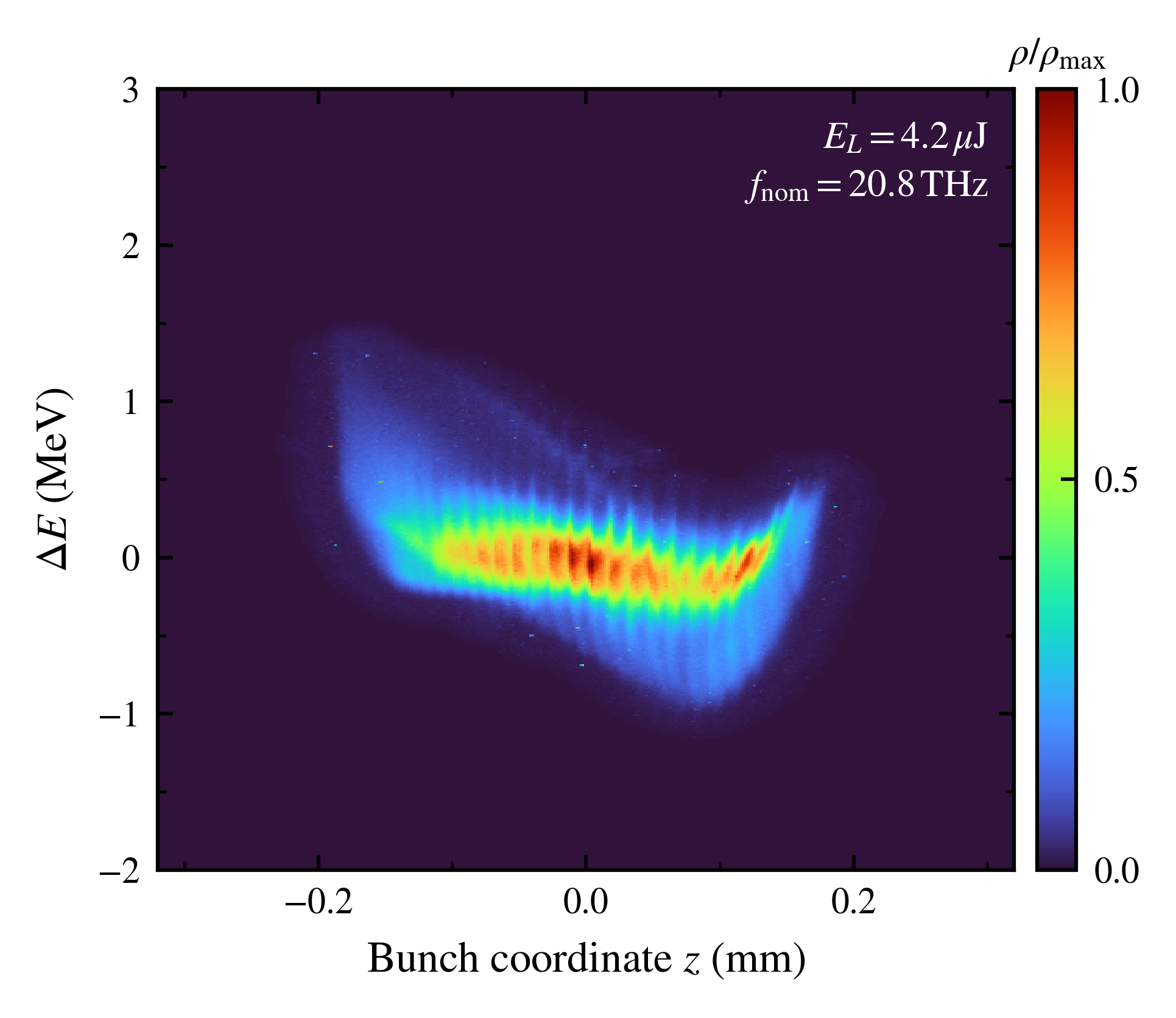}}\hspace{0.01\textwidth}%
\subfloat[]{\includegraphics[width=0.24\textwidth]{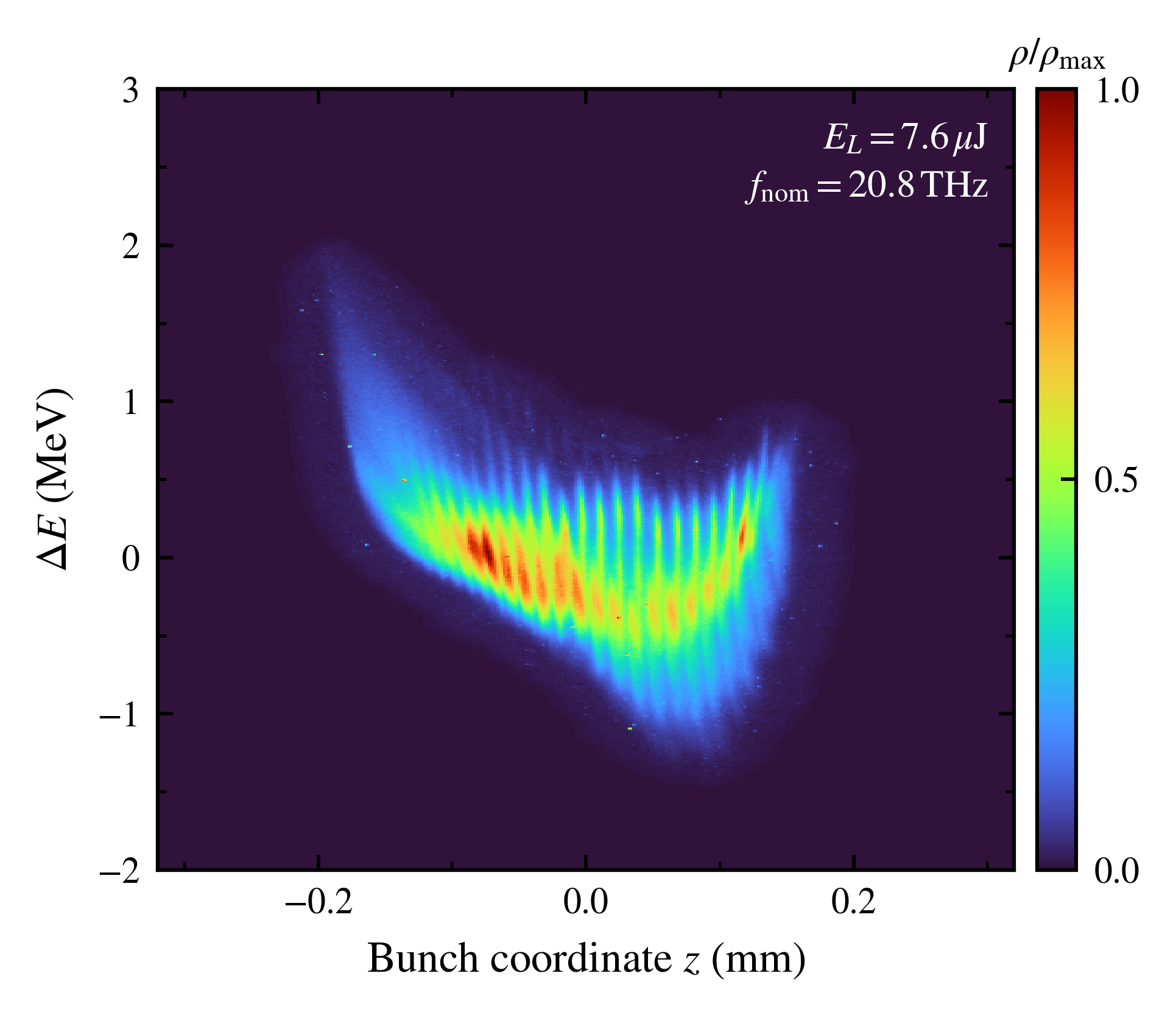}}
\caption{TDX-measured linac-exit longitudinal phase spaces for $(E_L,f_b)=(9.8\,\mu\mathrm{J},1.43\,\mathrm{THz})$, $(9.8\,\mu\mathrm{J},2.08\,\mathrm{THz})$, $(4.2\,\mu\mathrm{J},2.08\,\mathrm{THz})$, and $(7.6\,\mu\mathrm{J},2.08\,\mathrm{THz})$ in panels (a)--(d), respectively. Background-subtracted phase-space densities are normalized to their respective maxima. The labels $f_{\rm nom}$ denote the nominal beam modulation frequencies for a total compression of ten, $f_{\rm nom}=10f_b$.}
\label{fig:experiment_frequency_phase_space}
\end{figure*}

Representative measured phase spaces are shown in Fig.~\ref{fig:experiment_frequency_phase_space}. Panels (a) and (b) compare $f_b=1.43$ and $2.08\,\mathrm{THz}$ at $E_L=9.8\,\mu\mathrm{J}$; the longitudinal modulation period is shortened as the beat frequency is increased. Panels (c), (d), and (b) compare $E_L=4.2$, $7.6$, and $9.8\,\mu\mathrm{J}$ at fixed $f_b=2.08\,\mathrm{THz}$. A larger energy excursion and stronger modulation contrast are observed at higher pulse energy. The modulation spacing remains approximately unchanged. The labels $f_{\rm nom}=10f_b$ denote the nominal beam modulation frequencies after compression, $14.3\,\mathrm{THz}$ in (a) and $20.8\,\mathrm{THz}$ in (b)--(d), based on a total compression of ten. The labels indicate nominal frequencies; the spectral-peak wavelengths are extracted from the measured distributions. 

\begin{figure*}[width=\textwidth,pos=tp,align=\centering]
\centering
\subfloat[]{\includegraphics[width=0.32\textwidth]{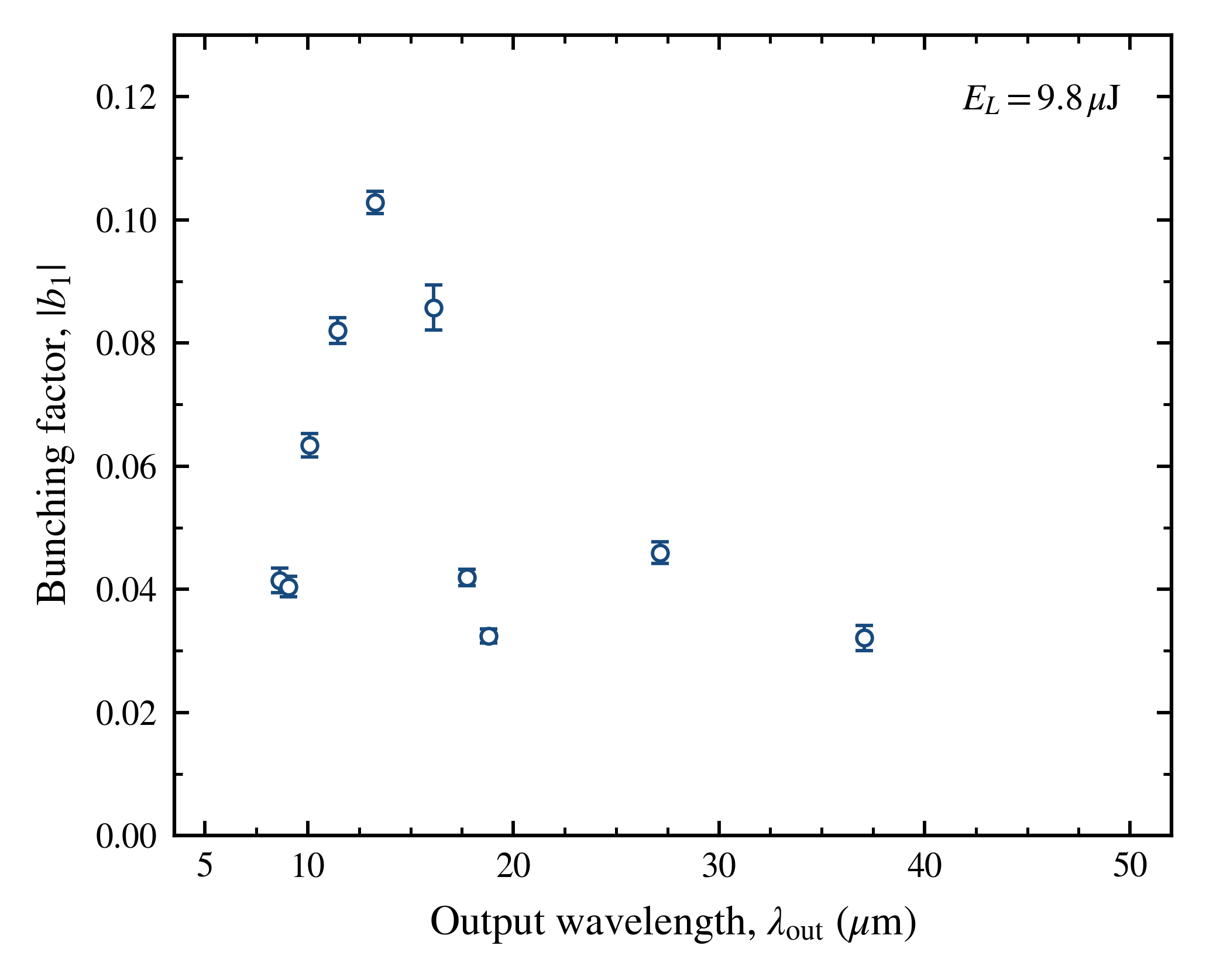}}\hspace{0.01\textwidth}%
\subfloat[]{\includegraphics[width=0.32\textwidth]{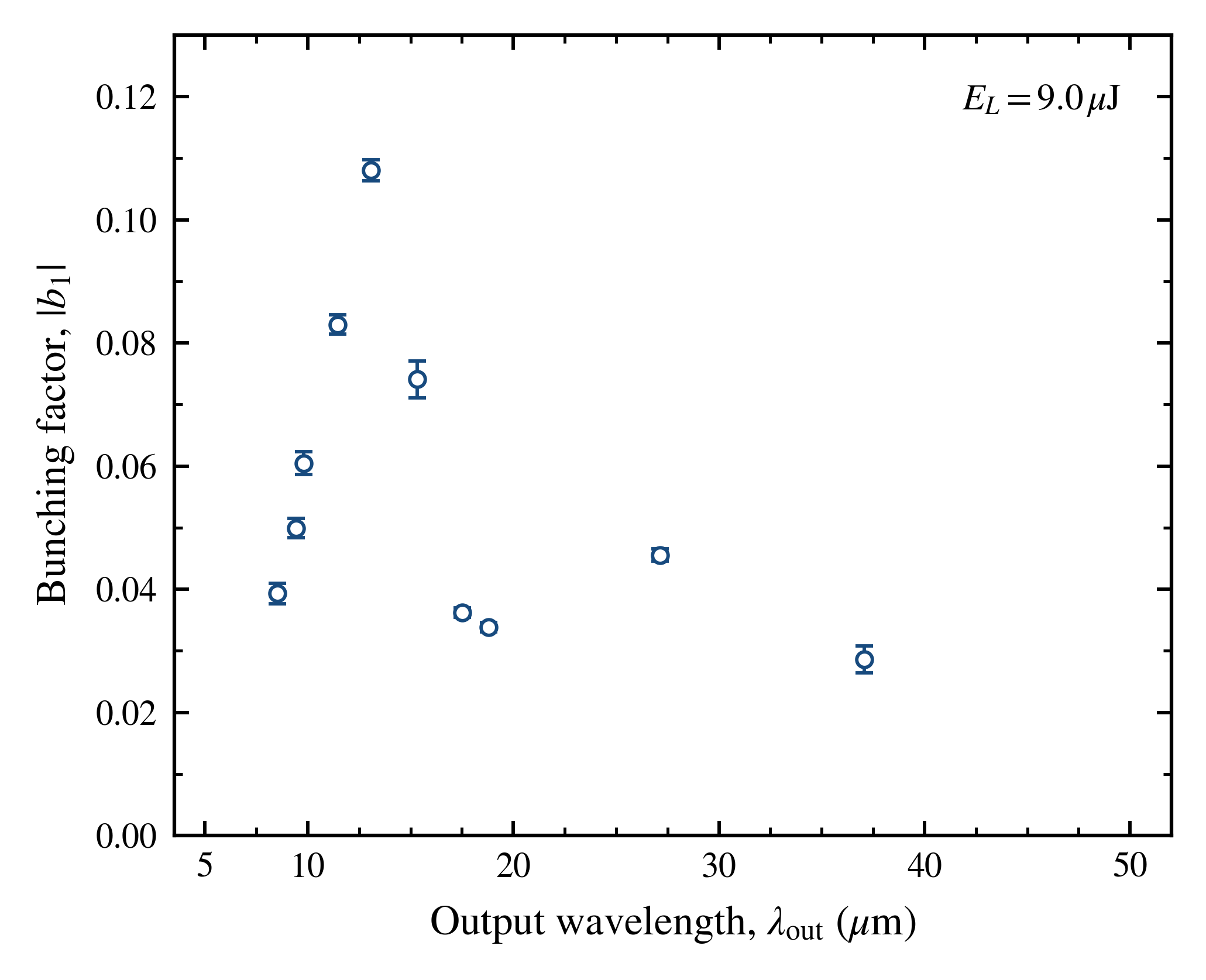}}\hspace{0.01\textwidth}%
\subfloat[]{\includegraphics[width=0.32\textwidth]{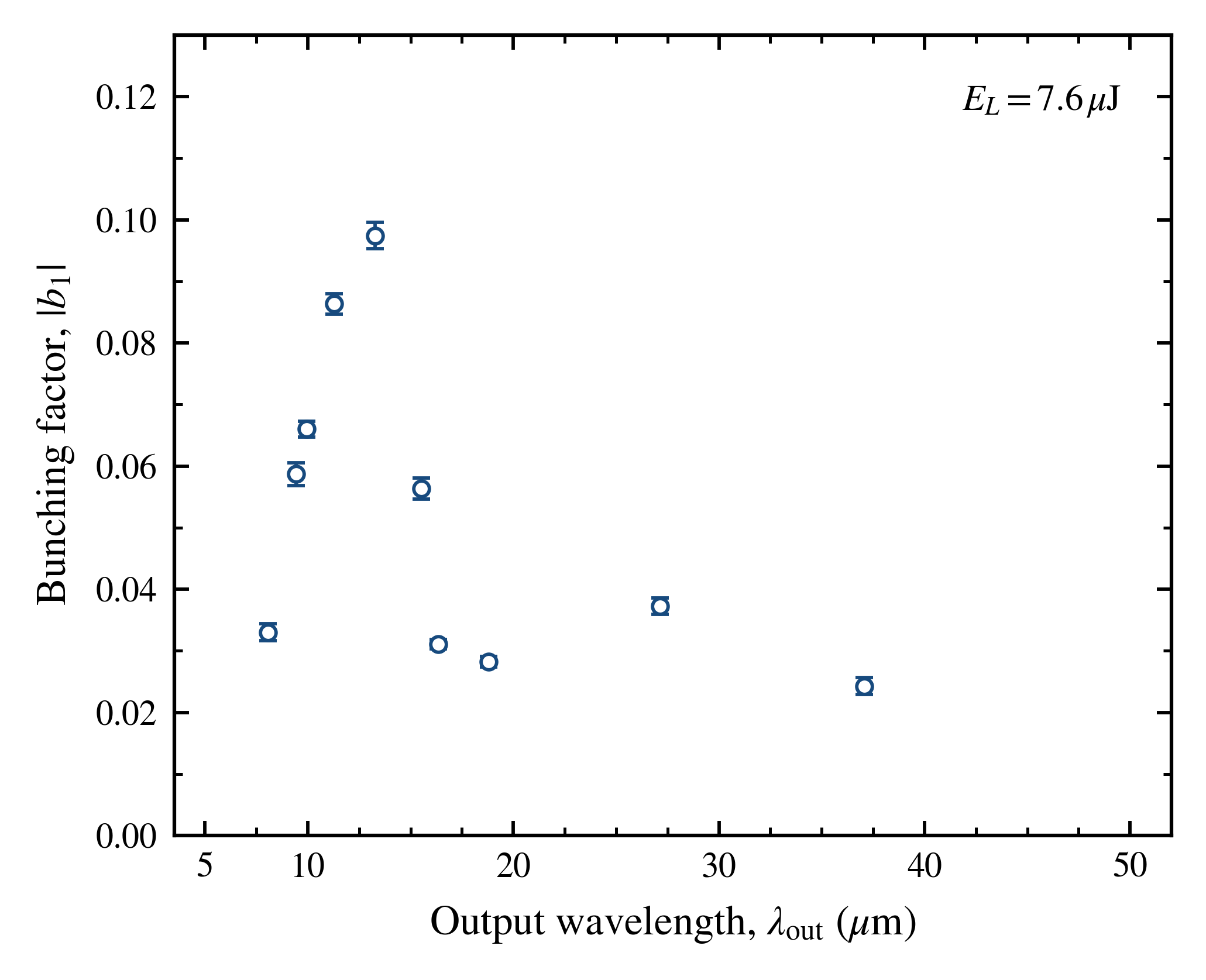}}\\[0.5em]
\subfloat[]{\includegraphics[width=0.32\textwidth]{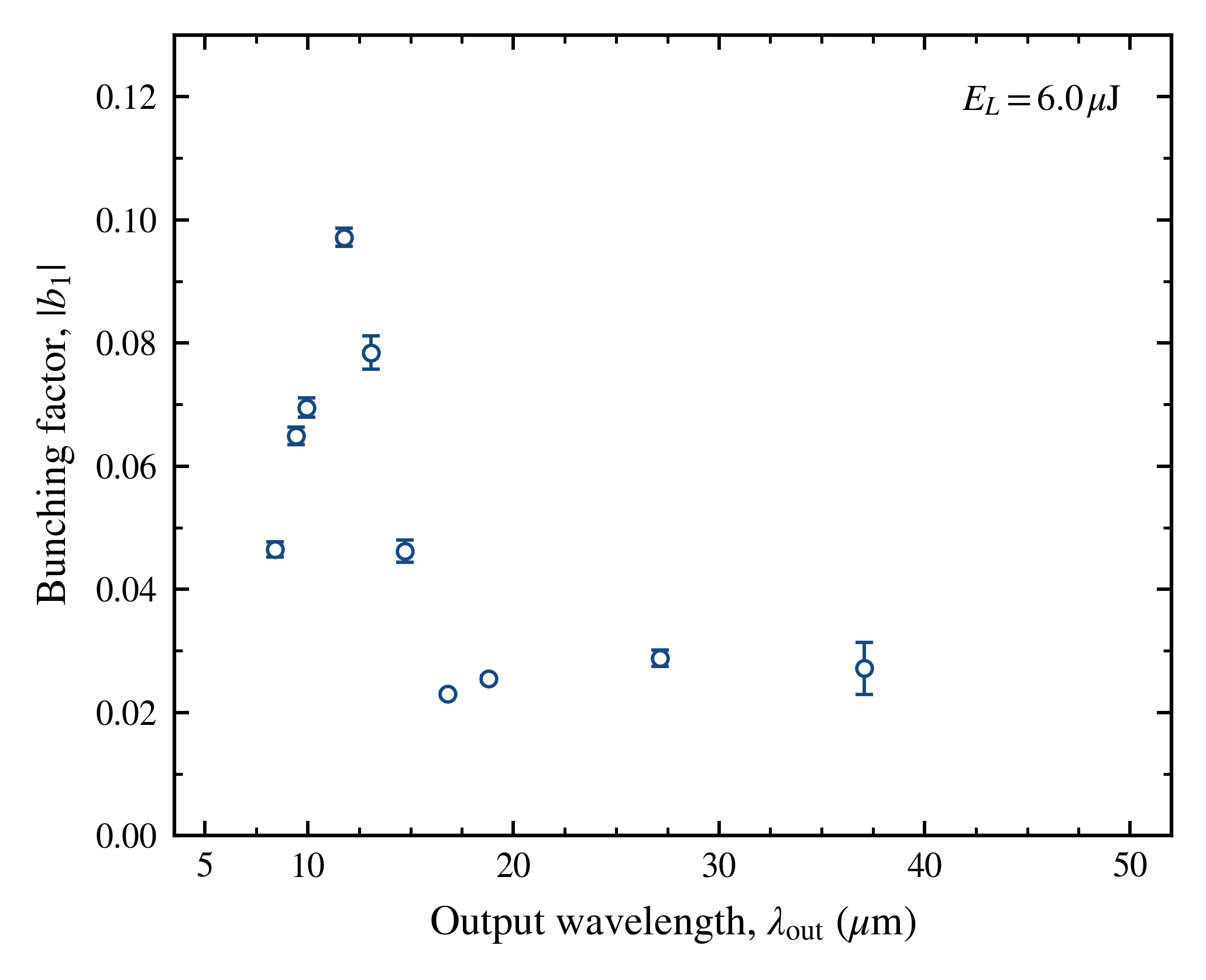}}\hspace{0.01\textwidth}%
\subfloat[]{\includegraphics[width=0.32\textwidth]{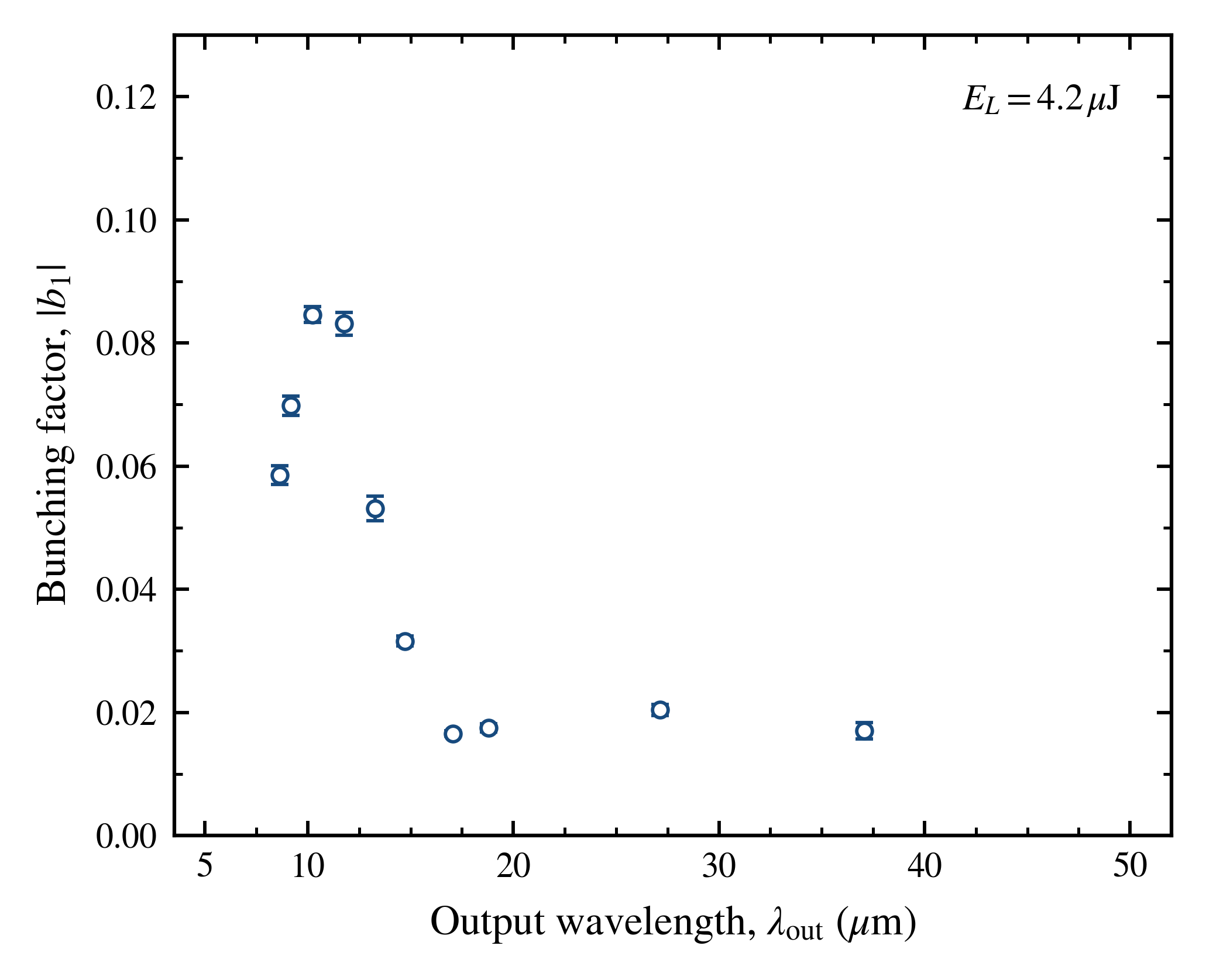}}\hspace{0.01\textwidth}%
\subfloat[]{\includegraphics[width=0.32\textwidth]{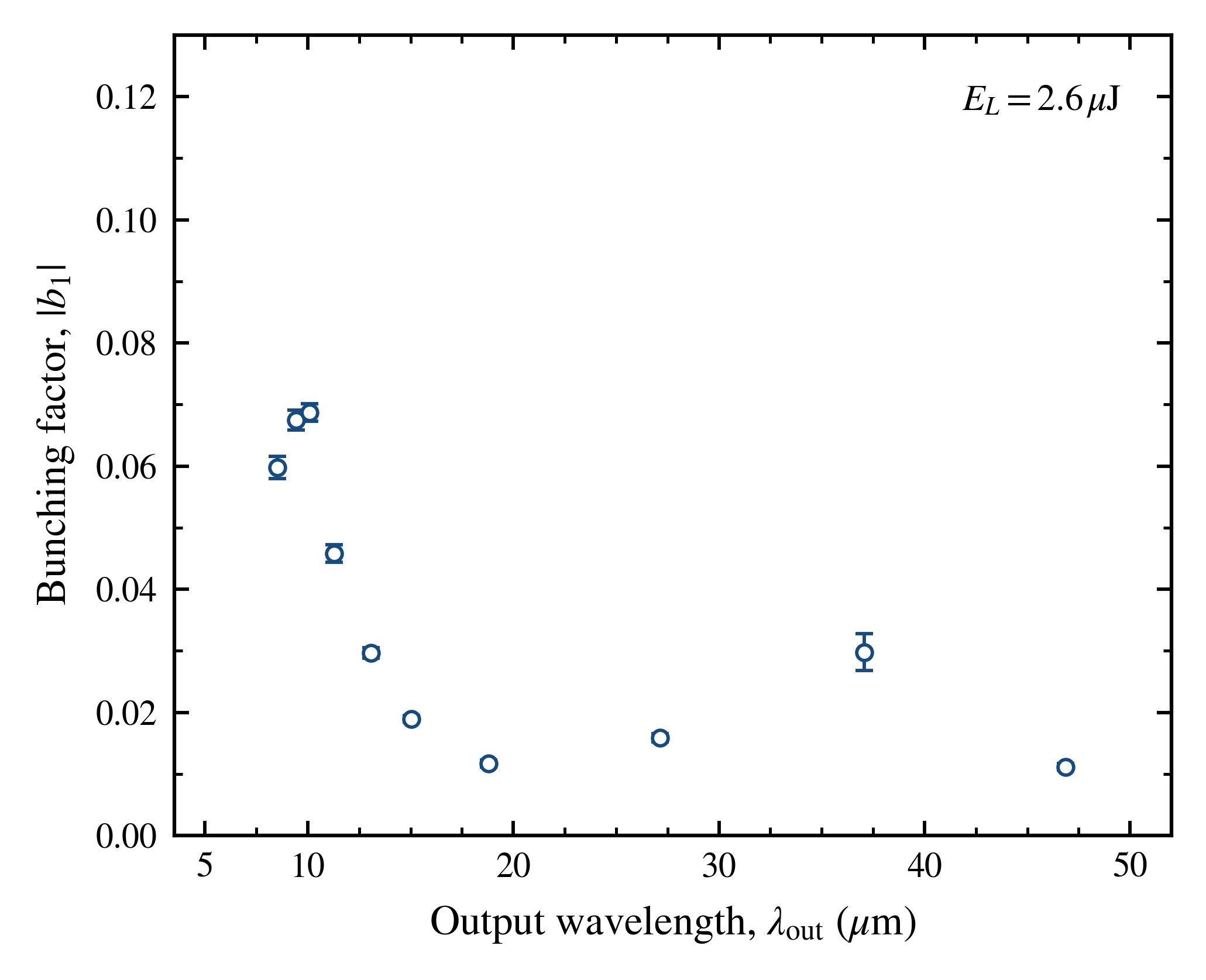}}
\caption{Experimental linac-exit bunching factor $|b_1|$ versus modulation wavelength for beat-frequency scans at $E_L=(9.8,9.0,7.6,6.0,4.2,2.6)\,\mu\mathrm{J}$ in panels (a)--(f), respectively. Open circles denote repeated-shot means, and error bars indicate the standard error of the mean.}
\label{fig:experiment_frequency_spectra}
\end{figure*}

Time-resolved characterization of microbunching has previously been enabled by longitudinal phase-space imaging, while density and energy modulations can be systematically resolved by Fourier-domain methods~\cite{PhysRevSTAB.18.030704,Brynes2020}. Each measured phase-space image is analyzed using the same sequence illustrated for the simulation in Fig.~\ref{fig:exit_case_diagnostic}. The phase-space density is first projected onto the longitudinal coordinate to obtain $I(z)$. A third-order polynomial $I_0(z)$ is fitted over a common central interval, and the slowly varying current envelope is removed according to
\begin{equation}
 \Delta_I(z)=\frac{I(z)-I_0(z)}{I_0(z)}.
 \label{eq:experimental_relative_current}
\end{equation}
The analysis interval is selected such that complete modulation periods are retained and the TDX temporal resolution remains sufficient for the wavelength range of interest. After application of the same window $w(z)$ at every operating point, the experimental bunching spectrum is evaluated as
\begin{equation}
 b_{\rm exp}(k)=
 \frac{\left|\int w(z)\Delta_I(z)e^{-ikz}\,dz\right|}
 {\int w(z)\,dz},
 \qquad \lambda=\frac{2\pi}{k}.
 \label{eq:experimental_fft}
\end{equation}
The standard bunching factor $|b_1|$ is obtained with this normalization; if the discrete one-sided FFT is expressed as a sinusoidal modulation depth, its amplitude is divided by two. Each shot is processed independently, and the mean, sample standard deviation, and standard error are then evaluated from the repeated measurements. In Figs.~\ref{fig:experiment_frequency_spectra} and~\ref{fig:three_way_comparison}, error bars denote the standard error of the mean, $s/\sqrt{N}$, where $s$ is the sample standard deviation and $N$ is the number of repeated shots at each setting. The longitudinal calibration, analysis window, detrending order, and spectral sampling are kept unchanged within each scan.

The wavelength dependence reconstructed from beat-frequency scans at six laser pulse energies is shown in Fig.~\ref{fig:experiment_frequency_spectra}. A maximum at an intermediate wavelength is resolved in each scan, with lower bunching factors on both sides. For $E_L=7.6$--$9.8\,\mu\mathrm{J}$, the largest measured mean bunching factors are approximately $0.10$--$0.11$ at $\lambda_{\rm out}\simeq13\,\mu\mathrm{m}$. As the pulse energy is reduced further, the maximum shifts toward shorter wavelengths and its amplitude decreases overall. At $E_L=\SI{2.6}{\micro\joule}$, the largest measured value is $|b_1|=0.0687\pm0.0014$ at $\lambda_{\rm out}=\SI{10.08}{\micro\metre}$, where the uncertainty denotes the standard error of the mean. The measured response thus depends on both the input beat frequency and the laser pulse energy.

\begin{figure*}[width=\textwidth,pos=tp,align=\centering]
\centering
\subfloat[]{\includegraphics[width=0.24\textwidth]{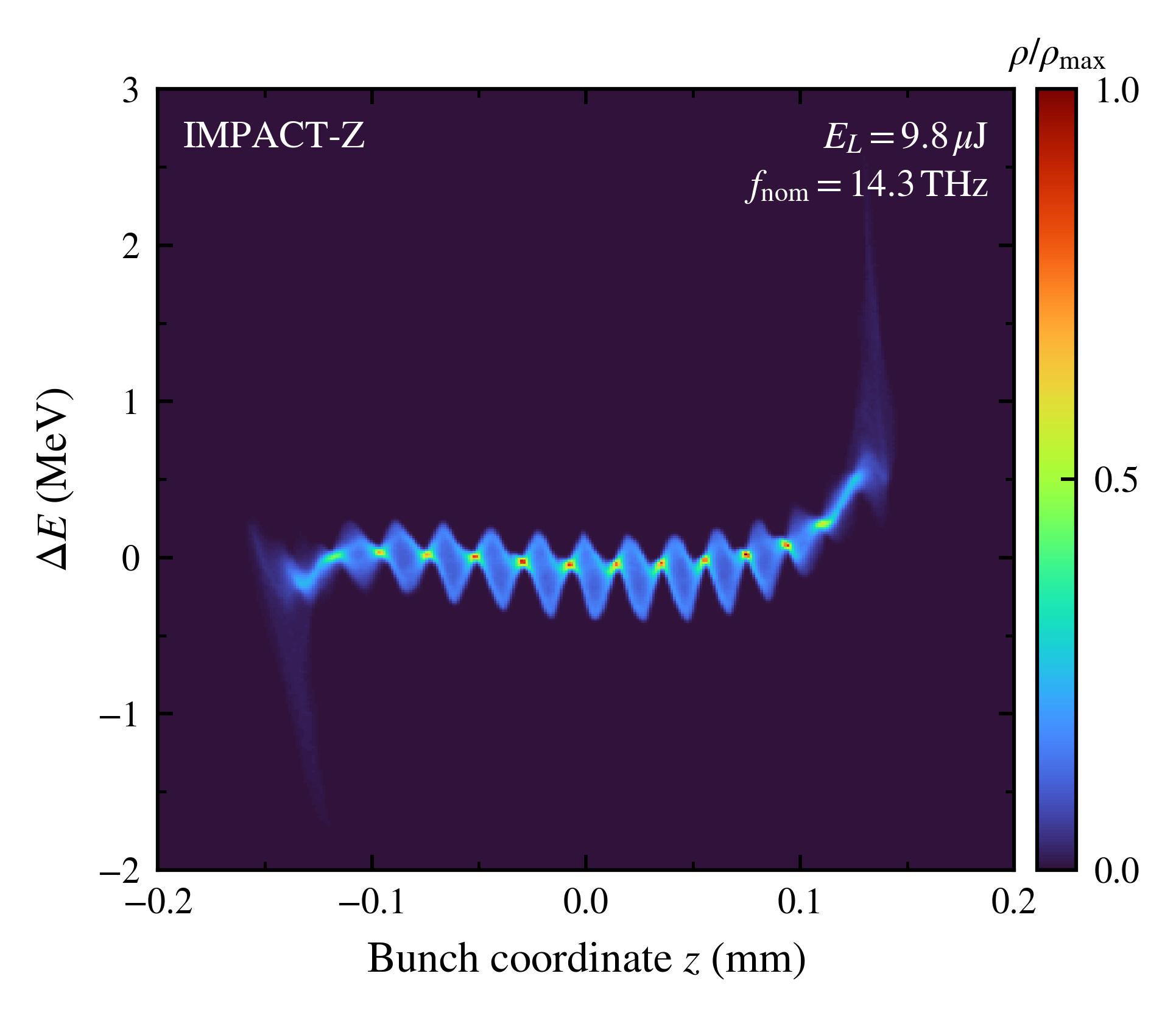}}\hspace{0.01\textwidth}%
\subfloat[]{\includegraphics[width=0.24\textwidth]{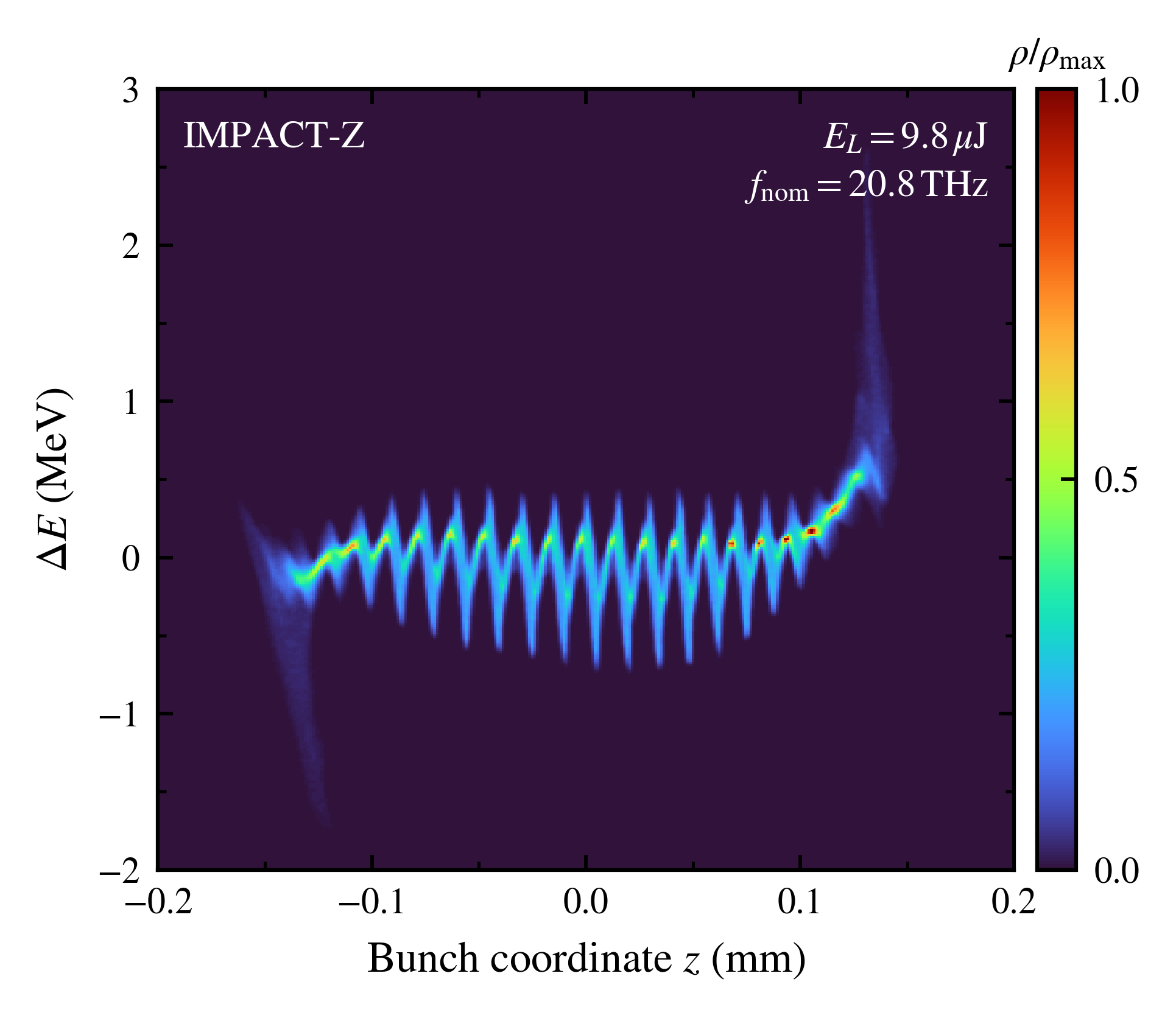}}\hspace{0.01\textwidth}%
\subfloat[]{\includegraphics[width=0.24\textwidth]{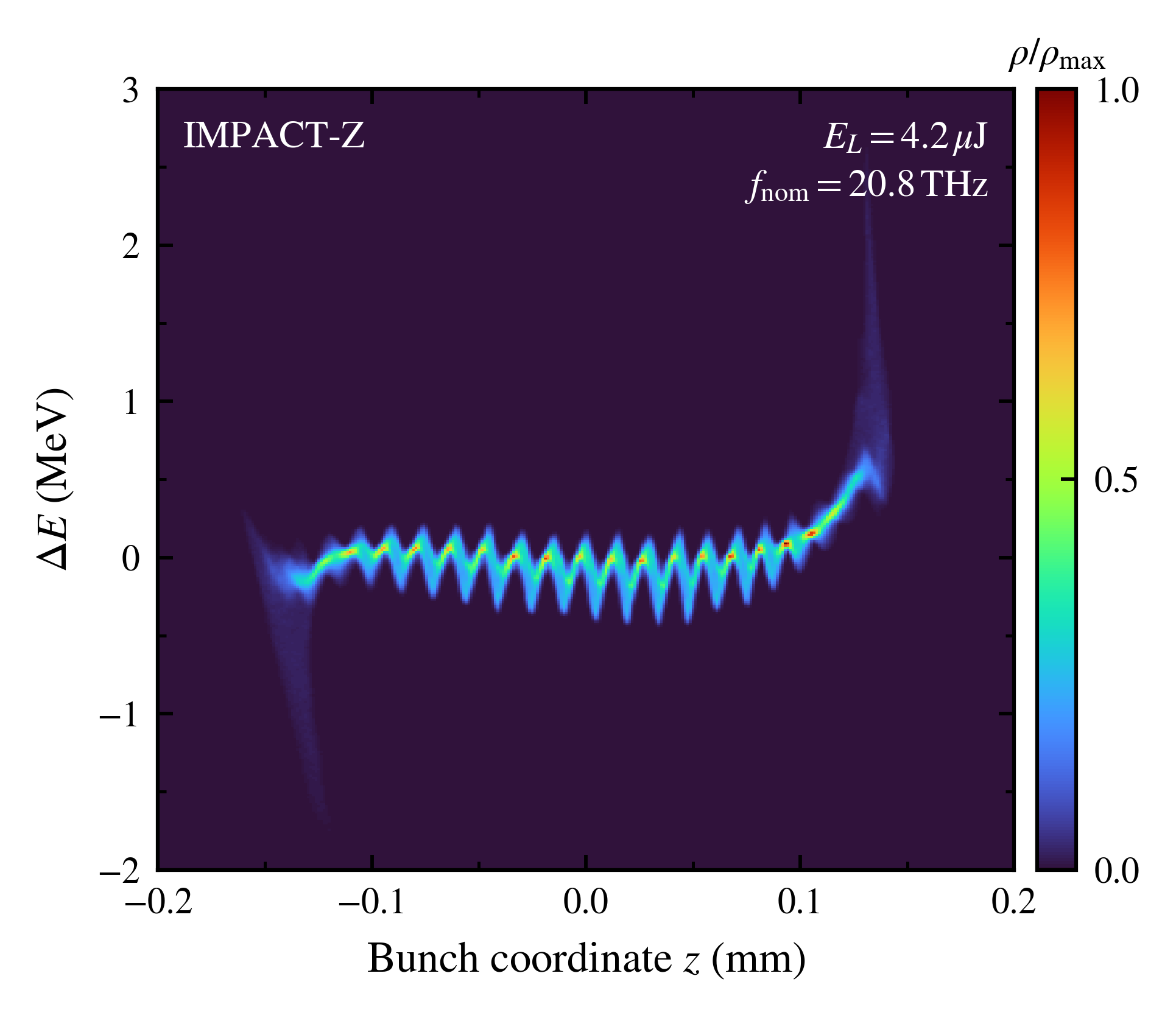}}\hspace{0.01\textwidth}%
\subfloat[]{\includegraphics[width=0.24\textwidth]{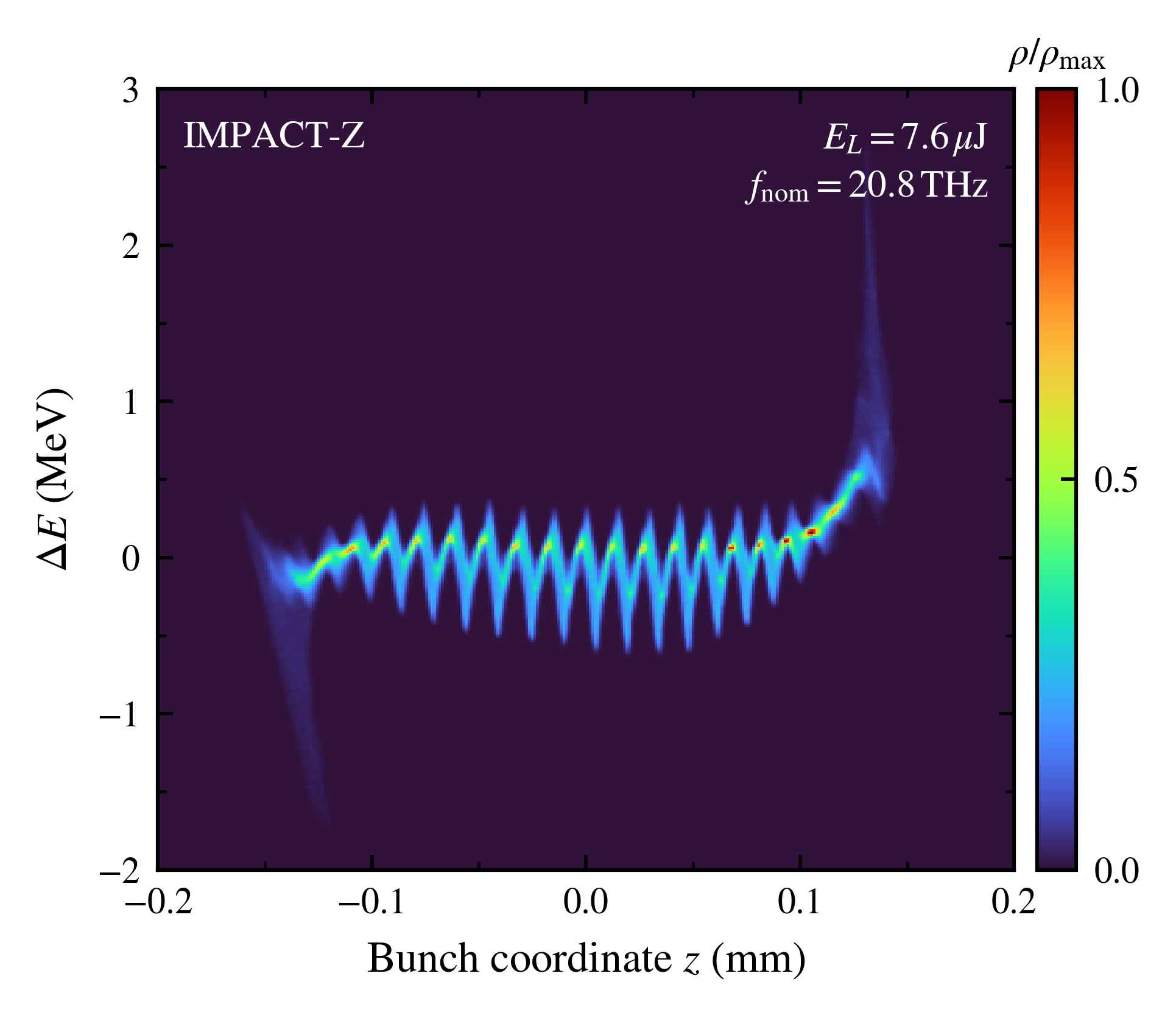}}
\caption{IMPACT-Z linac-exit longitudinal phase spaces at the four operating points in Fig.~\ref{fig:experiment_frequency_phase_space}, in the same panel order, with nominal $C_1=6.5$ and $C_2=10/6.5$. Each density map is normalized to its maximum and displayed with a linear color scale. The labels $f_{\rm nom}=10f_b$ indicate nominal beam modulation frequencies after compression: $14.3\,\mathrm{THz}$ in (a) and $20.8\,\mathrm{THz}$ in (b)--(d).}
\label{fig:simulation_frequency_phase_space}
\end{figure*}

\begin{figure*}[width=\textwidth,pos=tp,align=\centering]
\centering
\subfloat[]{\includegraphics[width=0.32\textwidth]{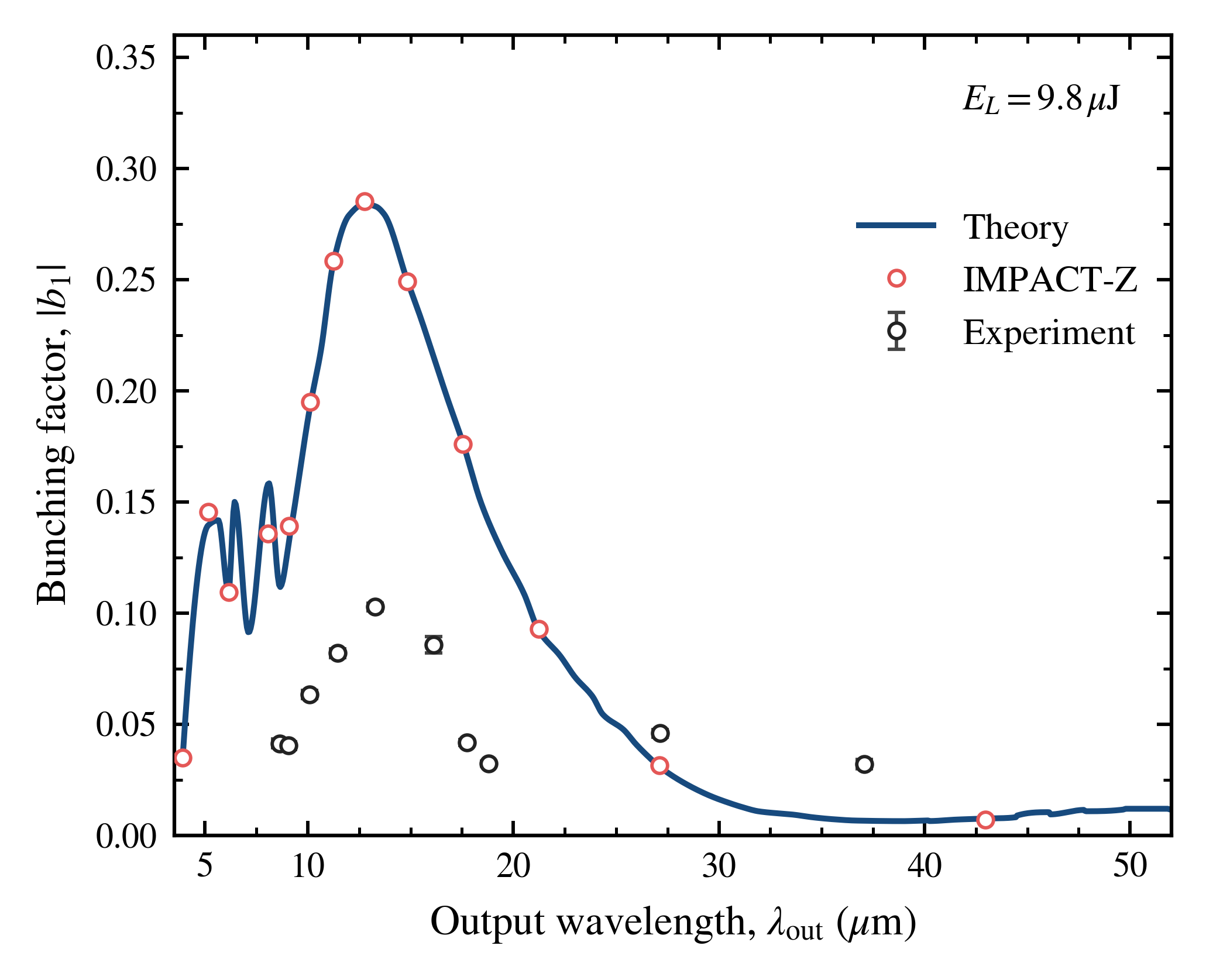}}\hspace{0.01\textwidth}%
\subfloat[]{\includegraphics[width=0.32\textwidth]{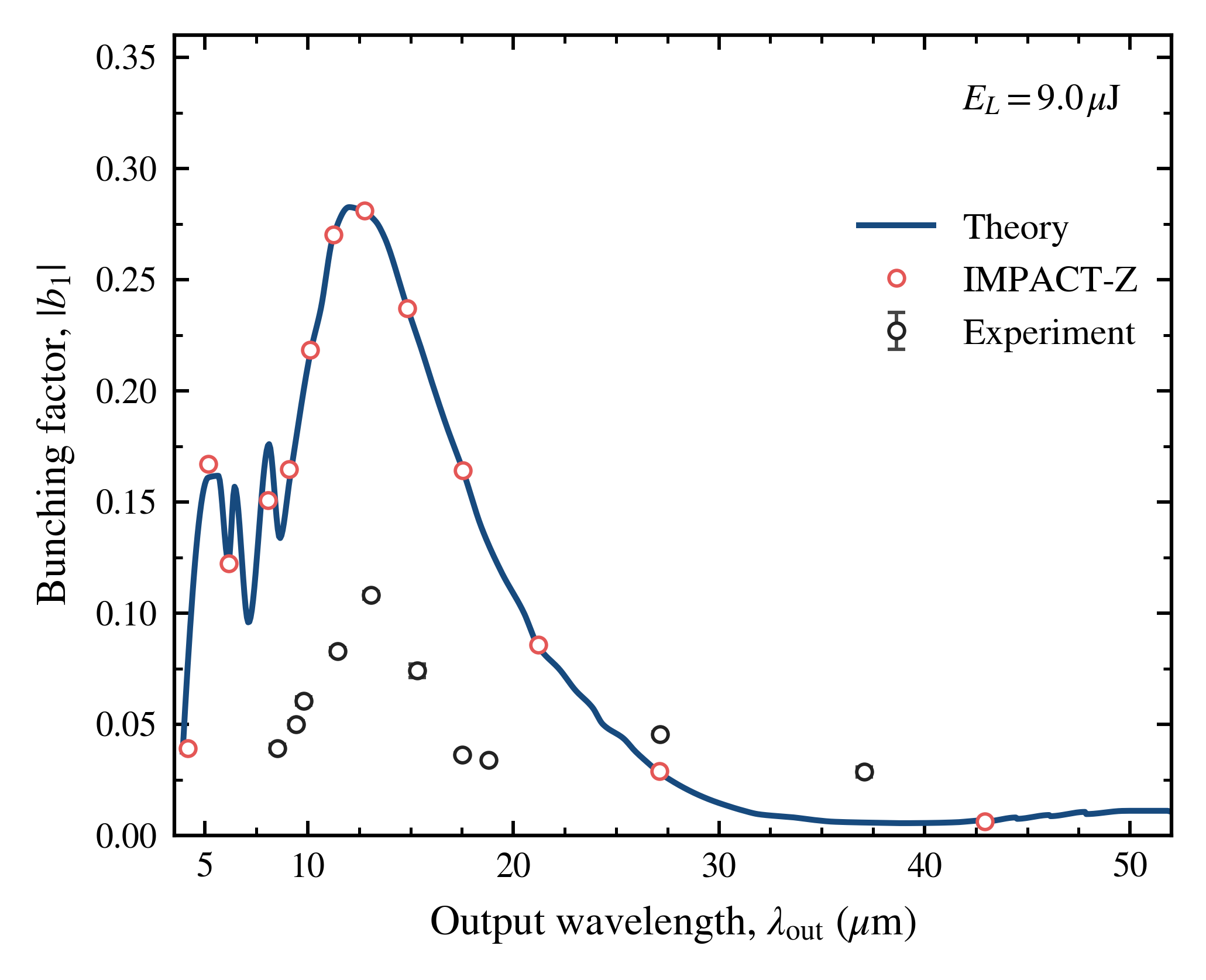}}\hspace{0.01\textwidth}%
\subfloat[]{\includegraphics[width=0.32\textwidth]{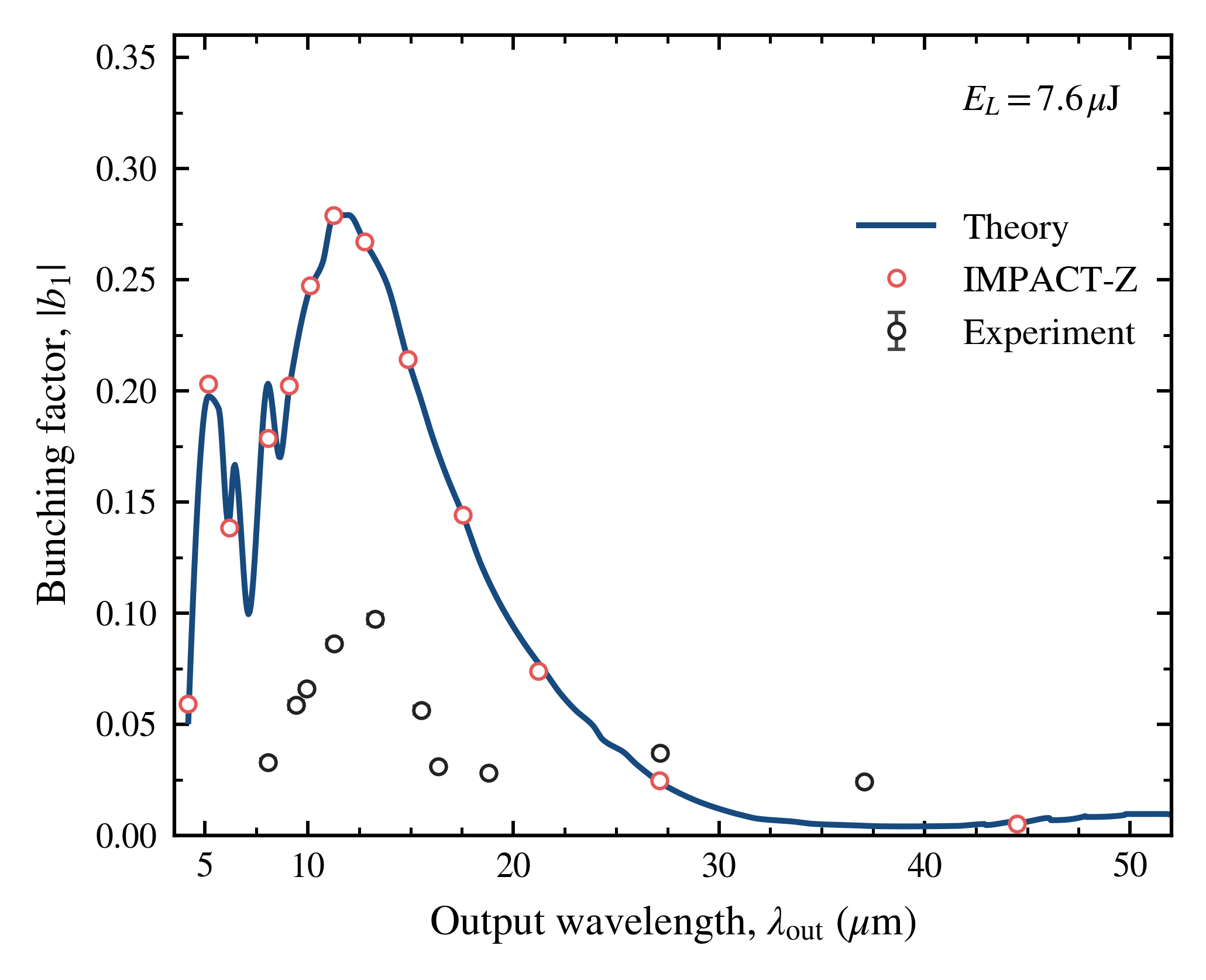}}\\[0.5em]
\subfloat[]{\includegraphics[width=0.32\textwidth]{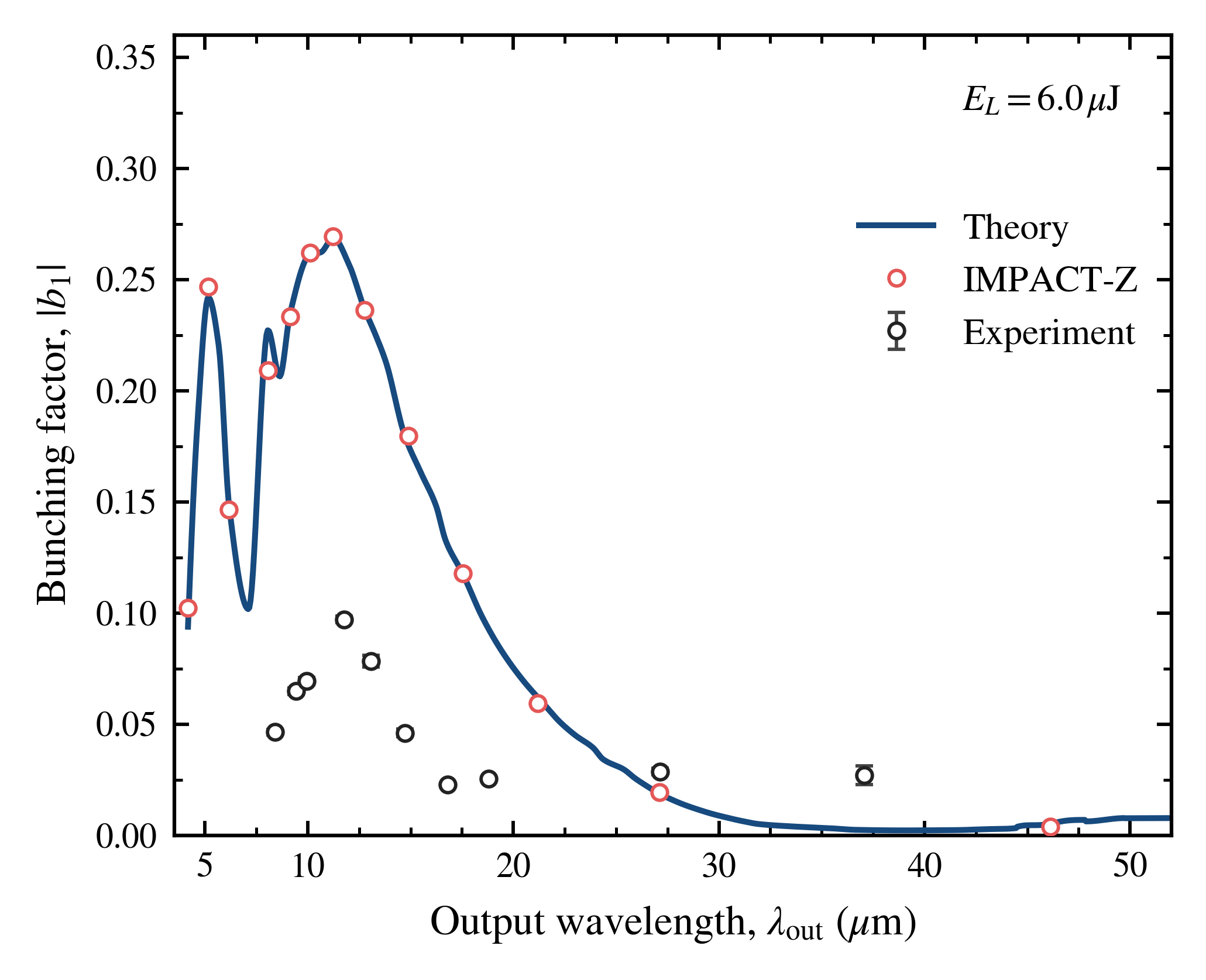}}\hspace{0.01\textwidth}%
\subfloat[]{\includegraphics[width=0.32\textwidth]{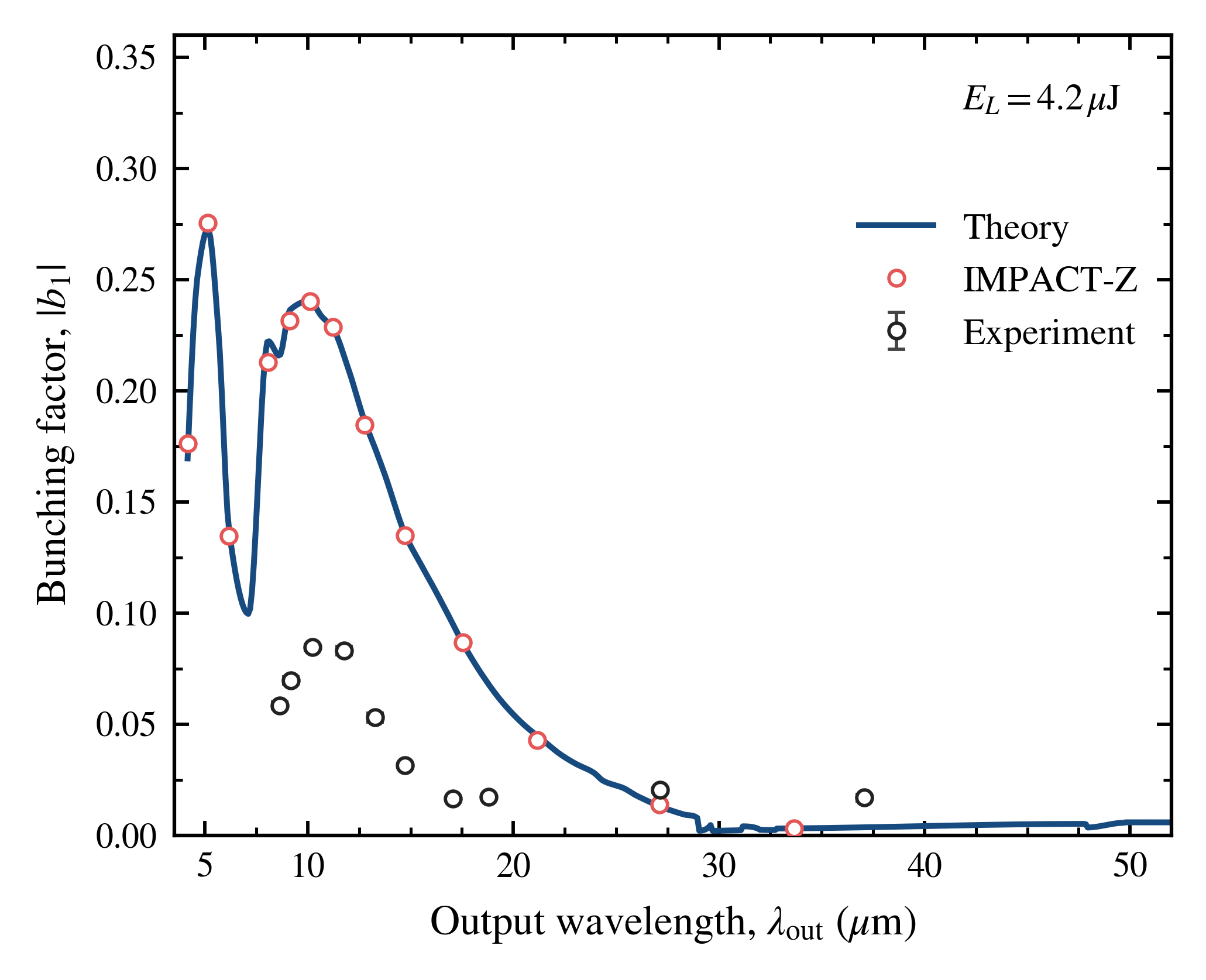}}\hspace{0.01\textwidth}%
\subfloat[]{\includegraphics[width=0.32\textwidth]{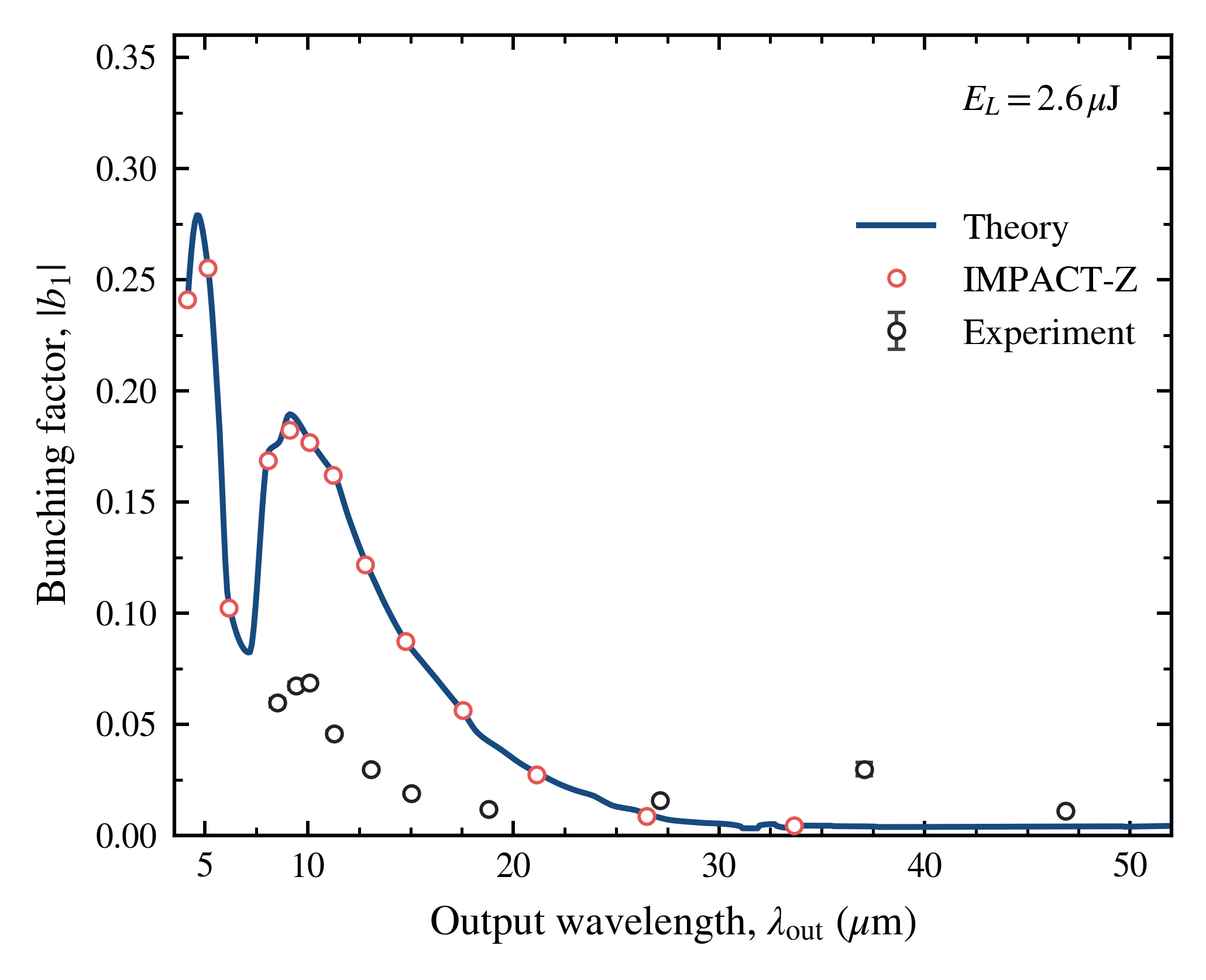}}
\caption{Linac-exit spectral-peak bunching factor $|b_1|$ versus output wavelength from the theory (blue curves), IMPACT-Z (red open circles), and TDX measurements (black open circles). Panels (a)--(f) correspond to $E_L=(9.8,9.0,7.6,6.0,4.2,2.6)\,\mu\mathrm{J}$ at nominal $C_1=6.5$ and $C_2=10/6.5$. Experimental points denote repeated-shot means, with error bars indicating the standard error of the mean.}
\label{fig:three_way_comparison}
\end{figure*}

The corresponding IMPACT-Z phase spaces are shown in Fig.~\ref{fig:simulation_frequency_phase_space}, with the same operating-point order and linear density normalization as the measurements. The nominal compression settings are $C_1=6.5$ and $C_2=10/6.5$. A shorter modulation period is obtained from (a) to (b). A larger energy excursion is obtained from (c) through (d) to (b), consistent with the pulse-energy trend in the measurements. More uniform modulation is obtained in the simulation, whereas the measured distributions exhibit broader and less regular structure. The comparison of modulation spacing and contrast is made with the experimental energy scale cross-calibrated against simulation.

The wavelength-dependent comparison among the theory, IMPACT-Z, and experiment is shown in Fig.~\ref{fig:three_way_comparison} for six laser pulse energies. The theoretical and simulated spectral-peak bunching factors are closely matched at the sampled operating points, including the short-wavelength structure. A broad maximum near $10$--$13\,\mu\mathrm{m}$ is observed in the measurements and is also obtained in the calculations, with its amplitude reduced overall toward lower laser energies. Additional structure is predicted below this wavelength range, where experimental coverage is limited. The measured bunching factors near the broad maximum are systematically lower than the calculated values, so the agreement with experiment concerns the wavelength and energy trends rather than the absolute amplitude.

The broad bunching maximum is increased as the laser pulse energy is raised, but progressively smaller gains are obtained at the highest energies in Fig.~\ref{fig:three_way_comparison}. This approach to saturation is consistent with stronger Landau damping from the increased laser-induced energy spread. The stronger seed modulation is increasingly offset by phase mixing during dispersive transport, which limits further growth of the bunching factor. A higher laser pulse energy therefore does not necessarily yield stronger bunching at the wavelength of interest.

Several experimental effects can contribute to the lower measured bunching amplitude. A stronger and more uniform modulation is obtained in the simulated phase spaces than in the measurements [Figs.~\ref{fig:experiment_frequency_phase_space} and~\ref{fig:simulation_frequency_phase_space}]. Stable transverse overlap between the laser and electron beam is assumed in the calculation, whereas the experimental overlap is affected by beam motion. Fine-scale modulation can also be attenuated by the finite TDX resolution and the background removal used to extract the longitudinal profile. Timing, laser-energy, RF, and orbit fluctuations provide additional sources of variation, while finite detector dynamic range can reduce the measured contrast at large modulation amplitudes. The separate contributions of these effects are not resolved by the present comparison.

\section{Conclusion}
\label{sec:conclusion}

A nonlinear model has been developed to connect laser-heater source formation with collective microbunching evolution through multistage compression. The non-Gaussian source distribution and its phase-space correlations are retained in a common description of optical modulation, dispersive conversion, and collective amplification. The source modulation is closely reproduced relative to Elegant, and good agreement with IMPACT-Z is obtained for the wavelength-dependent bunching response. Efficient parameter scans are enabled by optical phase averaging and reuse of the external transport maps. The framework thus combines agreement with particle tracking and efficient exploration of laser and compression settings. It provides a practical basis for identifying accelerator operating points with enhanced bunching and guiding experimental optimization of coherent THz radiation.

The comparisons establish LSC as the dominant mechanism responsible for the short-wavelength multiple-peak structure under the investigated conditions. RF wakefields and CSR modify the response but do not produce the pronounced double peak on their own. Bunching near a selected wavelength can also be enhanced by redistributing compression between BC1 and BC2 at fixed nominal total compression. The compression partition therefore provides control of the modulation amplitude beyond the wavelength scaling set by total compression. The wavelength-dependent trends at different laser pulse energies are also observed in the SXFEL measurements. The broad bunching maximum initially grows with laser energy and then approaches saturation, consistent with stronger Landau damping from the increased energy spread. Laser strength and compression partition must therefore be considered together when optimizing the bunching response.

In future experiments, this validation will be extended from the electron-beam bunching spectrum to the emitted THz radiation. A controlled two-dimensional scan of the delivered laser pulse energy and the BC1--BC2 compression partition is planned, with bunch charge, final current, and transverse optics constrained. The laser pulse shape and focal size will also be controlled so that changes in delivered energy correspond to changes in heater intensity. The predicted microbunching gain will be tested with the electron-beam diagnostics. Calibrated measurements of the THz pulse energy and spectrum will then quantify the dependence of radiation output, central frequency, and bandwidth on laser strength and compression partition. A direct relation between experimentally adjustable parameters and THz radiation can thereby be established, enabling model-guided optimization of narrow-band, continuously tunable THz generation.

\section*{Acknowledgments}
The authors thank the SXFEL operation and beam-diagnostics teams for their support during the beam experiments. This work was supported by the National Natural Science Foundation of China (No. 12505172, No. 12675417, No. 12275340, No. 12125508, and No. 12541503) and the National Key Research and Development Program of China (No. 2024YFA1612104).

\section*{Declaration of generative AI and AI-assisted technologies in the manuscript preparation process}

During the preparation of this work, the authors used ChatGPT for language editing, assistance with reference and LaTeX formatting, and assistance with figure formatting and visualization of simulation and experimental data.

All simulation and experimental data used in the figures were generated by the authors. The authors reviewed and edited the output as needed and take full responsibility for the content of the published article.

\section*{DATA AVAILABILITY}

The data that support the findings of this article are available from the corresponding author upon reasonable request.

\bibliography{ASTA}

\end{document}